\documentclass[12pt,a4paper]{article}
\pdfoutput=1
\usepackage{graphicx}
\usepackage{tikz}
\usetikzlibrary{decorations.pathreplacing,decorations.markings,snakes}
\usetikzlibrary{decorations.pathmorphing}
\tikzset{snake it/.style={decorate, decoration=snake}}
\usepackage{jheppub}
\usepackage{amsmath,amssymb,euscript,array,mathrsfs,appendix,ctable,mathabx}
\usepackage{bm}
\usepackage[small]{caption}
\usepackage{xcolor}
\usepackage{float}
\usepackage{braket}
\usepackage{comment}
\usepackage{xcolor}
\usepackage{empheq}
\definecolor{lightblue}{RGB}{100,180,255}
\newcommand \arXiv [1]{\href{http://arxiv.org/abs/#1}{\tt arXiv:#1}}

\def\ben
{\begin{equation}}
\def\een{\end{equation}}

\let\a=\alpha    
   
   \let\x=\xi 
   
\let\y=\psi
    \let\L=\Lambda
   
 \let\W=\mu

\def\W={\cal W}
\def\L ={\cal L}

\def\be{\begin{equation}}
\def\ee{\end{equation}}
\def\ba{\begin{array}}
\def\ea{\end{array}}

\def\dalemb#1#2{{\vbox{\hrule height .#2pt
        \hbox{\vrule width.#2pt height#1pt \kern#1pt
                \vrule width.#2pt}
        \hrule height.#2pt}}}

\newcommand{\bea}{\begin{eqnarray}}
\newcommand{\eea}{\end{eqnarray}}

\newcommand{\bQ}{{\bf Q}}

\title{Probing Evaporating Black Holes with Modular Flow in SYK }
\author{Nicol\`o Bragagnolo and S. Prem Kumar}
\affiliation{Centre for Quantum Fields and Gravity, \\
Department of Physics,\\
Swansea University,\\
Singleton Park, Swansea,\\
SA2 8PP, U.K.}
\emailAdd{nicolo.bragagnolo@swansea.ac.uk, s.p.kumar@swansea.ac.uk}
\abstract{We study the effect of modular flow on correlation functions of fermions in the Sachdev-Ye-Kitaev (SYK) model coupled weakly to a bath, which we take to be another SYK model. The system and bath, together are prepared in the thermofield double (TFD) state, and we focus on the effect of modular flow generated by the reduced density matrix for the SYK system, obtained by tracing out the bath. We show, in the late time limit, that modular flowed correlators of two Majorana fermions, single-sided and two-sided, exhibit non-trivial singularities. Beyond a critical value of the modular parameter, the ``modular scrambling time", the singularity structure shows correlations being transferred from one boundary to the other. The calculations are performed by employing  the replica trick in Euclidean time and appropriately analytically continuing to real time. Exploiting the connection between modular flow generators and SL$(2,{\mathbb R})$ boosts we use the  microscopic picture to reconstruct the dual bulk modular flow in two-sided AdS$_2$ black hole spacetime. Fixed points of the flow allow to identify quantum extremal surfaces (QES) demarcating the entanglement wedge of the boundary system and the island. We show that bulk modular flow can move fermion insertions near the right  boundary past the horizon leading to lightcone singularities in appropriately smeared boundary correlators, probing physics beyond the horizon.}

\begin{document}
\maketitle
\flushbottom
\section{Introduction}
Exploring the physics behind the horizon of a black hole has long been viewed as the route to providing a glimpse into some of the mysteries of quantum gravity, the nature of the black hole singularity and its resolution. With the discovery of  holographic dualities and the AdS/CFT correspondence \cite{Maldacena:1997re, Gubser:1998bc, Witten:1998qj} some of these questions were brought into sharp focus by framing them in terms correlators in dual conformal field theory (CFT) \cite{Fidkowski:2003nf, Festuccia:2005pi}. More recent developments underlining the nontrivial entanglement structure of old evaporating black holes, point to new perspectives on this question. 
The entanglement structure of evaporating black holes undergoes a transition at the Page time \cite{Page:1993wv, Page:1993df} when correlations between early and late time Hawking quanta build up causing the von Neumann entropy of entanglement of the radiation to decrease  and follow the Page curve. In a post Page time black hole (coupled to a reservoir), semiclassical gravity accounts for this by the appearance of a quantum extremal surface (QES) demarcating an island inside the black hole, the island itself having been scrambled and evaporated away into the radiation \cite{Penington:2019npb, Almheiri:2019qdq, Penington:2019kki}.\footnote{The appearance of QES and associated islands in the semiclassical description is established in the presence of absorptive radiation baths coupled to the system. The relevance of the Page curve and islands for black holes in asymptotically flat space and theories with a massless graviton is an intensely studied question \cite{Antonini:2025sur, Geng:2025byh}. } A diary carrying a small entropy and energy when thrown into such a  black hole, will be scrambled and radiated away as soon as it reaches the island i.e within a scrambling time \cite{Hayden:2007cs, Penington:2019npb}. It is  natural to ask if the interior of the black hole has changed in some way that can be probed either from the perspective of the infalling diary (observer) itself, or from the viewpoint of the dual microscopic system. This question is of particular interest, since post Page time, the dimension of the microscopic Hilbert space associated to the black hole is much smaller than the dimension of the Hilbert space of semiclassical QFT excitations in the black hole interior \cite{Akers:2022qdl, Iliesiu:2024cnh}.

In situations without evaporation, a  powerful framework to reconstruct the experience of an infalling observer within the   holographic boundary CFT is offered by modular flow \cite{Jafferis:2020ora, deBoer:2022zps, Gao:2021tzr}. Modular flow in this case is generated by the modular Hamiltonian $K=-\log\rho_r$, specified by the reduced density matrix $\rho_r$ for the black hole, obtained by tracing out a reference system (the infalling observer) to which the black hole is coupled. For an evaporating black hole, the system must be further coupled to an absorptive reservoir.
As a first step toward extending  to the evaporating case, in this paper we pose a simpler question (without infalling observer): can  boundary correlators see behind the  horizon of an evaporating black hole when subjected to  modular flow? 

We take two Sachdev-Ye-Kitaev (SYK) models \cite{Sachdev:1992fk, kitaev, Sachdev:2015efa, Kitaev:2017awl, Maldacena:2016hyu}  of Majorana fermions $\chi$ and $\psi$, coupled together weakly, and prepared in the thermofield double (TFD) state. Tracing out the SYK$_\psi$ system which plays the role of the bath, yields  the state of the SYK$_\chi$ system described by a reduced density matrix. Coupled SYK systems of this nature were considered in \cite{Gu:2017njx}, and specifically in  \cite{Penington:2019npb} to understand the emergence of a Page curve for the R\'enyi 2-entropy within a UV complete microscopic framework (see also \cite{Dadras:2020xfl}).

The non-zero interaction between system and bath implies that the reduced density matrix is non-trivial and that the correlation functions of Majorana fermions, both single-sided and two-sided will exhibit departures from those in the pure TFD state. We want to understand  the effect of modular flow on generic fermion correlators 
\be
W^{(s)}(t_1, t_2)={\rm Tr}\left[{\rho}_r\, \rho_r^{-is}\,\chi(t_1)\,\rho_r^{is}\chi(t_2)\right]\,,\label{2ptsimple}
\ee
generated by the reduced density matrix. In practice, there is no obvious direct route to compute such correlation functions involving complex powers of the density matrix. We must instead employ a replica trick to first evaluate
correlation functions of the type,
\be
W^{(k, s)}(t_1, t_2)={\rm Tr}\left[ \rho_r^{k-s}\,\chi(t_1)\,\rho_r^{s}\chi_1(t_2)\right]\,,\qquad {s,k}\in{\mathbb Z}\,,\label{correx}
\ee
and eventually set $k=1$ and perform the analytic continuation $s\to i s$, from integers to the complex plane. We therefore need an analytical handle on the  coupled SYK systems. 
\begin{figure}
\begin{center}
\begin{tikzpicture}[scale=1.2, >=stealth]

  \draw[->, very thick] (-5.5,-0.5)--(0.5,5.5) node[anchor=west] {$V$};
  \draw[->, very thick] (0.5,-0.5)--(-5.5,5.5) node[anchor=east] {$U$};

  \fill[yellow, opacity=0.5] (-2,2.5)--(0,0.5)--(-0.4,1.4)--(-0.8, 2.5)--(-0.4,3.6)--(0,4.5)--cycle;

  \draw[thick, dashed, blue] (0,0.5)--(-5.3,5.8);
  \draw[thick, dashed, blue] (0,4.5)--(-5.3,-0.8);

  \draw [gray, line width=0.7mm, dashed] plot[smooth]coordinates {(0.3, 0) (0,0.5) (-0.8, 2.5) (0, 4.5) (0.3, 5)};
  \draw [gray, line width=0.7mm, dashed] plot[smooth]coordinates {(-5.3,0) (-4.3,2.5) (-5.4,5)};

  \draw [green, line width=0.7mm, postaction={decorate},
    decoration={markings, mark=at position 0.6 with {\arrow{>}}}] plot[smooth]coordinates {(0,0.5) (-1,2.5) (0, 4.5)};
  \draw (-1,2.5) node[circle,fill,inner sep=1pt]{};

  \draw[red, line width =0.7mm, postaction={decorate},
    decoration={markings, mark=at position 0.5 with {\arrow{<}}}] (0,4.5)--(-0.5,4);
  \draw[red, line width =0.7mm, postaction={decorate},
    decoration={markings, mark=at position 0.5 with {\arrow{<}}}] (-0.5,4)--(-2,2.5);
  \draw (-0.5,4) node[circle,fill,inner sep=1pt]{};

  \draw [orange, line width=0.7mm, postaction={decorate},
    decoration={markings, mark=at position 0.6 with {\arrow{<}}}] plot[smooth]coordinates {(0, 4.5) (-0.2,4.6) (0.5,5.7)};
  \draw (-0.2,4.6) node[circle,fill,inner sep=1pt]{};
  \draw [orange, line width=0.7mm, postaction={decorate},
    decoration={markings, mark=at position 0.6 with {\arrow{<}}}] plot[smooth]coordinates {(-5.5,0) (-4.5,2.5) (-5.6,5)};
  \draw [orange, line width=0.7mm, postaction={decorate},
    decoration={markings, mark=at position 0.5 with {\arrow{<}}}] plot[smooth]coordinates {(0.4, -0.6) (0.2,-0.2) (0,0.5)};

  \draw [blue, line width=0.7mm, postaction={decorate},
    decoration={markings, mark=at position 0.85 with {\arrow{<}}}] plot[smooth]coordinates {(0,4.5) (-0.4,4.25) (-0.5,4.2)};
  \draw (-0.5,4.2) node[circle,fill,inner sep=1pt]{};
  \draw [blue, line width=0.7mm, postaction={decorate},
    decoration={markings, mark=at position 0.8 with {\arrow{<}}}] plot[smooth]coordinates {(-0.5,4.2) (-2.3,3.5) (-5.1,5.8)};
  \draw [blue, line width=0.7mm, postaction={decorate}, decoration={markings, mark=at position 0.63 with {\arrow{<}}}] plot[smooth]coordinates {(0.6, 0.1) (0.2,0.3) (0,0.5)};

  \draw (-2,2.5) node[circle,fill,inner sep=1.8pt]{};
  \node at (-1.5,2.5) {$a_{R}$};
  
\end{tikzpicture}
\end{center}
\caption{\footnotesize Different trajectories under modular flow (generated by the reduced density matrix) of a right Majorana fermion inserted at a cutoff radial coordinate in AdS$_2$, at different temporal regimes $t_R$, in the sector with twist field $T_{R}$ in the right copy. The AdS$_{2}$ conformal boundaries are the dashed gray lines, and the yellow region represents the entanglement wedge of the right boundary points within a scrambling time. The location of the right QES, the fixed point of the flow, is labeled as $a_R$. For further details, we refer the reader to the summary of results below and to Section \ref{TRflownew}.}
\label{figRflow}
\end{figure}
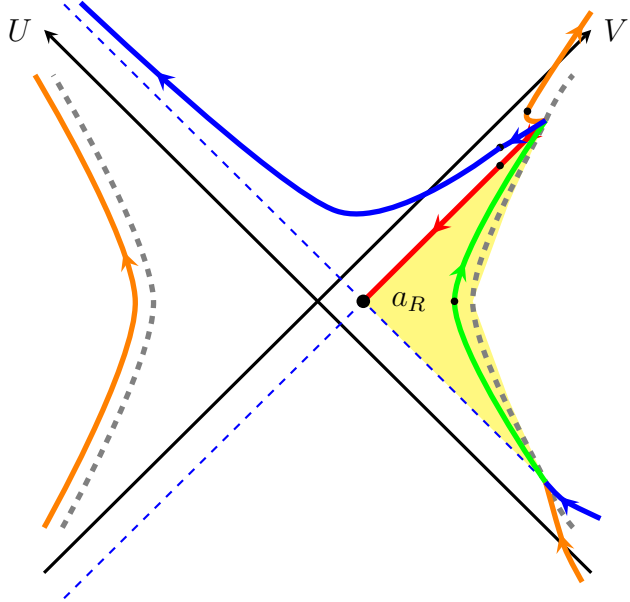
As in the case of the usual SYK model, in the limit of large number of fermion flavors ($N_{\chi,\psi}\to \infty$) in both systems, the path integral for the coupled system localizes onto a saddle point. Solving the coupled saddle point equations analytically is, however, hard. We can make these tractable if we take each SYK to be built from  $q$-fermion random interactions \cite{Maldacena:2016hyu}, {\em and} the coupling between the two systems also $q$-fermion random (treating the two systems democratically), and then  further taking the limit of large $q$. In this limit, each SYK system displays the well known  emergent IR conformal symmetry, and the interaction between the two systems gets neatly packaged into a new effective coupling for each. At first sight the effect of coupling the two systems might not appear to have produced a nontrivial effect. This is indeed the case for the replica diagonal sector of the replica partition function. 

\begin{figure}
\begin{center}
\begin{tikzpicture}[scale=1, >=stealth]

  \draw[->, very thick] (-5.5,-0.5)--(0.5,5.5) node[anchor=west] {$V$};
  \draw[->, very thick] (0.5,-0.5)--(-5.5,5.5) node[anchor=east] {$U$};

  \fill[yellow, opacity=0.5] (-3.5,2.5)--(-5.5,0.5)--(-5,1.5)--(-4.7,2.5)--(-5,3.5)--(-5.5,4.5)--cycle;

  \draw (-3.5,2.5) node[circle,fill,inner sep=1.8pt]{};
  \node at (-3,2.5) {$a_{L}$};
  \draw[thick, dashed, blue] (-5.5,0.5)--(0,6);
  \draw[thick, dashed, blue] (-5.5,4.5)--(0, -1);

  \draw [gray, line width=0.7mm, dashed] plot[smooth]coordinates {(0.3, 0) (0,0.5) (-0.8, 2.5) (0, 4.5) (0.3, 5)};
  \draw [gray, line width=0.7mm,dashed] plot[smooth]coordinates {(-5.8,0) (-5.5,0.5) (-4.7,2.5) (-5.5, 4.5) (-5.8,5)};

  \draw [red, line width=0.7mm, postaction={decorate},
    decoration={markings, mark=at position 0.35 with {\arrow{<}}}] plot[smooth]coordinates {(0.2, -0.8) (-1.2, 2.5) (0.2,5.8)};
  \draw (-1.2,2.5) node[circle,fill,inner sep=1pt]{};
  \draw [red, line width=0.7mm, postaction={decorate},
    decoration={markings, mark=at position 0.5 with {\arrow{<}}}] plot[smooth]coordinates {(-6, 0) (-5.7, 0.2) (-5.5,0.5)};
  \draw [red, line width=0.7mm, postaction={decorate},
    decoration={markings, mark=at position 0.7 with {\arrow{<}}}] plot[smooth]coordinates {(-5.5,4.5) (-5.7, 4.8) (-6, 5.1)};
  \end{tikzpicture}
\end{center}
\caption{Modular flowed trajectory (red line) of a Majorana fermion inserted on the right boundary of the cut off AdS$_2$ geometry at $t_R=0$ in the sector with twist field $T_{L}$ on the left copy. The positions of the AdS$_{2}$ conformal boundaries are denoted in dashed gray and the yellow region is the entanglement wedge of the left boundary points within a scrambling time. $a_L$  denotes the location of the left QES, which is the fixed point of the flow. For further details, we refer the reader to the summary of results below and to Section \ref{TLflownew}.}
\label{figLflow}
\end{figure}
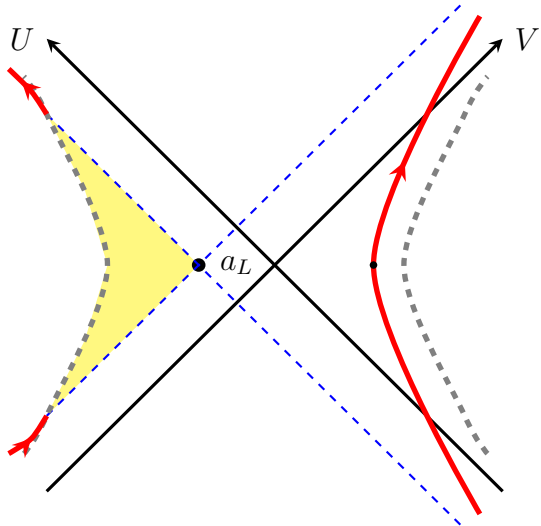

The replica partition function which yields moments of the reduced density matrix is obtained by gluing together copies of the SYK$_\chi$ system across a (complex) time contour with the boundary values of the fermions  in adjacent replicas identified at the gluing point, yielding the so-called twisted boundary conditions (figure \ref{fig3}). The reservoir system which is traced out is replicated trivially and the system-bath interactions are forced to be diagonal in replica space. Interestingly then, it is actually correlation functions of the type \eqref{correx} which straddle fermion insertions in different replicas (therefore replica non-diagonal), which sense the interaction between system and bath in a nontrivial way. This is due to the twist field boundary conditions which glue adjacent replicas. The problem of determining the replica non-diagonal contributions to fermion Green's functions is solved by imposing twisted boundary conditions on the solutions to the saddle point conditions which take the form of the Liouville equation. The solutions to the latter are known up to SL$(2,{\mathbb R})$ transformations whilst the boundary conditions themselves relate solutions in successive replica sheets by a M\"obius or SL$(2,{\mathbb R})$ transform. 

Two key ingredients make possible the  solution of replica non-diagonal contributions. The first is the late time limit, or post Page time limit, when the the two ``twist field" insertions (one for each copy of the TFD state) factorise. The replica boundary conditions can be thought of as insertions of local twist field operators, with  a correlator whose connected contribution decays exponentially, leaving behind a constant floor value at late times determined by the one-point functions of the twist fields \cite{Penington:2019npb, Chen:2020wiq, Sully:2020pza}. The factorisation of twist fields then allows us to recast the boundary conditions on the correlators as  nontrivial recursive relations that we are able to solve in the limit of prametrically small coupling between  bath and system, and perform the necessary analytic continuation from integer $s$ to complex values.
We summarise our results below:

\paragraph{Modular flowed correlators and singularities:} Under modular flow, we find that both single-sided and two-sided  SYK$_\chi$ correlators involving the two copies, left ($L$) and right ($R$), of the thermofield double reveal new real time singularity structures. Indeed, one of our inferences will be that modular flow transfers correlations from $R$ to $L$ (and vice versa) by ``pulling" operators from one side to the other.

Both correlators receive separate contributions from the factorised left and right twist field sectors. In the absence of bath interactions ($\beta {\cal V} \to 0$), the two contributions become identical and modular flow  coincides with time translation of the fermion coordinates in the two point correlator \eqref{2ptsimple}. Accordingly, in the {\em absence} of interactions, the single-sided correlator exhibits singularities when $t_2 = t_1+\beta s$ when the two fermion insertions are brought together by modular flow, while the two-sided correlator where each fermion is placed in one copy of the TFD exhibits no singularities for any $s$. 

With non-zero interaction strength $\beta {\cal V}$ between system and bath, the modular flow is non-trivial and the {\em single-sided} correlator exhibits {\em two distinct} singularities. It is instructive to take  the insertion time $t_1$ to vanish. Then, we find two non-trivial singularities at $t_2={t}_\pm(s)$. The location of one of these, at $t={t}_+(s)$, diverges as $|s|$ approaches a critical value which we call the modular scrambling time,
\be
s_{\rm scr}=\frac{1}{2\pi}\log\frac{4\pi^2}{\beta^2{\cal V}^2}\,.
\ee
This modular parameter scale also signals the onset of maximal modular chaos \cite{DeBoer:2019kdj, Chandrasekaran:2021tkb, Chandrasekaran:2022qmq} in a way that closely parallels the chaos bound of Maldacena-Shenker-Stanford for real time evolution \cite{Maldacena:2015waa}.
The  second singularity approaches a finite value $|{t}_-(s)|\to t_{\rm scr}$ as $|s| \to \infty$, which we will identify as the scrambling time of the system,
\be
t_{\rm scr}=\frac{\beta}{2\pi}\log\frac{4\pi^2}{\beta^2{\cal V}^2}\,.
\ee
Therefore this second singularity only appears when the two insertion points in the same copy are within a scrambling time of each other. 

Examination of the {\em two-sided} correlator reveals a complementary picture. Taking $t_1=0$ in the right ($R$) copy, and $t_2$ inserted in the left ($L$) copy, there is exactly {\em one} singularity in this correlator appearing only when $|s|> s_{\rm scr}$. This comes in  from infinity in the left copy and asymptotes to $|t_2|= t_{\rm scr}$ on the left copy as $|s| \to \infty$. Taking the behaviour of the one-sided and two-sided correlators together, we arrive at the natural implication: 
correlations are being transferred from one copy $(R)$ to the other $(L)$ after a modular scrambling time, and the fermion operator under modular flow has  effectively  been pulled from one copy into the other. Modular flow in the dual gravity setup whose reconstruction we  outline below corroborates this interpretation (see figure \ref{figRflow} and \ref{figLflow}). It is worth noting that the structure of the modular flowed correlators we find, closely resembles those found in \cite{Gao:2021tzr} and \cite{Chandrasekaran:2022qmq} within related, but different physical contexts.

\paragraph{Bulk reconstruction:} The replica approach to  modular flow in the SYK$_\chi$ system makes clear that the discrete M\"obius transformations between replicas labelled by integer $s$,  upon analytic continuation $s\to is$, can be  identified with the SL$(2, {\mathbb R})$ action of modular flow on the AdS$_2$ bulk dual geometry. In the Euclidean setting, the discrete steps ratchet through successive boundaries of replica wormhole geometries. The identification of the precise combination of generators of this flow proceeds by first constructing the SL$(2, {\mathbb R})$ charges associated to each replica in the SYK setup. This corresponds to the generator of the  specific $U(1)$ that leaves invariant the solution to the Liouville equation in a given replica \cite{Gao:2021tzr}. Matching these to  the corresponding isometries of AdS$_2$,  provides the dictionary between the requisite boost and rotation parameters in AdS$_2$ and corresponding SYK parameters. We find that the resulting combination of boosts in AdS$_2$ has fixed points which are immediately identified with QESs, one from each twist field sector. The Kruskal-Szekeres (KS) coordinates of the QESs are determined by the (weak) coupling between the SYK$_\chi$ system and SYK$_\chi$ bath. At $t=0$ these are given as,\footnote{We note that strictly speaking, the picture only applies at times after Page time when the factorised saddle point dominates. To get the QES coordinates at any time slice $t\neq 0$, we just perform the rescalings $(U_{\rm QES, R},V_{\rm QES, R})\to (e^{-2\pi t/\beta}U_{\rm QES, R}, e^{2\pi t/\beta}V_{\rm QES, R})$.}
\be
U_{\rm QES, j}= - V_{\rm QES, j}=-\tanh \frac{x_j}{2}\,,\qquad j=L,R
\ee
where $x_L=\frac{\beta^2 {\cal V}^2}{2\pi^2}$ and $x_R=-\frac{\beta^2 {\cal V}^2}{2\pi^2}$. As shown in figures \ref{figRflow} and \ref{figLflow} this  determines the entanglement wedge of the AdS$_2$ boundary point at the given time.

\paragraph{Bulk modular flow:} The action of the AdS$_2$ boosts generating the SYK$_\chi$ modular flow on points in bulk AdS$_2$ reveals a rich and intriguing family of trajectories of bulk points under modular flow. We find that 
points on the conformal boundary stay on the boundary, but exactly as anticipated from the boundary correlators, a point at the right boundary at $t_R=0$ exits the right boundary after a modular scrambling time $s_{\rm scr}$ and reappears on the left. 

The flows of points starting slightly away from the conformal boundary at a cutoff surface, resolve different possible trajectories depicted in figures \ref{figRflow} and  \ref{figLflow}. Operator insertions within a scrambling time  of the twist field/QES time slice, remain within the entanglement wedge (shaded yellow) of the boundary point approaching the tip of the wedge asymptotically for large modular parameter. A point starting at the edge of the wedge near the boundary, remains on the edge and approaches the QES. Perhaps most  interesting of all, is the trajectory depicted in blue in figure \ref{figRflow} of a point within a small region just outside the wedge at the cutoff surface, penetrating the horizon at finite modular time, diverging and reappearing on the left boundary. It is clear that such trajectories explore physics behind the horizon. Indeed, bulk-to-boundary propagators with one point on either boundary and the second point on the modular flowed trajectory exhibit null-cone singularities behind the horizon. Such propagators should be viewed as  modular flowed SYK$_\chi$ correlation functions with one fermion insertion smeared with HKLL propagators \cite{Hamilton:2006az, Hamilton:2006fh} to reconstruct a local bulk insertion slightly displaced from the boundary.

The paper is organised as follows. In section \ref{sec2} we provide a brief summary of the setup with the coupled SYK framework which is the focus of our work and define the observables/correlators of interest. Section \ref{sec3} sets up the replica computation for the coupled SYK system and the large-$N$ saddle point equations, along with the large-$q$ limit. For the sake of completeness, we also spend some time underlining the two equivalent ways of viewing the replicated system: as a single theory formulated on an appropriate $n$-fold cover of the complex time contour, or as  $n$-copies of the theory on a time contour with local twist field operator insertions. Sections \ref{sec4} and \ref{sec5} are devoted to the replica diagonal and replica non-diagonal saddle points  respectively and solving  the replica recursion relations, the latter leading  to the real time modular flowed SYK correlators in section \ref{sec6}. In section \ref{sec7} we explain how to infer the SL$(2,{\mathbb R})$ charge associated to each replica and use this to work out the corresponding picture of modular flow transformations in the AdS$_2$ bulk, and modular flowed boundary correlators matching the SYK results in section \ref{sec8}. Section \ref{sec9} is devoted to the bulk trajectories of modular flowed points, excursions into the horizon, QESs as fixed points of the flow. We finally summarise our conclusions and open questions in section \ref{sec10}. In the Appendix, we collect useful formulae,  AdS/EAdS coordinates and embeddings, technical details of microscopic-bulk matching, and finally some relevant results on modular flow in BCFT in the TFD state.

\section{Set-up}
\label{sec2}
Our goal in this paper is to study the effect of modular flow on the correlation functions of a system prepared in an appropriate mixed state, which provides a microscopic window into the physics behind the horizon of a putative holographic black hole dual. To this end, we consider two SYK models \cite{Sachdev:1992fk, kitaev, Sachdev:2015efa, Kitaev:2017awl}, namely SYK$_\chi$ and SYK$_\psi$ coupled via an interaction, prepared in a thermofield double (TFD) state,
\be\label{TFD_SYK2}
|{\rm TFD}\rangle \,=\,  \frac{1}{\sqrt{Z(\beta)}}\,e^{-\frac{\beta}{4}(H_{L}+H_{R})}\ket{0_{\chi}}_{\rm LR}\otimes\ket{0_{\psi}}_{\rm LR}\,,
\ee
with $\beta$ the inverse temperature parameter, and the vacuum state  $\ket{0_{\chi,\psi}}_{\rm LR}$ is the maximally entangled TFD state of two copies of the SYK$_{\chi,\psi}$ system.\footnote{$H_L$ and $H_R$ are the Hamiltonians of the left (L) and right (R) copies of the system and $Z(\beta)$ the partition function of the system.}
Viewing SYK$_\psi$ degrees of freedom as the ``bath" variables, we will trace them out to obtain the reduced density matrix for the SYK$_\chi$ system.

We take the two interacting subsystems to be SYK$_q$ models \cite{Maldacena:2016hyu}  with Hamiltonian,
\begin{equation}\label{Hamiltonian-model}
\begin{gathered}
    H= i^{q/2}\sum_{i_{1},\dots, i_{q}}^{N_{\chi}}J^{\chi}_{i_{1},\dots, i_{q}}\chi^{i_{1}}\cdots\chi^{i_{q}} + i^{q/2}\sum_{i_{1},\dots, i_{q}}^{N_{\psi}}J^{\psi}_{i_{1},\dots, i_{q}}\psi^{i_{1}}\cdots\psi^{i_{q}} + \\
    +i^{q/2}\sum_{i_{1},\dots, i_{q/2}}^{N_{\chi}}\sum_{i'_{1},\dots, i'_{q/2}}^{N_{\psi}}V_{i_{1},\dots, i_{q/2}; i'_{1}, \dots, i'_{q/2}}\,\chi^{i_{1}}\cdots\chi^{i_{q/2}}\,\psi^{i'_{1}}\cdots\psi^{i'_{q/2}}\,,
\end{gathered}
\end{equation}
describing $N_\chi$ and $N_\psi$ Majorana fermions $\{\chi^i\}_{i=1\ldots N_\chi}$ and $\{\psi^i\}_{i=1\ldots N_\psi}$ respectively, with random interactions involving $q$ fermions at a time ($q$ even). The fermions satisfy the usual anti-commutation relations
\begin{equation}\label{anticomm_rel}
    \begin{gathered}
        \{\chi^{j},\chi^{k}\}=\delta^{jk}\,,\\\{\psi^{j},\psi^{k}\}=\delta^{jk}\,.
    \end{gathered}
\end{equation}
The disorder couplings are random variables drawn from Gaussian ensembles with  zero mean,
\be
\overline{J^{\psi,\chi}_{I_{q}}}=0\,,\qquad \qquad\overline{V_{I_{q/2}, I^\prime_{q/2}}}=0\,,
\ee
and non-vanishing variance,
\begin{align}
&\overline{\left(J^{\psi,\chi}_{I_{q}}\right)^{2}}\quad=\quad\frac{J^{2}(q-1)!}{N_{\psi,\chi}^{q-1}}\quad =\quad\frac{2^{q-1}}{q}\cdot\frac{\mathcal{J}^{2}(q-1)!}{N_{\psi,\chi}^{q-1}}\\\nonumber\\
    & \overline{\left(V_{I_{q/2}, I'_{q/2}}\right)^{2}}\quad=\quad\frac{V^{2}(q/2)!(q/2-1)!}{N_{\psi}^{\frac{q-1}{2}}N_{\chi}^{\frac{q-1}{2}}}\quad=\quad\frac{2^{q-1}}{q}\cdot\frac{\mathcal{V}^{2}(q/2)!(q/2-1)!}{N_{\psi}^{\frac{q-1}{2}}N_{\chi}^{\frac{q-1}{2}}}
\end{align}
where we introduced the shorthand notation $I_{q}=i_{1}i_{2}\cdots i_{q}$, and the couplings $J, V$, and ${\cal J}, {\cal V}$ are defined with appropriate $N$- and $q$-dependent factors allowing for smooth large-$N$ and large-$q$ limits.

To set the system up in the TFD state, we take two copies of the interacting SYK$_{q}$ subsystems described above, a left copy with $\chi_{L}$ and $\psi_{L}$ fields and a right one with $\chi_{R}$ and $\psi_{R}$ fermions.  We take the vacuum $\ket{0} = \ket{0_{\chi}}_{\rm LR}\otimes\ket{0_{\psi}}_{\rm LR}$ where $\ket{0_{\psi, \chi}}_{\rm LR}$ are the maximally entangled states between left and right copies, satisfying 
\bea
&&(\chi_{L}^{j}+i\chi_{R}^{j})\ket{0_{\chi}}_{\rm LR}=0\,,\qquad
j=1,\ldots,N_{\chi}\,,\label{MES_chi}\\\nonumber\\
&&(\psi_{L}^{j}+i\psi_{R}^{j})\ket{0_{\psi}}_{\rm LR}=0\,,\label{MES_psi}
\qquad j\prime = 1,\ldots N_\psi\,.
\eea
The TFD state \eqref{TFD_SYK2} is annihilated by $H_L-H_R$,
\be
(H_L-H_R)\ket{{\rm TFD}}=0\,,
\ee
and its time evolution is generated by the Hamiltonian $H_{L}+H_{R}$ so that, 
\begin{equation}\label{TFD_SYK}
    |{\rm TFD}(t)\rangle\,=\,\frac{1}{\sqrt{Z(\beta)}}\,
    e^{-2itH_{L}}e^{-\frac{\beta}{2}H_{L}}\ket{0_{\chi}}_{\rm LR}\otimes\ket{0_{\psi}}_{\rm LR}\,.
\end{equation}
We are interested in modular flows generated by the  reduced density matrix of the subsystem $\chi_{L} \cup \chi_{R}$, given by
\begin{equation}
    \rho_{\chi_{L}\cup\chi_{R}}(t)= \text{Tr}_{\psi_{L}, \psi_{R}}\ket{\text{TFD}(t)}\bra{\text{TFD}(t)}
\end{equation}
In what follows we will denote this reduced density matrix simply as $\rho$.
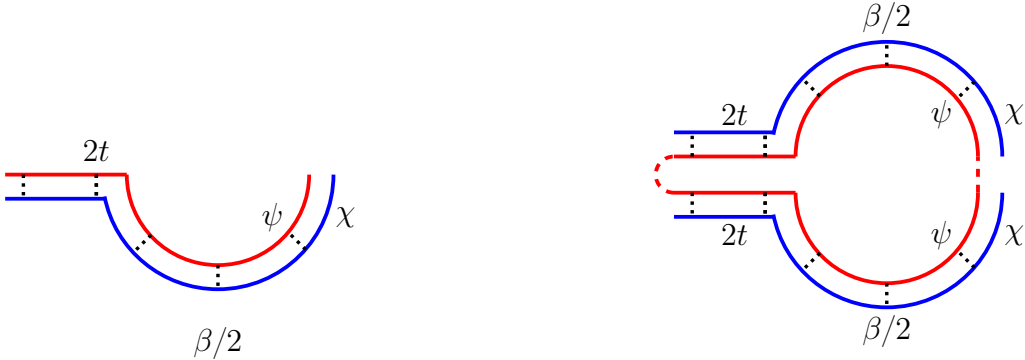
\begin{figure}[ht]
\begin{center}
\begin{tikzpicture}[scale=0.8]
\node at (-3.5, 0.4) {$2t$};
\draw[red, line width=0.5mm] (0,0) arc (360:180:1.5);
\node at (-1.5, -2.8) {$\beta/2$ };
\draw[red, line width=0.5mm] (-5,0) -- (-3,0);
\draw[blue, line width=0.5mm] (0.4,0) arc (360:191:1.9);
\draw[blue, line width=0.5mm] (-5,-0.4) -- (-3.35,-0.4);
\node at (-0.6, -0.7) {$\psi$ };
\node at (0.6, -0.7) {$\chi$ };
\draw[line width=0.45mm, dotted] (-4.7,0) -- (-4.7, -0.4) ;
\draw[line width=0.45mm, dotted] (-3.5,0) -- (-3.5, -0.4) ;
\draw[line width=0.45mm, dotted] (-3.5,0) -- (-3.5, -0.4) ;
\draw[line width=0.45mm, dotted] (-1.5,-1.5) -- (-1.5, -1.9) ;
\draw[line width=0.45mm, dotted] (-2.6,-1.0) -- (-2.9, -1.3) ;
\draw[line width=0.45mm, dotted] (-0.3,-1.0) -- (0.0, -1.3) ;
\begin{scope}[yscale=-1,xscale=1]
\draw[red, line width=0.5mm] (11,-0.3) arc (360:180:1.5);
\draw[red, line width=0.5mm] (6,-0.3) -- (8,-0.3);
\draw[blue, line width=0.5mm] (11.4,-0.3) arc (360:191:1.9);
\draw[blue, line width=0.5mm] (6,-0.7) -- (7.65,-0.7);
\node at (10.4, -1.0) {$\psi$ };
\node at (11.6, -1.0) {$\chi$ };
\draw[line width=0.45mm, dotted] (6.3,-0.3) -- (6.3, -0.7) ;
\draw[line width=0.45mm, dotted] (7.5,-0.3) -- (7.5, -0.7) ;
\draw[line width=0.45mm, dotted] (7.5,-0.3) -- (7.5, -0.7) ;
\draw[line width=0.45mm, dotted] (9.5,-1.8) -- (9.5, -2.2) ;
\draw[line width=0.45mm, dotted] (8.4,-1.3) -- (8.1, -1.6) ;
\draw[line width=0.45mm, dotted] (10.7,-1.3) -- (11.0, -1.6) ;
\node at  (7.0,-1.0)  {$2t$};
\node at  (9.5,-2.6)  {$\beta/2$};
\end{scope}
\draw[red, line width=0.5mm] (11,-0.3) arc (360:180:1.5);
\draw[red, line width=0.5mm] (6,-0.3) -- (8,-0.3);
\draw[blue, line width=0.5mm] (11.4,-0.3) arc (360:191:1.9);
\draw[blue, line width=0.5mm] (6,-0.7) -- (7.65,-0.7);
\node at (10.4, -1.0) {$\psi$ };
\node at (11.6, -1.0) {$\chi$ };
\draw[line width=0.45mm, dotted] (6.3,-0.3) -- (6.3, -0.7) ;
\draw[line width=0.45mm, dotted] (7.5,-0.3) -- (7.5, -0.7) ;
\draw[line width=0.45mm, dotted] (7.5,-0.3) -- (7.5, -0.7) ;
\draw[line width=0.45mm, dotted] (9.5,-1.8) -- (9.5, -2.2) ;
\draw[line width=0.45mm, dotted] (8.4,-1.3) -- (8.1, -1.6) ;
\draw[line width=0.45mm, dotted] (10.7,-1.3) -- (11.0, -1.6) ;
\draw[red, line width=0.5mm, dashed] (11.0,-0.3) -- (11.0, 0.3) ;
\draw[red, line width=0.5mm, dashed] (6.0,-0.3) arc (270:90: 0.3) ;
\node at  (7.0,-1.0)  {$2t$};
\node at  (9.5,-2.6)  {$\beta/2$};
\end{tikzpicture}
\caption{\footnotesize On the left is the depiction of the state $\ket{{\rm TFD}(t)}$, obtained by evolution by $H_L$ in Euclidean time $\beta/2$ followed by evolution in real time $2t$. Dotted lines represent interactions between the $\chi$ and $\psi$ systems. The figure on the right represents the reduced density matrix $\rho$ obtained by tracing over the SYK$_\psi$ bath degrees of freedom.}
\label{fig2} 
\end{center}
\end{figure}
\subsection{The reduced density matrix and replica observables}

\paragraph{Rényi entropies:} The reduced density matrix $\rho$ encodes the entanglement structure of the SYK$_\chi$ system  with the bath, quantified by the $n$-th R\'enyi entropies,
\begin{equation}
    S_{\chi_{L}\cup\chi_{R}}^{(n)}(t)= \frac{1}{1-n}\text{log\,Tr}\rho^{n}(t)\,.
\end{equation}
The trace of the $n$th power of $\rho$ can be computed via a path integral representation using $n$ replicas of the original $\chi-\psi$ system sewn together with appropriate boundary conditions that swap the replica indices of $\chi_{L,R}$ fields. Specifically, employing  the replica trick, 
\begin{equation}
    {\rm Tr}\rho^{n}=\frac{Z_{n}}{Z_{\beta}^{n}}\,,
\end{equation}
where $Z_{n}$ is the partition function evaluated on the $n$-fold cover of spacetime. In particular, the replicated theory describes fermions $\chi^{\alpha}_{j}$, $\psi^{\alpha}_{j}$  with index $\alpha=1,\dots, n$ labeling the replica interval in which the  Majorana fermions live. The replicas are joined so that the fermions have their replica indices cyclically permuted across adjacent copies.
\paragraph{Modular flowed correlators:} The reduced density matrix $\rho$ allows to define the modular Hamiltonian  $K=-\log\rho$ which generates an automorphism of the algebra of operators, via modular flow. Our focus will be on correlation functions of modular flowed operators in the SYK$_\chi$ subsystem,
\begin{equation}\label{modular_flow_without probe}
    W_{RL}^{(s)}(t_1, t_2)\,=\,\frac{1}{N}\sum_{j=1}^{N}\text{Tr}[\rho^{1-is}\chi_{R}^{j}(t_{1})\rho^{is}\,\chi_{L}^{j}(t_{2})]\,.
\end{equation}
This is the two-point correlator between a right Majorana fermion $\chi_{R}(t_1)$, evolved forward in modular time by an amount $s$ with the unitary operator $\rho^{is}=e^{-i K s}$,  and a left  Majorana fermion $\chi_{L}(t_2)$ inserted in the left copy of the SYK$_\chi$ subsystem. In the AdS$_2$ gravity dual, this corresponds to the insertion of fermion operators on the left and right boundaries and examining how modular flow allows to explore and probe the bulk geometry behind the horizon.

Our strategy will be to employ the replica trick to compute the  modular flowed correlators \eqref{modular_flow_without probe}. In particular, we will focus on correlation functions of the form,
\begin{equation}
 W_{RL}^{k,s}(t_1, t_2) \,=\, \frac{1}{N}\sum_{j=1}^{N}\text{Tr}[\rho^{k-s}\chi_{R}^{j}(t_{1})\rho^{s}\chi_{L}^{j}(t_{2})]
\end{equation}
where $k$ and $s$ are integers, with $0\leq s \leq k$, and subsequently,  analytically  continue $k\to 1$ and $s\to is$ to get \eqref{modular_flow_without probe}.

\section{Replica partition function and saddle points}
\label{sec3}
The calculation of the replicated partition function relies on the path integral representation of the density matrix and its saddle point evaluation in the large-$N_{\chi,\psi}$ limit. There are two equivalent ways to formulate the replica problem for the quantum mechanical problem at hand. We can weither have a single set of $\chi$ fermions living on a replicated contour ${\cal C}$ as we describe in detail below, or we can take $n$ sets of fermions $\{\chi^\alpha\}$ with replica index $\alpha=1,\ldots n$ with certain twisted boundary conditions around the time contour.

\subsection{Coupled SYK on the replica contour}
\begin{figure}[ht]
\begin{center}
\begin{tikzpicture}[scale=0.7]
\draw[blue, line width=0.5mm, dashed] (-0.28,11.5) arc (120:60:6.60);
\draw[blue, line width=0.5mm, dashed] (5.7,11.5) arc (120:60:6.65);
\draw[blue, line width=0.5mm, dashed] (-0.64,6.0) arc (230:310:5.7);
\draw[blue, line width=0.5mm, dashed] (5.36,6.0) arc (230:310:5.7);
\draw[blue, line width=0.5mm, dashed] (0.31,11.5) arc (140:40:7.40);
\draw[blue, line width=0.5mm, dashed] (0.68,6.0) arc (225:315:7.44);
\begin{scope}[rotate=90]
\begin{scope}[yscale=-1,xscale=1]
\draw[red, line width=0.5mm] (11,-0.3) arc (360:180:1.5);
\draw[red, line width=0.5mm] (6,-0.3) -- (8,-0.3);
\draw[blue, line width=0.5mm] (11.4,-0.3) arc (360:191:1.9);
\draw[blue, line width=0.5mm] (6,-0.7) -- (7.65,-0.7);
\node at (10.4, -1.0) {$\psi$ };
\node at (11.6, -1.0) {$\chi$ };
\draw[line width=0.45mm, dotted] (6.3,-0.3) -- (6.3, -0.7) ;
\draw[line width=0.45mm, dotted] (7.5,-0.3) -- (7.5, -0.7) ;
\draw[line width=0.45mm, dotted] (7.5,-0.3) -- (7.5, -0.7) ;
\draw[line width=0.45mm, dotted] (9.5,-1.8) -- (9.5, -2.2) ;
\draw[line width=0.45mm, dotted] (8.4,-1.3) -- (8.1, -1.6) ;
\draw[line width=0.45mm, dotted] (10.7,-1.3) -- (11.0, -1.6) ;
\node at  (6.2,-1.6)  {${\tau}=0$};
\end{scope}
\draw[red, line width=0.5mm] (11,-0.3) arc (360:180:1.5);
\draw[red, line width=0.5mm] (6,-0.3) -- (8,-0.3);
\draw[blue, line width=0.5mm] (11.4,-0.3) arc (360:191:1.9);
\draw[blue, line width=0.5mm] (6,-0.7) -- (7.65,-0.7);
\node at (10.4, -1.0) {$\psi$ };
\node at (11.6, -1.0) {$\chi$ };
\draw[line width=0.45mm, dotted] (6.3,-0.3) -- (6.3, -0.7) ;
\draw[line width=0.45mm, dotted] (7.5,-0.3) -- (7.5, -0.7) ;
\draw[line width=0.45mm, dotted] (7.5,-0.3) -- (7.5, -0.7) ;
\draw[line width=0.45mm, dotted] (9.5,-1.8) -- (9.5, -2.2) ;
\draw[line width=0.45mm, dotted] (8.4,-1.3) -- (8.1, -1.6) ;
\draw[line width=0.45mm, dotted] (10.7,-1.3) -- (11.0, -1.6) ;
\draw[red, line width=0.5mm, dashed] (11.0,-0.3) -- (11.0, 0.3) ;
\draw[red, line width=0.5mm, dashed] (6.0,-0.3) arc (270:90: 0.3) ;
\end{scope}
\begin{scope}[xshift=6cm]
\begin{scope}[rotate=90]
\begin{scope}[yscale=-1,xscale=1]
\draw[red, line width=0.5mm] (11,-0.3) arc (360:180:1.5);
\draw[red, line width=0.5mm] (6,-0.3) -- (8,-0.3);
\draw[blue, line width=0.5mm] (11.4,-0.3) arc (360:191:1.9);
\draw[blue, line width=0.5mm] (6,-0.7) -- (7.65,-0.7);
\node at (10.4, -1.0) {$\psi$ };
\node at (11.6, -1.0) {$\chi$ };
\draw[line width=0.45mm, dotted] (6.3,-0.3) -- (6.3, -0.7) ;
\draw[line width=0.45mm, dotted] (7.5,-0.3) -- (7.5, -0.7) ;
\draw[line width=0.45mm, dotted] (7.5,-0.3) -- (7.5, -0.7) ;
\draw[line width=0.45mm, dotted] (9.5,-1.8) -- (9.5, -2.2) ;
\draw[line width=0.45mm, dotted] (8.4,-1.3) -- (8.1, -1.6) ;
\draw[line width=0.45mm, dotted] (10.7,-1.3) -- (11.0, -1.6) ;
\node at  (6.2,-2.3)  {${\tau}=4t+\beta$};
\end{scope}
\draw[red, line width=0.5mm] (11,-0.3) arc (360:180:1.5);
\draw[red, line width=0.5mm] (6,-0.3) -- (8,-0.3);
\draw[blue, line width=0.5mm] (11.4,-0.3) arc (360:191:1.9);
\draw[blue, line width=0.5mm] (6,-0.7) -- (7.65,-0.7);
\node at (10.4, -1.0) {$\psi$ };
\node at (11.6, -1.0) {$\chi$ };
\draw[line width=0.45mm, dotted] (6.3,-0.3) -- (6.3, -0.7) ;
\draw[line width=0.45mm, dotted] (7.5,-0.3) -- (7.5, -0.7) ;
\draw[line width=0.45mm, dotted] (7.5,-0.3) -- (7.5, -0.7) ;
\draw[line width=0.45mm, dotted] (9.5,-1.8) -- (9.5, -2.2) ;
\draw[line width=0.45mm, dotted] (8.4,-1.3) -- (8.1, -1.6) ;
\draw[line width=0.45mm, dotted] (10.7,-1.3) -- (11.0, -1.6) ;
\draw[red, line width=0.5mm, dashed] (11.0,-0.3) -- (11.0, 0.3) ;
\draw[red, line width=0.5mm, dashed] (6.0,-0.3) arc (270:90: 0.3) ;
\end{scope}
\end{scope}
\begin{scope}[xshift=12cm]
\begin{scope}[rotate=90]
\begin{scope}[yscale=-1,xscale=1]
\draw[red, line width=0.5mm] (11,-0.3) arc (360:180:1.5);
\draw[red, line width=0.5mm] (6,-0.3) -- (8,-0.3);
\draw[blue, line width=0.5mm] (11.4,-0.3) arc (360:191:1.9);
\draw[blue, line width=0.5mm] (6,-0.7) -- (7.65,-0.7);
\node at (10.4, -1.0) {$\psi$ };
\node at (11.6, -1.0) {$\chi$ };
\draw[line width=0.45mm, dotted] (6.3,-0.3) -- (6.3, -0.7) ;
\draw[line width=0.45mm, dotted] (7.5,-0.3) -- (7.5, -0.7) ;
\draw[line width=0.45mm, dotted] (7.5,-0.3) -- (7.5, -0.7) ;
\draw[line width=0.45mm, dotted] (9.5,-1.8) -- (9.5, -2.2) ;
\draw[line width=0.45mm, dotted] (8.4,-1.3) -- (8.1, -1.6) ;
\draw[line width=0.45mm, dotted] (10.7,-1.3) -- (11.0, -1.6) ;
\node at  (6.2,-2.3)  {${\tau}=8t+2\beta$};
\end{scope}
\draw[red, line width=0.5mm] (11,-0.3) arc (360:180:1.5);
\draw[red, line width=0.5mm] (6,-0.3) -- (8,-0.3);
\draw[blue, line width=0.5mm] (11.4,-0.3) arc (360:191:1.9);
\draw[blue, line width=0.5mm] (6,-0.7) -- (7.65,-0.7);
\node at (10.4, -1.0) {$\psi$ };
\node at (11.6, -1.0) {$\chi$ };
\draw[line width=0.45mm, dotted] (6.3,-0.3) -- (6.3, -0.7) ;
\draw[line width=0.45mm, dotted] (7.5,-0.3) -- (7.5, -0.7) ;
\draw[line width=0.45mm, dotted] (7.5,-0.3) -- (7.5, -0.7) ;
\draw[line width=0.45mm, dotted] (9.5,-1.8) -- (9.5, -2.2) ;
\draw[line width=0.45mm, dotted] (8.4,-1.3) -- (8.1, -1.6) ;
\draw[line width=0.45mm, dotted] (10.7,-1.3) -- (11.0, -1.6) ;
\draw[red, line width=0.5mm, dashed] (11.0,-0.3) -- (11.0, 0.3) ;
\draw[red, line width=0.5mm, dashed] (6.0,-0.3) arc (270:90: 0.3) ;
\end{scope}
\end{scope}
\end{tikzpicture}
\caption{\footnotesize Depiction of the $n=3$ replica contour ${\cal C}$ with blue dashed lines showing cyclic identification of contours across different copies. Black dotted lines represent the interactions between the SYK$_\chi$ (blue) and SYK$_\psi$ (red) systems.}
\label{fig3} 
\end{center}
\end{figure}
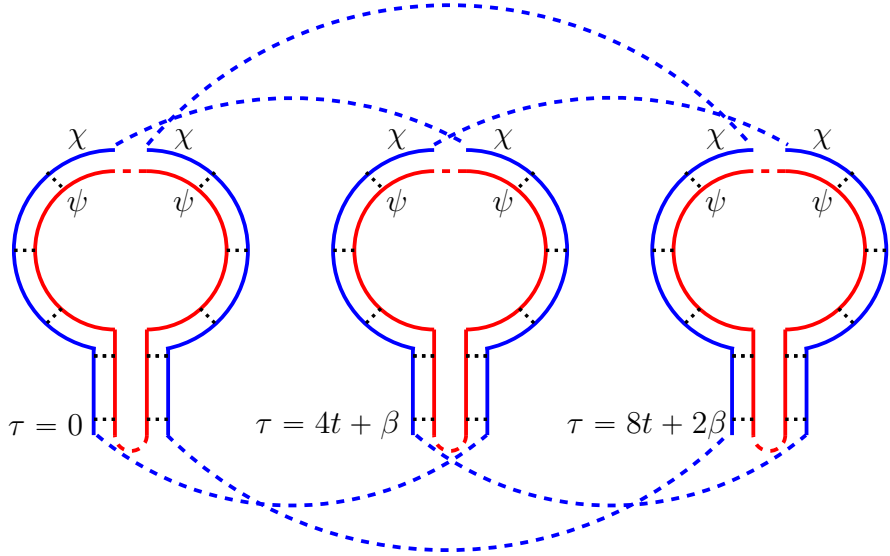

We write the path integral for the SYK$_\chi$ system, with $\psi$ degrees of freedom integrated out, over a single contour ${\cal C}$ parametrised by a single real variable ${\tau},$\footnote{We adopt the conventions in \cite{Chen:2020wiq} where the $n=2$ case with SYK coupled to a spin chain bath was studied.} 
\be
0\leq {\tau}\leq n(4t+\beta)\,.\ee
To compute the reduced density matrix and associated  R\'enyi entropies at $t$, the parameter ${\tau}$ accounts for both real time and Euclidean evolution along every replica.
For the $\alpha^{\rm th}$ replica ($\alpha=1,2,\ldots n$), the forward/backward real time evolution and the imaginary time evolution correspond to specific ranges of the ${\tau}$ parameter,
\bea
{\tau}\in
\begin{cases}
  &[(2\alpha-2)2t + (\alpha-1)\beta, (2\alpha-1)2t+ (\alpha-1)\beta]\qquad \text{\rm forward real time}\\
    & [(2\alpha-1)2t + (\alpha-1)\beta, (2\alpha-1)2t +\alpha\beta]\qquad\qquad\,\,\text{Euclidean time}\\
    & [(2\alpha-1)2t +\alpha\beta, 4\alpha t +\alpha\beta]\qquad\qquad\qquad\qquad\qquad\text{backward real time}
\end{cases}
\eea
This is shown in figure \ref{fig3} which also depicts how the contour ${\cal C}$ is obtained by identifying appropriate segments of the time evolution, in one-to-one correspondence with the cyclic contractions of time evolved bra and ket states in the explicit representantion of ${\rm Tr}\rho^n$. 

Crucially, the boundary conditions corresponding to the identifications depicted in figure \ref{fig3} are distinct for the $\chi$ and $\psi$ fermions. The trace over the SYK$_\psi$ bath degrees of freedom is performed in each replica copy independently. Figure \ref{fig3} also indicates interactions between the $\chi$ and $\psi$ systems, which are key to yielding a nontrivial reduced density matrix.

Following \cite{Chen:2020wiq}, in order to capture the contributions from  different segments of ${\cal C}$ we introduce the function $f(\tau)$
\begin{equation}
    f({\tau}) =  
    \begin{cases}
      i & \qquad \tau \in {\cal C}^+_{\alpha}\\
      -i & \qquad \tau \in {\cal C}^-_{\alpha}\\
      1 & \qquad \tau \in {\cal C}^\beta_{\alpha}
    \end{cases}
\end{equation}
where ${\cal C}^\pm_\alpha$ and ${\cal C}^\beta_\alpha$ are forward, backward and imaginary time portions, respectively, of the contour in the $\alpha^{\rm th}$ replica.

The path integral  representation of the replicated partition function is then,
\be
    Z_{n}=\int \prod_{j=1}^{N_\chi}\mathcal{D}\chi^j\prod_{j=1}^{N_\psi}\mathcal{D}\psi^{j}\,\exp(-I_{\cal C})\,,
    \ee
    where
    \be
    I_{\cal C}\,=\,\int_{\mathcal{C}} d\tau \left[\frac{1}{2}\sum_{i=1}^{N_{\chi}}\chi^{i}\partial_{\tau}\chi^{i} \,+\,\frac{1}{2}\sum_{i=1}^{N_{\psi}}\psi^{i}\partial_{\tau}\psi^{i} \,+\, f(\tau)H(\tau)\right]\,,
    \ee
    with $H$ given by the interaction Hamiltonian in \eqref{Hamiltonian-model}.
To compute the quenched average of the Rényi entropy $S^{(n)}_{\chi_{L}\cup \chi_{R}}$, over disordered couplings $J_{I_{q/2}}$ and $V_{I_{q/2}, I'_{q/2}}$, one should compute the disorder average $\overline{Z^{k}_{n}}$ for a general integer $k$, and then analytically continue to the $k\to1$ limit. In the large-$N$ SYK model, it is known that at leading and next-to-leading order in $1/N$, $\overline{Z^{k}_{n}}\sim \overline{Z_{n}}^{k}$ \cite{Maldacena:2016hyu,Gu:2016oyy, Cotler:2016fpe}, therefore we will simply compute $\overline{Z_{n}}$ and take the logarithm to obtain the quenched average of the R\'enyi entropy.

After averaging over disordered couplings, we introduce the bilocal fields
\begin{equation}
    G_{\chi}(\tau_{1},\tau_{2})\equiv \frac{1}{N_{\chi}}\sum_{j}^{N_{\chi}}\chi^{j}(\tau_{1})\chi^{j}(\tau_{2}), \quad\quad\quad G_{\psi}(\tau_{1},\tau_{2})\equiv \frac{1}{N_{\psi}}\sum_{j}^{N_{\psi}}\psi^{j}(\tau_{1})\psi^{j}(\tau_{2})\,,\label{bilocal}
\end{equation}
and the respective Lagrange multiplier fields $\Sigma_{\chi,\psi}(\tau_1,\tau_2)$ which enforce the constraints \eqref{bilocal}, and finally integrate out fermions $\chi^{j}$ and $\psi^{j}$ , to get the large-$N$ effective action (where we have set $N_\chi=N_\psi=N$),
\begin{align}\label{replica action}
    \frac{I_{\cal C}}{N} =& -\frac{1}{2}\text{log\,det}(\partial_{\tau}-\Sigma_{\chi}) -\frac{1}{2}\text{log\,det}(\partial_{\tau}-\Sigma_{\psi})+ \\\nonumber
    &+\frac12\sum_{a=\psi,\chi}\int_{\mathcal{C}} d\tau_{1}d\tau_{2}f(\tau_{1})f(\tau_{2})\left(\Sigma_a(\tau_1,\tau_2)G_a(\tau_1,\tau_2)-\frac{\mathcal{J}^{2}}{2q^{2}}\left[2G_{a}(\tau_{1},\tau_{2})\right]^{q}\right)\\\nonumber
    &-\frac{\mathcal{V}^{2}}{2q^{2}}\int_{\mathcal{C}} d\tau_{1}d\tau_{2}f(\tau_{1})f(\tau_{2})\left[2G_{\chi}(\tau_{1},\tau_{2})\right]^{q/2}\left[2G_{\psi}(\tau_{1},\tau_{2})\right]^{q/2}\,.
\end{align}
Below, we will work mainly in imaginary (Euclidean) time and subsequently analytically continue back to real time by sending $\tau\to 2 it $.
 In imaginary time, each  factor of $f(\tau)=1$.
 
The saddle point equations follow immediately upon varying the action above with respect to the Lagrange multiplier fields $\Sigma_a$ and the bilocal operators $G_a$,\footnote{The expectation values of the fields $G_{a}$ and $\Sigma_{a}$ must be viewed as matrices in the replica time domain $(\tau_{1},\tau_{2})$.}
\bea
&& G_{a}= (\partial_{\tau} - \Sigma_{a})^{-1} \,,
\label{saddle-point_equations_coupledSYK}\\\nonumber
&& \Sigma_{a}(\tau_{1},\tau_{2})=\frac{\mathcal{J}^{2}}{q}\,\left[2G_{a}(\tau_{1},\tau_{2})\right]^{q-1} \,+\, \frac{\mathcal{V}^{2}}{q}\,\left[2G_{a}(\tau_{1},\tau_{2})\right]^{\frac{q}{2}-1}\left[2G_{\Bar{a}}(\tau_{1},\tau_{2})\right]^{\frac{q}{2}}
\eea
where all above relations are understood to apply to expectation values of the respective fields in the vacuum state, and  we have introduced a complementary index notation $(\ldots)_{\Bar{\chi}}\equiv(\ldots)_{ \psi}$ and $ (\ldots)_{\Bar{\psi}}\equiv(\ldots)_{ \chi}$. 
\subsubsection{Large $q$ limit}
For any $q$, in the infrared conformal limit, for the uncoupled ${\cal V}=0$ theories, the expectation values of  $G_{\chi,\psi}(\tau_1,\tau_2)$ and $\Sigma_{\chi,\psi}(\tau_1,\tau_2)$ behave like correlators of dimension $1/q$ and $(q-1)/q$ operators. This motivates the large-$q$ limit \cite{Maldacena:2016hyu} and corresponding ansatz,
\begin{equation}\label{largeqg}
    G_{a}(\tau_1,\tau_2)=G_{a,0}\,e^{g_{a}/q}\,= \,G_{a,0}(\tau_1, \tau_2)\left(1+\frac{g_{a}(\tau_1,\tau_2)}{q}+\ldots\right), \quad\quad\quad a=\chi,\psi\,.
\end{equation}
$G_{a,0}$ is the free fermion Green's function. For the replica theory we will require,
\bea
&&        G_{\psi, 0}(\tau_{1},\tau_{2})=
        \begin{cases}
            \frac{1}{2}\text{sgn}(\tau_{1}-\tau_{2}) \qquad \tau_{1},\tau_{2} \in&\text{same replica}\\\\
            0 \qquad &\text{otherwise}
        \end{cases}\label{Gpsi0}\\\nonumber\\
&&        G_{\chi,0}(\tau_{1},\tau_{2})=\frac{1}{2}\text{sgn}(\tau_{1}-\tau_{2})\,.
\eea
The choice above reflects that the Green's function for the bath fermions $\{\psi_j\}$ is replica-diagonal. On the other hand, $G_{\chi,0 }(\tau_{1},\tau_{2})$, representing the UV two-point function of the $\chi$ fermions is defined on a nontrivial replica contour, with the definition of the time ordering of the $\chi$ fermions following straightforwardly from the   parameterization of  the replicas in the two saddle points  we will analyze below.

The saddle point equations in the large-$q$ limit become,
\bea\label{SD-equations}
&&\Sigma_{a}(\tau_{1},\tau_{2})=-\frac{1}{q}\partial_{\tau_{1}}\partial_{\tau_{2}}\left[G_{a,0}g_{a}(\tau_{1},\tau_{2})\right]\\\nonumber\\\nonumber
&&\Sigma_{a}(\tau_{1},\tau_{2}) = \frac{\mathcal{J}^{2}}{q}(2G_{a,0})^{q-1}e^{g_{a}(\tau_{1},\tau_{2})} +\frac{\mathcal{V}^{2}}{q}(2G_{a,0})^{q/2-1}(2G_{\bar{a},0})^{q/2}e^{\frac{1}{2}(g_{a}(\tau_{1},\tau_{2})+g_{\bar{a}}(\tau_{1},\tau_{2}))}
\eea
where the first equation can be obtained from the operator equation for $G_a$ in eq.\eqref{saddle-point_equations_coupledSYK} and approximating $G_a^{-1}$ employing \eqref{largeqg} in the large-$q$ limit. Combining  the two equations in \eqref{SD-equations}, we find the $q$-independent condition,
\bea
\label{saddle-point-eq-fin}
\partial_{\tau_{1}}\partial_{\tau_{2}}\left[G_{a,0}\,g_{a}(\tau_{1},\tau_{2})\right]&=& -\mathcal{J}^{2}\,(2G_{a,0})^{q-1}\,\exp[g_{a}(\tau_{1},\tau_{2})]\\\nonumber\\\nonumber &-&\mathcal{V}^{2}\,(2G_{a,0})^{q/2-1}\,(2G_{\bar{a},0})^{q/2}\,\exp[\tfrac{1}{2}(g_{a}(\tau_{1},\tau_{2})+g_{\bar{a}}(\tau_{1},\tau_{2}))]\,.
\eea
\subsection{Alternative formulation: $n$ copies of coupled SYK's}\label{Alternative formulation}
\begin{figure}[ht]
\centering
\begin{tikzpicture}[scale=0.9]
  \coordinate (L)  at (5,-0.1);       
  \coordinate (R)  at (11.4,-0.2);    
  \coordinate (MidA) at (7.63,-0.1);  
  \coordinate (Ltop) at (5,-0.3);      
  \coordinate (Rtop) at (11.4,0.3);   
  \coordinate (MidAtop) at (7.63,-0.3);

  \begin{scope}[yscale=-1]
    \draw[blue, line width=0.5mm] (R) arc (360:183:1.9);
    \draw[blue, line width=0.5mm] (L) -- (MidA);
    \node at (4.8,-0.4) {$0$};
    \node at (13,-0.15) {$\beta/2\,+2it$};
    \node at (7.3,-2) {\color{blue}$C_1$};
  \end{scope}

  \draw[cyan, line width=0.5mm] (R) arc (360:183:1.9);
  \draw[cyan, line width=0.5mm] (Ltop) -- (MidAtop);
  \node at (4.2,-0.8) {$\beta+4it$};
  \node at (7.3,-2) {\color{cyan}$C_2$};

   \node[red] at (5,-0.2) {\Large $\times$};
   \node[red] at (R) {\Large$\times$};


\end{tikzpicture}
\caption{{\small The time contour $C_{*}$, divided into $C_1$ and $C_2$. One swap operator is inserted at $\tau = 0$ and another is inserted at $\tau = \beta/2+it$.}}
\label{time_contour_replica}
\end{figure}
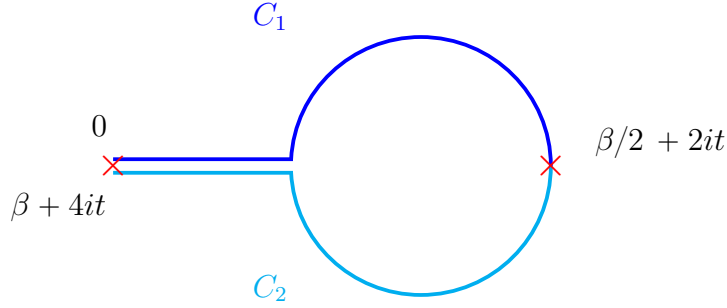

An alternative useful  way to think of the replica theory is  by taking $n$ copies of all the Majorana fermions on a complex time contour with twist operator insertions whose effect is to cyclically permute the replica index on the fermions.\footnote{This picture is familiar from 2D CFT where the replica formulation of R\'enyi entropies can be viewed as insertions  of twist field operators  introducing a branch cut in the complex plane. Across the branch cut joining  successive Riemann sheets, the replica indices on fields is cyclically permuted.} For notational compactness in this subsection we define 
\be
\chi_{j, a=1}^\alpha\equiv \chi^\alpha_j \,,\qquad \chi_{j, a=2}^\alpha\equiv \psi^\alpha_j\,,\qquad \alpha=1\,\ldots n
\ee
for the system and bath fermions respectively with replica index $\alpha$. Then the replica partition function $Z_{n}$ can then be represented as,
\bea
 &&   Z_{n}=\int\prod_{a,j,\alpha}\left[ {\cal D}\chi^{\alpha}_{a,j}\right]\,e^{-I}\,,\\\nonumber
 &&   I^{(n)}\,=\,\sum_{\alpha}\int_{C_*}d\tau\left[\sum_{a,j}\frac{1}{2}\chi^{\alpha}_{a,j}\partial_{\tau}\chi^{\alpha}_{a,j}
    -\sum_{a=1,2}i^{q/2}\sum_{I_{q}}^{N}J_{I_{q}}\chi^{\alpha}_{a,i_{1}}\cdots\chi^{\alpha}_{a,i_{q}}\right. \\\nonumber
    &&\left.\hspace{1.5in}-i^{q/2}\sum_{I_{q/2}}^{N}\,\sum_{I'_{q/2}}^{N}V_{I_{q/2}; I'_{q/2}}g^{\alpha\beta}(\tau)\chi^{\alpha}_{1,i_{1}}\cdots\chi^{\alpha}_{1,i_{q/2}}\chi^{\beta}_{2,i'_{1}}\cdots\chi^{\beta}_{2,i'_{q/2}}\right]\,,
\eea
where $C_{*}=C_{1}\cup C_{2}$ is the complex time contour depicted in figure \ref{time_contour_replica}. The contour is split in two components by the twist operators inserted at $\tau=0$ and $\tau=\beta/2+2it$. The effect of the twist fields is to modify the coupling $V_{I_{q/2}, I'_{q/2}}$ by the matrix $g^{\alpha\beta}(\tau)$, 
\begin{equation}\label{twists}
    g^{\alpha\beta}(\tau)=
    \begin{cases}
        \delta^{\alpha+1,\beta}\, \quad \text{for}\; \tau\in C_{2} \\
        \delta^{\alpha,\beta}\, \quad\quad \text{for}\; \tau\in C_{1} 
    \end{cases}
\end{equation}
where $\alpha,\beta$ are defined mod $n$ so $\alpha,\beta=n+1$ is identified with $\alpha,\beta=1$. This picture arises directly from  the identifications for the $\chi$ and $\psi$ contours in figure \ref{fig3}. Both the system and bath fermions reside in  two separate sets of $n$ disconnected contours, so naturally we have $n$ copies each of the $\chi$ and $\psi$ fermions, labelled by the respective replica indices. The interactions between the system and bath fermions  are, however, not always diagonal in replica space and  in fact offset by one replica unit when the time coordinate is in $C_2$.

As before, after  averaging over the  disordered couplings, and introducing the bilocal fields,
\begin{equation}
    G_{a}^{\alpha\alpha'}(\tau,\tau')\equiv \frac{1}{N}\sum_{j}^{N}\chi^{\alpha}_{a,j}(\tau)\chi^{\alpha'}_{a,j}(\tau')\,,
\end{equation}
and  corresponding Lagrange multipliers $\Sigma_{a}^{\alpha\alpha'}(\tau,\tau')$, we  integrate out fermions $\chi^{\alpha}_{a,j}$, to get the effective action
\bea\label{In-2}
 \frac{I^{(n)}}{N} = &&\sum_{a=1,2}\left[-\log\text{Pf}\left(\partial_\tau \delta^{\alpha\alpha'}-\Sigma_a^{\alpha\alpha'}\right) + \right.\\\nonumber
  +&&\left.  \frac{1}{2}\,\int_{C_{*}}\mathrm{d}\tau\mathrm{d}\tau'\sum_{\alpha,\alpha'}\left(\Sigma_a^{\alpha\alpha'}(\tau,\tau')G_a^{\alpha\alpha'}(\tau,\tau') - \frac{\mathcal{J}^2}{2q^{2}}\,\left[2G_{a}^{\alpha\alpha'}(\tau,\tau')\right]^q\right)\right]\\\nonumber
   -&& \frac{\mathcal{V}^2}{2q^{2}}\int_{C_{*}}\mathrm{d}\tau\mathrm{d}\tau' \sum_{\alpha\alpha'\beta\beta'}\left[2G_1^{\alpha\alpha'}(\tau,\tau')\right]^{q/2} \ g^{\alpha\beta}(\tau)g^{\alpha'\beta'}(\tau')  \left[2G_2^{\beta\beta'}(\tau,\tau')\right]^{q/2}\,.
\eea
We then obtain saddle point equations for all the replica fields,\footnote{For fixed $a$, the fields $G_{a}$ and $\Sigma_{a}$ are viewed as matrices, acting on the vector space parametrized by $\tau, \alpha$, and $\partial_{\tau}$ is also viewed as a diagonal matrix.}
\bea
&&G_a = (\partial_\tau - \Sigma_a)^{-1}\label{saddlept}\\\nonumber\\\nonumber
&&\Sigma_a^{\alpha\alpha'}(\tau,\tau') = \frac{{\mathcal{J}^2}}{q}\left[2G_a^{\alpha\alpha'}(\tau,\tau')\right]^{q-1} +\\\nonumber
&&\hspace{1.5in}+\,\frac{{\mathcal{V}^2}}{q}\,\left[2G_a^{\alpha\alpha'}(\tau,\tau')\right]^{q/2-1}\sum_{\beta\beta'}g^{\alpha\beta}(\tau)g^{\alpha'\beta'}(\tau')\left[2G_{\bar{a}}^{\beta\beta'}(\tau,\tau')\right]^{q/2}\,.
\eea
In the final line we use $\bar{a}$ to mean the complementary physical subsystem, so if $a =1$ then $\bar{a} = 2$. 

\section{Replica-diagonal saddle point}
\label{sec4}
 In the absence of twist field boundary conditions, the saddle-point solution to the Schwinger-Dyson equation is diagonal in replica space. To set the stage,  we  first review some details of  the replica-diagonal solution and its implications, and subsequently turn our attention to the replica non-diagonal saddle point.
 
We begin with the theory with $\mathcal{V}=0$, which reduces to a theory of decoupled SYK subsystems and the saddle-point solution is diagonal in replica space. When a small $\mathcal{V}$ is gradually turned on, we could consider a perturbative solution to the Schwinger-Dyson (SD) equations \eqref{saddle-point_equations_coupledSYK} or \eqref{saddlept}. However, the saddle-point  equations imply that the structure of  $\Sigma_{a}(\tau_{1},\tau_{2})$ in replica space  is the same as that of $G_{a}(\tau_{1},\tau_{2})$. This means that if we start with the $\mathcal{V}=0$ replica-diagonal solution and we solve the SD equations perturbatively, we will find that the solution stays diagonal to all orders in  the coupling $\mathcal{V}$.

In a replica-diagonal setting, the $\chi$ and $\psi$ systems are identical and the  solutions to  the SD equations for both are identical. Focussing on the  $\chi$ subsystem for specificity, the untwisted saddle point condition,
\begin{equation}
    \partial_{\tau_{1}}\partial_{\tau_{2}}g_{\chi}(\tau_{1},\tau_{2})= -2(\mathcal{J}^{2} +\mathcal{V}^{2})e^{g_{\chi}(\tau_{1},\tau_{2})}\,,
\end{equation}
is Liouville's equation. The solution  was described in \cite{Maldacena:2016hyu}. In the absence of twist fields,  time translation invariance  ensures that $g_{\chi}(\tau_{1},\tau_{2})$ depends only on time differences $(\tau_{1}-\tau_{2})$, so the Liouville equation becomes
\begin{equation}
    \partial_{\tau_{1}}^{2}g_{\chi}(\tau_{1}-\tau_{2})= 2(\mathcal{J}^{2} +\mathcal{V}^{2})e^{g_{\chi}(\tau_{1}-\tau_{2})}\,.
\end{equation}
The general solution to this equation is
\begin{equation}
    e^{g_{\chi}(\tau_{1}-\tau_{2})}=\frac{1}{\mathcal{J}_{0}^{2}}\,\frac{\omega^2}{\cos^{2}\left(\omega(|\tau_{1}-\tau_{2}|+\tau_{0})\right)}\,,\label{szerosol}
\end{equation}
where $\omega, \tau_0$ are integration constants, and 
\begin{equation}
\mathcal{J}_{0}\equiv\sqrt{\mathcal{J}^{2}+\mathcal{V}^{2}}\,.
\end{equation} 
The dependence on the absolute value   $|\tau_{1}-\tau_{2}|$ ensures that  $g_{\chi}(\tau_{1}-\tau_{2})$ is  symmetric under the $\tau_1\leftrightarrow \tau_2$ exchange, which is in turn required for antisymmetry of the fermion correlator  $G_{\chi}=G_{\chi,0} \,e^{g_{\chi}/{q}}$. The integration constants are determined by imposing periodicity in imaginary time and requiring that short distance behavior be that of free fermions,  
\begin{equation}
    g_{\chi}(0)=g_{\chi}(\beta)=0\,,
\end{equation}
yielding,
\begin{equation}\label{RP-sol}
    G_{\chi}(\tau_{1},\tau_{2}) = \frac{1}{2}\text{sgn}(\tau_{1}-\tau_{2})\left[\frac{\omega}{\mathcal{J}_{0}\cos{(\omega(|\tau_{1}-\tau_{2}|-\beta/2)}}\right]^{2/q}
\end{equation}
where the parameter $\omega$ is implicitly a solution to the equation
\begin{equation}
    \omega=\mathcal{J}_{0}\cos{\frac{\omega\beta}{2}}\,.
\end{equation}
In the strong coupling (IR) limit $\omega/{\cal J}_0\to 0$, $\omega$ approaches $\pi/\beta$ to yield  expected conformal scaling behaviour.
\subsection{Replica-diagonal R\'enyi entropies}
Employing the formulation in terms of $n$-copies of the coupled SYK system (outlined in section \ref{Alternative formulation}), with replica-diagonal ansatz
\begin{equation}\label{replica-diag-ansatz}
    G_{a}^{\alpha\beta}(\tau_{1},\tau_{2})= \delta^{\alpha\beta}G_{a}(\tau_{1},\tau_{2})\,,\qquad \Sigma_{a}^{\alpha\beta}(\tau_{1},\tau_{2})= \delta^{\alpha\beta}\Sigma_{a}(\tau_{1},\tau_{2})\,, \qquad a=\chi,\psi\;,
\end{equation}
the $n$-replica effective action with twist field couplings can be evaluated and expressed as the sum of two contributions
\begin{equation}
    I= I_{0} + \Delta I\,,
\end{equation}
where the first term is the $n$ times the {\em unreplicated} action of the coupled SYK systems, 
\bea\label{unreplicated action}
    I_{0}&=&n\,N \sum_{a=\chi,\psi}\left\{-\frac{1}{2}\text{log\,det}(\partial_{\tau}-\Sigma_{a}) + \frac{1}{2}\int_{C_{*}}\int_{C_{*}} d\tau_{1}d\tau_{2}\left[\Sigma_{a}(\tau_{1},\tau_{2})G_{a}(\tau_{1}, \tau_{2})\right.\right.\nonumber\\\nonumber\\
  &-&  \left.\left. \frac{\mathcal{J}^{2}}{2q^{2}}\left[2G_{a}(\tau_{1},\tau_{2})\right]^{q}
    -\frac{\mathcal{V}^{2}}{q^{2}}\left[2G_{\chi}(\tau_{1},\tau_{2})\,2G_{\psi}(\tau_{1},\tau_{2})\right]^{q/2}\right]\right\}\,,
\eea
and the second term $\Delta I$ is the extra portion needed to account  for the effect of  twist fields which couple $\chi$ and $\psi$ fermions differently across replicas as in \eqref{twists}. In the replica diagonal saddle this extra portion is,
\begin{align}\label{DeltaI}
    {\Delta I}=&n N\frac{\mathcal{V}^{2}}{2q^{2}}\left(\int_{C_{1}}d\tau_{1}\int_{C_{2}}d\tau_{2}+\int_{C_{2}}d\tau_{1}\int_{C_{1}}d\tau_{2}\right)(2G_{\chi}(\tau_{1},\tau_{2}))^{q/2}(2G_{\psi}(\tau_{1},\tau_{2}))^{q/2}\notag\\\\
    =&nN\frac{\mathcal{V}^{2}}{q^{2}}\int_{\epsilon}^{\tau_{0}-\epsilon}d\tau_{1}\int_{\tau_{0}+\epsilon}^{\beta-\epsilon}d\tau_{2}(2G_{\chi}(\tau_{1},\tau_{2}))^{q/2}(2G_{\psi}(\tau_{1},\tau_{2}))^{q/2}\,.\notag
\end{align}
Here $\tau_{0}=\beta/2 -2it$ and $\epsilon$  a small UV regulator. This extra piece ensures that when only one of $\tau_{1}$ and $\tau_{2}$ is on the twisted contour $C_2$, the $\mathcal{V}^{2}$ term must vanish as it only couples fermions on different replica copies \eqref{twists}.
The $n$-th R\'enyi entropy is thus
\begin{equation}\label{renyin}
    S^{(n)}_{\chi}=\frac{1}{1-n}\log\frac{Z_n}{Z_1^n}= \frac{\Delta I}{n-1} 
\end{equation}
A straightforward evaluation  
gives the replica diagonal contribution to the Rényi entropy,
\begin{equation}
    S^{(n)}_{\chi}=\,\frac{n}{n-1}\cdot\frac{N}{q^{2}}\cdot\frac{\mathcal{V}^{2}}{\mathcal{J}_{0}^{2}}\,\,\,\log\, \frac{\left|\cosh\omega (2t + i\beta)\right|^2}
    {\cos\omega(\tfrac{1}{2}{\beta} + 2\epsilon)\,\cos\omega\left(\tfrac{3}{2}\beta - 2\epsilon\right)}\,.
\end{equation}
This grows linearly for times $t\gg\beta$, 
\begin{equation}
    S^{(n)}_{\chi}(t)=\frac{n}{n-1}\cdot\frac{2N}{q^{2}}\cdot\frac{\mathcal{V}^{2}}{\mathcal{J}_{0}^{2}}\,\left(2\omega t + \log \frac{\beta {\cal J}_0}{2\pi}\right)\,,\label{linear}
\end{equation}
where in the IR limit $\omega \to \pi/\beta$ and the constant term is evaluated in the same limit taking the cutoff $\epsilon \sim {\cal J}_0^{-1}$. 

It is worth remarking that  the result above is not well defined in the limit $n\to 1$, relevant for entanglement entropy, and therefore although perfectly well defined for all integer $n\neq 1$, it is not able to capture the linear growth of the entanglement entropy. This  problem was already identified in \cite{Gu:2017njx}. However, the authors of \cite{Dadras:2020xfl}, in a slightly different setting\footnote{They considered a product of two thermofield doubles as initial state, rather than the thermofield double of two \textit{interacting} subsystems.}, were able to show that, in a perturbative weak $\mathcal{V}$ coupling regime, the expected growth of entanglement entropy, for $t\gg\beta$, is
\begin{equation}
    \frac{dS^{(1)}_{\chi}(t)}{dt}\simeq-N\frac{x\log(x)}{\beta}, \quad\quad\quad x=\frac{1}{q^{2}}\cdot\frac{\mathcal{V}^{2}}{\mathcal{J}^{2}}\,.\label{linearEE}
\end{equation}
The finite result was found by identifying an additional replica diagrammatic contribution to the R\'enyi entropy, which is higher order in $\mathcal{V}$, but becomes important in the $n\to 1$ limit.


\section{Replica non-diagonal saddle}
\label{sec5}
Following fairly general arguments \cite{Penington:2019kki},  we expect that the early and late time behaviours of R\'enyi entropies are captured by two different saddle point configurations. In particular, early time linear growth as in \eqref{linear} (for the $n=2$ case, reflecting the exponential decay of purity) is governed by the replica diagonal saddle. But as in the case of purity ($n=2$), where its exponential decay leads  to a finite, small floor value, the correct late time asymptotics should be dictated by a replica-non-diagonal saddle (the so-called wormhole saddle in gravity). This  observation is supported by general arguments and numerical studies  \cite{Penington:2019kki, Chen:2020wiq, Almheiri:2019qdq, Goto:2020wnk, Hollowood:2024uuf}. 

The gist of the argument is that at late times, the forward and backward portions of the real time contours  of successive replicas of the system, approximately cancel against each other, leaving behind endpoint contributions originating from the insertions of the twist fields (i.e. the nontrivial boundary conditions implied by them). This has the net effect of the two twist field insertions effectively factorising at late times. Indeed, viewing the twist fields as local operator insertions in the left and right copies, on general grounds we expect at late times,
\be
\langle T_{L}(t_L)T_R(t_R)\rangle \approx \langle T_{L}(t_L)\rangle \langle T_R(t_R)\rangle + {\cal O}(e^{-2\pi(t_L+t_R)\Delta_n/\beta})\,,
\ee
that the connected contribution is exponentially small. Here $\Delta_n$ can be read off from the gradient of the R\'enyi entropy growth at early times.

The key ingredient characterizing the replica non-diagonal or wormhole saddle, expected to describe an old black hole, is the two-point correlator of the $\chi$ fermions $G_\chi(\tau_1, \tau_2)$. At late real times, the effect of the twist operators is purely local and $G_\chi$ is schematically  the sum of three distinct pieces:
\be
G_\chi=G_{\chi, {\rm diag}} + G_{\chi,T_{L}}+G_{\chi,T_{R}}\,,
\label{Gfactorised}
\ee
where the first piece is the disconnected contribution \eqref{RP-sol} accounting for the situation where both fermion  insertions are in the same replica. The remaining two pieces, the ``twist contributions" arise from considering the effect of only a single twist operator ($T_L$ or $T_R$) at a time, ignoring the effect of the other, with individual fermion insertions in {\em different} replicas.  This is the result of late (real) time factorisation of twist field insertions, whose backreaction becomes significant only locally near their insertion points in the $\tau_1-\tau_2$ plane. The twist-factorisation assumption can be explicitly verified a posteriori once the replica correlation functions in the two twist sectors are obtained, where it indeed proves to be well-justified.

Despite the simplifications following from late time factorisation of the twist fields, the calculation of  $G_\chi$, as indicated in \eqref{Gfactorised} will be somewhat involved in the replicated interacting SYK system. The calculation with the  two insertion points in two distinct  replica copies (separated by say $s$ copies) will allow us to infer the modular flowed correlator \eqref{modular_flow_without probe} which is the central object of interest in this paper.

\subsection{Twist contributions to $G_\chi$}
Technically, the Euclidean time calculation is simpler to follow through, so we will implement the appropriate twist boundary conditions in the Euclidean time setting and analytically continue our final result to Lorentzian signature.

For simplicity, we take the the twist operators $T_L$ and $T_R$ to sit at Euclidean times $\tau=0$ and $\tau =\beta/2$ respectively, implementing non-trivial twisted boundary conditions on $k$ copies of the time circle. In the end, these operators must be translated to late real times to describe the late time state when the factorised description dominates.
Assuming such factorisation, we consider each twist field  contribution  to \eqref{Gfactorised} in turn, pretending that the other twist is effectively absent. 

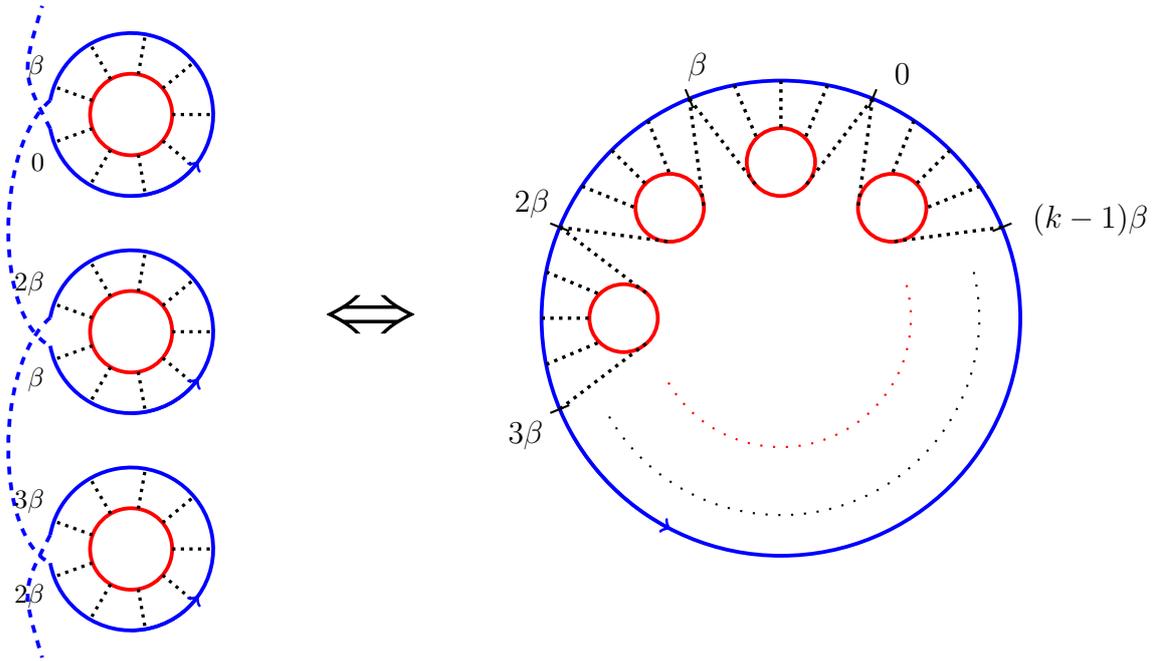
\begin{figure}[ht]
\begin{tikzpicture}[thick,scale=0.9]
\tikzset{
    redcircle/.style={draw=red,  line width=0.5mm},
    outercircle/.style={draw=blue,  line width=0.5mm},
    greenrays/.style={black, line width=0.43mm, dotted},
    bluerays/.style={blue!60!black, thin, dashed},
    bluearr/.style={blue, ->, very thick},
    label/.style={ font=\footnotesize}
}

\foreach \i in {0,1,2} {
    \def\yshift{-\i*3.2}
    
    \draw[outercircle]
    (-6,\yshift) ++(190:1.2) 
    arc[start angle=190,end angle=530,radius=1.2];
    \draw[redcircle]   (-6,\yshift) circle (0.6);
    
    \foreach \a in {0,40,...,320} {
        \draw[greenrays] 
            ({-6+0.6*cos(\a)},{\yshift+0.6*sin(\a)}) -- 
            ({-6+1.2*cos(\a)},{\yshift+1.2*sin(\a)});
    }

    \draw[bluearr] (-6,{\yshift-1.2}) arc[start angle=270,end angle=325,radius=1.2];
}

    \draw[blue, dashed,  line width=0.5mm]
  ({-6 + 1.2*cos(530)},{0 + 1.2*sin(530)})
  .. controls (-8, -0.5) and (-8, -3) ..
  ({-6 + 1.2*cos(190)},{-3.2 + 1.2*sin(190)});
  \draw[blue, dashed,  line width=0.5mm]
  ({-6 + 1.2*cos(530)},{-3.2 + 1.2*sin(530)})
  .. controls (-8, -3.7) and (-8, -6.2) ..
  ({-6 + 1.2*cos(190)},{-6.4 + 1.2*sin(190)});
  
  \draw[blue, dashed,  line width=0.5mm]
  ({-6 + 1.2*cos(530)},{-6.4 + 1.2*sin(530)})
  .. controls (-7.6, -7)..
  (-7.3,-8);

  \draw[blue, dashed,  line width=0.5mm]
  ({-6 + 1.2*cos(190)},{0 + 1.2*sin(190)})
  .. controls (-7.6, 0.6)..
  (-7.3, 1.6);

    \node[label, anchor=east] at (-7.1,0.7) {\(\beta\)};
    \node[label, anchor=east] at (-7.1,-0.7) {\(0\)};
    \node[label, anchor=east] at (-7.1,-2.5) {\(2\beta\)};
    \node[label, anchor=east] at (-7.1,-3.9) {\(\beta\)};
    \node[label, anchor=east] at (-7.1,-5.7) {\(3\beta\)};
    \node[label, anchor=east] at (-7.1,-7.1) {\(2\beta\)};

   { \ifnum\i=0
        \node[font=\Medium] at (-6,\yshift-0.2) {\(\psi\)};
        \node[font=\Medium] at (-5.5,\yshift+1.5) {\(\chi\)};
           \fi
    }

\node at (-2.5, -3) {\scalebox{3}{{$\Leftrightarrow$}}};

\def\R{3.5cm}               
\def\ticklength{0.15} 
\def\labeldistance{0.3} 

\coordinate (C) at (3.5,-3);
\def\delta{0.15cm}
\draw[outercircle] (3.5,-3) circle (\R);
\draw[red, loosely dotted] ({1.9*cos(210)+3.5}, {1.9*sin(210)-3}) arc (210:375:1.9);
\draw[loosely dotted] ({2.9*cos(210)+3.5}, {2.9*sin(210)-3}) arc (210:375:2.9);
\draw[bluearr] ({3.5 + 3.5*cos(210)},{-3 + 3.5*sin(210)}) arc[start angle=210,end angle=242.5,radius=\R];
  \path (C) ++(22.5:\R) coordinate (P);
  \path (C) ++(22.5:{\R - \delta}) coordinate (A);
  \path (C) ++(22.5:{\R + \delta}) coordinate (B);
  \draw[black, thick] (A) -- (B);
  \node[anchor=west] at ({3.5 + 3.8*cos(22.5)},{-3 + 3.8*sin(22.5)}) {$(k-1)\beta$};

  \path (C) ++(67.5:\R) coordinate (P0);
  \path (C) ++(67.5:{\R - \delta}) coordinate (A0);
  \path (C) ++(67.5:{\R + \delta}) coordinate (B0);
  \draw[black, thick] (A0) -- (B0);
  \node[anchor=west] at ({3.5 + 3.9*cos(67.5)},{-3 + 3.9*sin(67.5)}) {$0$};

  \path (C) ++(112.5:\R) coordinate (P1);
  \path (C) ++(112.5:{\R - \delta}) coordinate (A1);
  \path (C) ++(112.5:{\R + \delta}) coordinate (B1);
  \draw[black, thick] (A1) -- (B1);
  \node[anchor=west] at ({3.5 + 4*cos(112.5)},{-3 + 4*sin(112.5)}) {$\beta$};

  \path (C) ++(157.5:\R) coordinate (P2);
  \path (C) ++(157.5:{\R - \delta}) coordinate (A2);
  \path (C) ++(157.5:{\R + \delta}) coordinate (B2);
  \draw[black, thick] (A2) -- (B2);
  \node[anchor=west] at ({3.5 + 4.4*cos(157.5)},{-3 + 4.4*sin(157.5)}) {$2\beta$};
  
  \path (C) ++(202.5:\R) coordinate (P3);
  \path (C) ++(202.5:{\R - \delta}) coordinate (A3);
  \path (C) ++(202.5:{\R + \delta}) coordinate (B3);
  \draw[black, thick] (A3) -- (B3);
  \node[anchor=west] at ({3.5 + 4.5*cos(202.5)},{-3 + 4.5*sin(202.5)}) {$3\beta$};

\def\punctureRadius{0.5}       
\def\punctureDist{2.3}         

\def\lineMaxRadius{2.9}        

\def\labelRadius{3.8}          

    \pgfmathsetmacro{\angle}{0 + 45}
    \pgfmathsetmacro{\x}{3.5 + \punctureDist*cos(\angle)}
    \pgfmathsetmacro{\y}{-3 + \punctureDist*sin(\angle)}
    \draw[redcircle] (\x,\y) circle (\punctureRadius);
    \draw[line width=0.45mm, dotted]
    ({3.5 + 3.5*cos(67.5)},{-3 + 3.5*sin(67.5)})--({\x + \punctureRadius*cos(180)},{\y + \punctureRadius*sin(180)});
    \draw[line width=0.45mm, dotted]
    ({3.5 + 3.5*cos(45)},{-3 + 3.5*sin(45)})--({\x + \punctureRadius*cos(45)},{\y + \punctureRadius*sin(45)});
    \draw[line width=0.45mm, dotted]
    ({3.5 + 3.5*cos(22.5)},{-3 + 3.5*sin(22.5)})--({\x + \punctureRadius*cos(-90)},{\y + \punctureRadius*sin(-90)});
    \draw[line width=0.45mm, dotted]
    ({3.5 + 3.5*cos(33.75)},{-3 + 3.5*sin(33.75)})--({\x + \punctureRadius*cos(0)},{\y + \punctureRadius*sin(0)});
    \draw[line width=0.45mm, dotted]
    ({3.5 + 3.5*cos(56.25)},{-3 + 3.5*sin(56.25)})--({\x + \punctureRadius*cos(90)},{\y + \punctureRadius*sin(90)});

    \pgfmathsetmacro{\angle}{0 + 90}
    \pgfmathsetmacro{\x}{3.5 + \punctureDist*cos(\angle)}
    \pgfmathsetmacro{\y}{-3 + \punctureDist*sin(\angle)}
    \draw[redcircle] (\x,\y) circle (\punctureRadius);
    \draw[line width=0.45mm, dotted]
    ({3.5 + 3.5*cos(67.5+45)},{-3 + 3.5*sin(67.5+45)})--({\x + \punctureRadius*cos(180+45)},{\y + \punctureRadius*sin(180+45)});
    \draw[line width=0.45mm, dotted]
    ({3.5 + 3.5*cos(45+45)},{-3 + 3.5*sin(45+45)})--({\x + \punctureRadius*cos(45+45)},{\y + \punctureRadius*sin(45+45)});
    \draw[line width=0.45mm, dotted]
    ({3.5 + 3.5*cos(22.5+45)},{-3 + 3.5*sin(22.5+45)})--({\x + \punctureRadius*cos(-90+45)},{\y + \punctureRadius*sin(-90+45)});
    \draw[line width=0.45mm, dotted]
    ({3.5 + 3.5*cos(33.75+45)},{-3 + 3.5*sin(33.75+45)})--({\x + \punctureRadius*cos(0+45)},{\y + \punctureRadius*sin(0+45)});
    \draw[line width=0.45mm, dotted]
    ({3.5 + 3.5*cos(56.25+45)},{-3 + 3.5*sin(56.25+45)})--({\x + \punctureRadius*cos(90+45)},{\y + \punctureRadius*sin(90+45)});

    \pgfmathsetmacro{\angle}{0 + 135}
    \pgfmathsetmacro{\x}{3.5 + \punctureDist*cos(\angle)}
    \pgfmathsetmacro{\y}{-3 + \punctureDist*sin(\angle)}
    \draw[redcircle] (\x,\y) circle (\punctureRadius);
    \draw[line width=0.45mm, dotted]
    ({3.5 + 3.5*cos(67.5+45*2)},{-3 + 3.5*sin(67.5+45*2)})--({\x + \punctureRadius*cos(180+45*2)},{\y + \punctureRadius*sin(180+45*2)});
    \draw[line width=0.45mm, dotted]
    ({3.5 + 3.5*cos(45+45*2)},{-3 + 3.5*sin(45+45*2)})--({\x + \punctureRadius*cos(45+45*2)},{\y + \punctureRadius*sin(45+45*2)});
    \draw[line width=0.45mm, dotted]
    ({3.5 + 3.5*cos(22.5+45*2)},{-3 + 3.5*sin(22.5+45*2)})--({\x + \punctureRadius*cos(-90+45*2)},{\y + \punctureRadius*sin(-90+45*2)});
    \draw[line width=0.45mm, dotted]
    ({3.5 + 3.5*cos(33.75+45*2)},{-3 + 3.5*sin(33.75+45*2)})--({\x + \punctureRadius*cos(0+45*2)},{\y + \punctureRadius*sin(0+45*2)});
    \draw[line width=0.45mm, dotted]
    ({3.5 + 3.5*cos(56.25+45*2)},{-3 + 3.5*sin(56.25+45*2)})--({\x + \punctureRadius*cos(90+45*2)},{\y + \punctureRadius*sin(90+45*2)});

    \pgfmathsetmacro{\angle}{0 + 180}
    \pgfmathsetmacro{\x}{3.5 + \punctureDist*cos(\angle)}
    \pgfmathsetmacro{\y}{-3 + \punctureDist*sin(\angle)}
    \draw[redcircle] (\x,\y) circle (\punctureRadius);
    \draw[line width=0.45mm, dotted]
    ({3.5 + 3.5*cos(67.5+45*3)},{-3 + 3.5*sin(67.5+45*3)})--({\x + \punctureRadius*cos(180+45*3)},{\y + \punctureRadius*sin(180+45*3)});
    \draw[line width=0.45mm, dotted]
    ({3.5 + 3.5*cos(45+45*3)},{-3 + 3.5*sin(45+45*3)})--({\x + \punctureRadius*cos(45+45*3)},{\y + \punctureRadius*sin(45+45*3)});
   \draw[line width=0.45mm, dotted]
    ({3.5 + 3.5*cos(22.5+45*3)},{-3 + 3.5*sin(22.5+45*3)})--({\x + \punctureRadius*cos(-90+45*3)},{\y + \punctureRadius*sin(-90+45*3)});
    \draw[line width=0.45mm, dotted]
    ({3.5 + 3.5*cos(33.75+45*3)},{-3 + 3.5*sin(33.75+45*3)})--({\x + \punctureRadius*cos(0+45*3)},{\y + \punctureRadius*sin(0+45*3)});
    \draw[line width=0.45mm, dotted]
    ({3.5 + 3.5*cos(56.25+45*3)},{-3 + 3.5*sin(56.25+45*3)})--({\x + \punctureRadius*cos(90+45*3)},{\y + \punctureRadius*sin(90+45*3)});
\end{tikzpicture}
\caption{\footnotesize{Equivalent topologies of the replica theory with a single twist insertion: solid blue contours represent the $\chi$ fermion system, the $k$ disconnected solid red contours are the replicas of the bath fermion $\psi$ system. Blue dashed lines correspond to the boundary conditions introduced by  the twist field  and finally dotted black lines represent the interactions between the two SYK models. The replica label $s$ increases from the top to the bottom in the left hand side figure, and counter-clockwise on the right.}}
\label{fig4}
\end{figure}

Therefore, for instance if we ignore the effect of $T_R$, the insertion of $T_L$ at $\tau=0$ imposes  the cyclic identifications shown in figure \ref{fig4} on the $k$ replica circles for the $\chi$ fermions.  This results in a circle of  circumference $k\beta$ as shown.

\subsubsection{Contribution from $T_L$ insertion}
We write the fermion correlator in the presence of a single twist operator $T_L$ inserted at $\tau=0$, with fermion insertions at $\tau_1$ and $\tau_2$, in distinct replicas as,
\be
G_{\chi, T_L}^{(s)}(\tau_1, \tau_2) \equiv G_{\chi,0}(\tau_1, \tau_2)\exp\left(\tfrac{1}{q}\,g_{\chi, T_L}^{(s)}(\tau_1, \tau_2)\right)\,,\label{Gchis}
\ee
where $s=0, 1,\ldots k-1$, is the replica distance between the two insertions. Since the bath fermions $\psi$ are only trivially replicated with no correlation between different (disconnected) replicas, i.e. $G_\psi(\tau_1, \tau_2)=0$ for non-zero $s$, the saddle point equation \eqref{saddle-point-eq-fin} has no ${\cal V}$-dependent term,
\begin{equation}\label{saddle_p_e_replica_non_diagonal}
    \partial_{\tau_{1}}\partial_{\tau_{2}}g_{\chi, T_{L}}^{(s)}(\tau_{1},\tau_{2})=- 2\mathcal{J}^{2}\,e^{g_{\chi,T_{L}}^{(s)}(\tau_{1},\tau_{2})}\,,\qquad s\neq 0\,. 
\end{equation}
This follows immediately from the form of the fermion correlator \eqref{Gpsi0}, with $\tau_1,\tau_2$ in different replicas. Next we need to impose the boundary conditions on $g_{\chi, T_L}(\tau_1, \tau_2)$ implied by insertion of $T_L$. 
Without loss of generality, we assume that $\tau_2$ sits in the first replica and $\tau_1$ sits in the $s^{\rm th}$ replica, so that,\footnote{Strictly speaking, the correlator $G_{\chi,T_L}$ should be labelled by {\em two integers}, $G^{p,p^\prime}(\tau_1, \tau_2) = G_{\chi, T_L}(\tau_{1} +p\beta,\tau_{2}+p^\prime \beta)$. To avoid notational clutter we will assume that $p^\prime=0$ and $s=p-p^\prime$ and stick with a single superscript label.}
\begin{equation}
    G^{(s)}_{\chi, T_L}(\tau_{1},\tau_{2})=G_{\chi, T_L}(\tau_{1} +s\beta,\tau_{2}) \quad\quad\quad\quad \tau_{1},\tau_{2}\in [0,\beta).
\end{equation}
Then, continuity across distinct replicas requires,\footnote{It is worth remarking  that these twisted boundary conditions also appeared in \cite{Gao:2021tzr} in their  ``necklace" replica computation, but whereas in \cite{Gao:2021tzr} they constituted a special solution to the actual boundary conditions, in our case \eqref{twist bc, TR} is the immediate consequence of the local twist operator insertion implementing the replica computation.}
\begin{equation}\label{twist bc, TR}
    g_{\chi, T_{L}}^{(s)}(\beta,\tau_{2})=g_{\chi, T_{L}}^{(s+1)}(0,\tau_{2}), \quad\quad\quad g_{\chi, T_{L}}^{(s)}(\tau_{1},0)=g_{\chi,T_{L}}^{(s+1)}(\tau_{1},\beta)\,.\qquad
\end{equation}
Note that when $s=0$, instead of \eqref{saddle_p_e_replica_non_diagonal}, $g^{(0)}_{\chi,T_L}$ satisfies,
\be
\partial_{\tau_{1}}\partial_{\tau_{2}}g_{\chi, T_{L}}^{(0)}(\tau_{1},\tau_{2})=- 2\mathcal{J}_0^{2}\,e^{g_{\chi,T_{L}}^{(0)}(\tau_{1},\tau_{2})}\,,\qquad \mathcal{J}_0^{2}=\mathcal{J}^{2}+
\mathcal{V}^{2}\,.
\ee
This leads to the compact $s$-dependent form for the  Liouville equations for $g_{\chi, T_L}^{(s)} $, 
\begin{equation}\label{Liouville, gs}
    \partial_{\tau_{1}}\partial_{\tau_{2}}g_{\chi}^{(s)}(\tau_{1},\tau_{2})=- 2\mathcal{J}_{0}\mathcal{J}_{s}e^{g_{\chi}^{(s)}(\tau_{1},\tau_{2})},
\end{equation}
with an $s$-dependent coupling,
\begin{equation}
    \mathcal{J}_{s}\equiv\begin{cases}
        \mathcal{J}_{0}\quad\quad\quad s=0 \,{\rm mod} \,k\\
        \frac{\mathcal{J}^{2}}{\mathcal{J}_{0}} \quad\quad\quad s=1,\ldots k-1\,.\\
    \end{cases}
\end{equation}
The general solution to the Liouville equation  \eqref{Liouville, gs} is determined by a pair of  functions $\{f_{s}$, $h\}$  of $\tau_1$ and $\tau_2$ respectively \cite{tsutsumi, Eberlein:2017wah}, 
\begin{equation} \label{Liouville_sol}
    \exp\left(g_{\chi, T_L}^{(s)}(\tau_{1},\tau_{2})\right)=\frac{f_{s}'(\tau_{1})h'(\tau_{2})}{\mathcal{J}_{0}\mathcal{J}_{s}(1+f_{s}(\tau_{1})\,h(\tau_{2}))^{2}}\,.
\end{equation}
In keeping with our convention where the fermion insertion at $\tau_2$ is kept fixed in the reference or zeroth replica, $h(\tau_2)$ will have no dependence on $s$, whilst $f_s(\tau_1)$ will depend nontrivially on the replica number.  The pair of functions satisfying the Liouville equation are determined up to ${\rm SL}(2,\mathbb{R})$ transformations
\begin{equation}
    f_{s}\to {\rm sl}(f_{s})\equiv\frac{a+b\,f_s}{c+d\,f_s}, \quad\quad -h^{-1}\to {\rm sl}(-h^{-1})\equiv\frac{a - b \,h^{-1}}{c- d\,h^{-1}}\,, \quad\quad ad-bc=1\,.
\end{equation}
The $s=0$ solution with the two Majorana insertions in the same replica \eqref{szerosol} corresponds to the following choice of the pair of functions (up to the SL$(2, {\mathbb R})$ freedom),
\begin{equation}
    f_{0}(\tau)=\tan\omega\left(\tau-\tfrac{\beta}{2}\right), \quad\quad\quad h(\tau)=\tan\omega\tau\,.
\end{equation}
\subsubsection{Ansatz and recursive relations}
\label{recursive}
Our goal  now is to find the functions $f_{s}$ for all $s$,  $0\leq s\leq k-1$ using the continuity conditions \eqref{twist bc, TR}. Since the fermion insertion at $\tau_2$ can be viewed as an insertion in the zeroth replica, the function $h(\tau_2)=\tan\omega \tau_2$ will be treated as fixed  in 
the modular flowed correlator \eqref{Liouville_sol}. All nontrivial analytic dependence on $s$ is then captured by $f_s(\tau_1)$. Given the general form  \eqref{Liouville_sol} it is easy to check that the boundary conditions \eqref{twist bc, TR} imply that the $f_s$ are related to $f_0$ by a M\"obius transformation.

The M\"obius transformation can be naturally characterised by the functional form (see e.g. \cite{Gao:2021tzr}),
\begin{equation}\label{anstaz,fs}
    f_{s}(\tau)=u_{s}+v_{s}\tan{(\omega\tau+\gamma_{s})}\,,
\end{equation}
in terms of three independent parameter sets $\{u_{s}, v_{s}, \gamma_{s}\}$. 
Plugging in the ansatz for $f_{s}(\tau_{1})$ \eqref{anstaz,fs} and $h(\tau_{2})=\tan\omega\tau_{2}$ in \eqref{Liouville_sol},  
\bea\label{Liouville,sol2}
&&\exp\left(g_{\chi, T_{L}}^{(s)}(\tau_1, \tau_2)\right)=\\\nonumber
&&\qquad\frac{\omega^{2}v_{s}}{\mathcal{J}_{0}\mathcal{J}_{s}}\big(\cos(\omega\tau_{1}+\gamma_{s})(\cos\omega\tau_{2}+u_{s}\sin\omega\tau_{2})+v_{s}\sin\omega\tau_{2}\sin(\omega\tau_{1}+\gamma_{s})\big)^{-2}\,.
\eea
Comparison with the $s=0$ solution \eqref{RP-sol} yields,
\begin{equation}
    u_{0}=0, \quad\quad  v_{0}=1, \quad\quad\gamma_{0}=-\frac{\omega\beta}{2}\,.
\end{equation}
Now we apply the twist boundary conditions \eqref{twist bc, TR} to the ansatz \eqref{Liouville,sol2} and obtain a set of algebraic conditions which can be solved recursively for $\{u_{s}, v_{s}, \gamma_{s}\}$,
\bea
&&u_{s+1} =\tan(\omega\beta+\gamma_{s})-\frac{1}{2}\alpha_{s}v_{s}\sin2\gamma_{s+1}\sec^{2}(\omega\beta+\gamma_{s})\,,\\
&&v_{s+1} =\alpha_{s}v_{s}\cos^{2}\gamma_{s+1}\sec^{2}(\omega\beta+\gamma_{s})\,, \label{first_rec_rel_TR}\\
&&\tan\gamma_{s+1}=\tan\gamma_{s}+\alpha_{s}v_{s}\sin\omega\beta\sec\gamma_{s}\sec(\omega\beta+\gamma_{s})\,,\label{second_rec_rel_TR}\\
&&u_{s} =(1-v_{s})\tan(\omega\beta+\gamma_{s})\,.\label{self_consistency_us}
\eea
Here we have introduced the parameter $\alpha_s$ for the ratio of the effective couplings in successive replicas,
\bea
\alpha_{s}&\equiv\frac{\mathcal{J}_{s+1}}{\mathcal{J}_{s}}=\begin{cases}
        \frac{\mathcal{J}^{2}}{\mathcal{J}_{0}^{2}}\quad\quad\quad s=0\,,\\
        1 \quad\quad\quad 0<s<k-1\,,\\
        \frac{\mathcal{J}_{0}^{2}}{\mathcal{J}^{2}}\quad\quad\quad s=k-1\,.\\
    \end{cases}
\eea
We note that in all formulae above we are not taking the strict IR limit i.e. we assume that $\omega\beta < \pi$. In the strict IR limit ($\omega/{\cal J}_0 \to 0$), $\omega\beta$ approaches $\pi$ and the recursion relations become trivial.
\subsubsection{KMS symmetries}
The reality of the Euclidean correlator for Majorana fermions implies,
\begin{equation}
\text{Tr}(\rho^{k-s}\chi_{i}(\tau_{1})\rho^{s}\chi_{j}(\tau_{2}))=\text{Tr}(\rho^{k-s}\chi_{j}(-\tau_{2})\rho^{s}\chi_{i}(-\tau_{1}))\,,
\end{equation}
where the right hand side is obtained by applying Hermitean conjugation inside the trace operation. This result, along with the replica symmetry under simultaneous  translation of both insertion points by $\beta$ (one replica unit), yields the  ``small" KMS symmetry condition,
\be
G^{(s)}_{\chi, T_L}(\tau_1, \tau_2) = G^{(s)}_{\chi, T_L}(\beta-\tau_2, \beta-\tau_1)\,.
\ee
This symmetry can be seen explicitly in  \eqref{Liouville,sol2}, using   \eqref{self_consistency_us} to rewrite $g_{\chi, T_{L}}^{(s)}$ as
\bea\label{replica_correlator_v-gamma}
&&\exp\left(g_{\chi, T_{L}}^{(s)}(\tau_1, \tau_2)\right)=\\\nonumber
&&\qquad
\frac{\omega^{2}v_{s}\cos^{2}(\omega\beta+\gamma_{s})}{\mathcal{J}_{0}\mathcal{J}_{s}\left[\cos(\omega\tau_{1}+\gamma_{s})\cos(\omega(\tau_{2}-\beta)-\gamma_{s})+v_{s}\sin\omega\tau_{2}\sin\omega(\tau_{1}-\beta)\right]^{2}}\,\,.
\eea
The small KMS symmetry is now manifest, so that 
\begin{equation}
    g^{(s)}_{\chi, T_L}(\tau_{1},\tau_{2})=g^{(s)}_{\chi,T_L}(\beta-\tau_{2},\beta-\tau_{1})\,.
\end{equation}
as required.

In addition to the small KMS symmetry, we expect the replica correlator to satisfy another symmetry for $s\neq0,k$, following from cyclicity of the trace,
\begin{equation}
\text{Tr}(\rho^{k-s}\chi_{i}(\tau_{1})\rho^{s}\chi_{j}(\tau_{2}))=\text{Tr}(\rho^{s}\chi_{j}(\tau_{2})\rho^{k-s}\chi_{i}(\tau_{1}))
\end{equation}
which implies 
\begin{equation}\label{big_KMS}
g^{s}(\tau_{1},\tau_{2})=g^{k-s}(\tau_{2},\tau_{1})\,,
\end{equation}
a so-called ``big" KMS symmetry.
This symmetry is not automatically guaranteed by the recurrence relations, thus we will check it a posteriori once we solve \eqref{first_rec_rel_TR}-\eqref{second_rec_rel_TR}. In particular, the big KMS symmetry is preserved as long as the SL$(2,\mathbb{R})$ parameters satisfy,
\begin{equation}\label{big_KMS2}
    v_{k-s}=v_{s}\frac{\cos^{2}(\gamma_{k-s})}{\cos^{2}(\gamma_{s})},\quad\quad\quad\quad \sin(\gamma_{s}+\gamma_{k-s}+\omega\beta)=v_{s}\sin(\omega\beta)\frac{\cos(\gamma_{k-s})}{\cos(\gamma_{s})}\,.
\end{equation}

\subsubsection{Solving $v_{s}$ and $\gamma_{s}$ recurrence relations}
\label{sec:recursionsolve}
The task of solving for the correlator in the replica geometry with factorised twist fields now reduces to solving the  two  recurrence relations  \eqref{first_rec_rel_TR} and \eqref{second_rec_rel_TR} for $v_s$ and $\gamma_s$. Finding closed form expressions for $\gamma_{s}$ and $v_{s}$ as functions of $s$ appears to be a nontrivial challenge. Instead we choose to work in the limit of small interactions between system and bath, to second order in  $\mathcal{V}/\mathcal{J}$. At this order, $v_s$ and $\gamma_s$ are given by relatively simple discrete sums, 
\bea
    &&\gamma_{s}=\frac{2s-1}{2}\omega\beta + \frac{\mathcal{V}^{2}}{\mathcal{J}^{2}}\left[\sum_{n=1}^{2s-1}(-1)^{n}(2s-n)\sin{(\omega\beta n)}\right]\,,\\
    &&v_{s}=1 + \frac{\mathcal{V}^{2}}{\mathcal{J}^{2}}\left[(2s-1)+2\sum_{n=1}^{2s-1}(-1)^{n}(2s-n)\cos{(\omega\beta n)}\right]\,.
\eea
The sums can be evaluated exactly (see Appendix \ref{app:sums}),
\begin{empheq}[box=\fbox]{align}
    &\gamma_{s}= \frac{2s-1}{2}\omega\beta - \frac{\mathcal{V}^{2}}{2\mathcal{J}^{2}}\frac{2s\sin{\omega\beta}+\sin{(2\omega\beta s)}}{1+\cos{\omega\beta}}\,,\label{gammas}\\
    &v_{s}= 1 - \frac{\mathcal{V}^{2}}{2\mathcal{J}^{2}}\frac{\cos{\omega\beta}+\cos{(2\omega\beta s)}}{\cos^{2}{\frac{\omega\beta}{2}}}\,.\label{vs}
\end{empheq}
It is important that  we are not in the strict IR limit so that  $\omega\beta < \pi $. Then the corrections at leading order  in ${{\mathcal V}^{2}}/{{\mathcal{J}}^{2}}$ are non-vanishing for any integer $s\geq 0$ and the results for $v_s$ and $\gamma_s$ can be analytically continued to non-integer $s$.

One can easily show that our $\{\gamma_{s},v_s\}$ solutions satisfy the big KMS symmetry \eqref{big_KMS2} in the limit of interest $k\to 1$,\footnote{In computing the replica non-diagonal correlations functions, we set from the beginning $k=1$, since we are ultimately interested in taking this limit at the end of the calculations. The results for general $k$ can be obtained via the following replacements
\begin{equation}
    \gamma_{0}\to -\frac{k\omega\beta}{2}, \quad\quad\quad \gamma_{s}\to \gamma_{s+\frac{1-k}{2}}\,,\quad\quad\quad v_{s}\to v_{s+\frac{1-k}{2}}\,,
\end{equation}
with $\{\gamma_{s+\frac{1-k}{2}},v_{s+\frac{1-k}{2}}\}$ still satisfying the recurrence relations up to $\mathcal{O}({{\mathcal V}^{2}}/{{\mathcal{J}}^{2}})$ and the big KMS symmetry up to the same order and for any integer $k$.} up to order ${{\mathcal V}^{2}}/{{\mathcal{J}}^{2}}$ and at any order in $\omega$.

To explore the effect of modular flow we first perform the analytic continuation $s\to is$ in \eqref{gammas} and \eqref{vs} to obtain the functions $v(s)$ and $\gamma(s)$ where $s$ is now the modular flow parameter. Following this we consider an IR limit, 
 \begin{equation}\label{omega_IR}
    \omega\simeq \frac{\pi}{\beta}\left(1-\frac{2}{\beta{\mathcal J}_0}+\mathcal{O}\left(\frac{1}{\beta^{2}\mathcal{J}^{2}_0}\right)\right)\,,\qquad \beta{\cal J}_0\gg1\,
\end{equation}
with 
\be
\beta {\cal V} \ll 1\,.\label{smallv}
\ee
This also implies $\beta\mathcal{J}_{0}\simeq\beta\mathcal{J}\gg1$. Physically, this limit keeps the effective coupling between system and bath weak, whilst the system itself remains  strongly coupled. 
Taking these limits we obtain the analytic continuations of  $\gamma_{s}$ and $v_{s}$, 
\bea
    &&v(s)= 1-\frac{1}{2}\left(\frac{\beta\mathcal{V}}{\pi}\right)^{2}(\cosh{(2\pi s)} -1)\;+\mathcal{O}\left(\frac{\beta\mathcal{V}^{2}}{\mathcal{J}}\right)\,,\label{v(s)}\\
    &&\gamma(s)=\pi\left(1-\frac{2}{\beta\mathcal{J}}\right)\left(i s -\frac{1}{2}\right)-i\left(\frac{\beta\mathcal{V}}{2\pi}\right)^{2}\sinh{(2\pi s)} \;+\mathcal{O}\left(\frac{\beta\mathcal{V}^{2}}{\mathcal{J}}\right)\,.\label{gamma(s)}
\eea

\subsubsection{$T_{R}$ twist contribution}
We can now turn to the effect of the the twist $T_R$ operator inserted at Euclidean time $\tau = \beta/2$, ignoring $T_L$ or tracing over the left side of the $\chi$ system within replicas. 
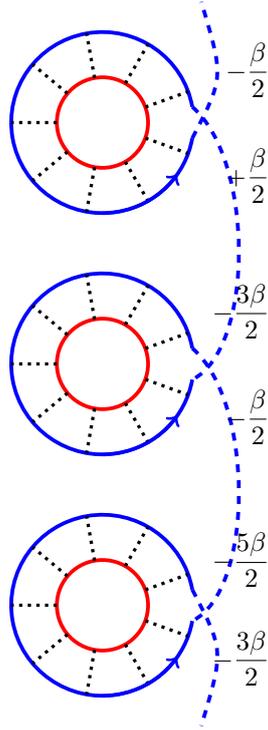
\begin{figure}[h]
    \centering
    \begin{tikzpicture}[thick]
\tikzset{
    redcircle/.style={draw=red, line width=0.5mm},
    outercircle/.style={draw=blue, line width=0.5mm},
    greenrays/.style={black, line width=0.43mm, dotted},
    bluerays/.style={blue!60!black, thin, dashed},
    bluearr/.style={blue, ->, very thick},
    label/.style={font=\footnotesize}
}

\foreach \i in {0,1,2} {
    \def\yshift{-\i*3.2}
   
    \draw[outercircle]
    (-6,\yshift) ++(10:1.2) 
    arc[start angle=10,end angle=350,radius=1.2];
    \draw[redcircle]   (-6,\yshift) circle (0.6);
   
    \foreach \a in {20,60,...,340} {
        \draw[greenrays] 
            ({-6+0.6*cos(\a)},{\yshift + 0.6*sin(\a)}) -- 
            ({-6+1.2*cos(\a)},{\yshift + 1.2*sin(\a)});
    }

    \draw[bluearr] (-6,{\yshift-1.2}) arc[start angle=270,end angle=325,radius=1.2];
}
    \draw[blue, dashed, line width=0.5mm]
  ({-6 + 1.2*cos(350)},{0 - 1.2*sin(350)})
  .. controls (-4, -0.5) and (-4, -3) ..
  ({-6 + 1.2*cos(10)},{-3.2 - 1.2*sin(10)});
  \draw[blue, dashed, line width=0.5mm]
  ({-6 + 1.2*cos(350)},{-3.2 - 1.2*sin(350)})
  .. controls (-4, -3.7) and (-4, -6.2) ..
  ({-6 + 1.2*cos(10)},{-6.4 - 1.2*sin(10)});
  
  \draw[blue, dashed, line width=0.5mm]
  ({-6 + 1.2*cos(350)},{-6.4 - 1.2*sin(350)})
  .. controls (-4.4, -7)..
  (-4.7,-8);

  \draw[blue, dashed, line width=0.5mm]
  ({-6 + 1.2*cos(10)},{0 - 1.2*sin(10)})
  .. controls (-4.4, 0.6)..
  (-4.7, 1.6);

    \node[label, anchor=east] at (-3.65,0.7) {\(-\displaystyle\frac{\beta}{2}\)};
    \node[label, anchor=east] at (-3.65,-0.7) {\(+\displaystyle\frac{\beta}{2}\)};
    \node[label, anchor=east] at (-3.65,-2.5) {\(-\displaystyle\frac{3\beta}{2}\)};
    \node[label, anchor=east] at (-3.65,-3.9) {\(-\displaystyle \frac{\beta}{2}\)};
    \node[label, anchor=east] at (-3.65,-5.8) {\(-\displaystyle\frac{5\beta}{2}\)};
    \node[label, anchor=east] at (-3.65,-7.1) {\(-\displaystyle\frac{3\beta}{2}\)};

   { \ifnum\i=0
        \node[red, font=\Medium] at (-6,\yshift-0.2) {\(\psi\)};
        \node[font=\Medium] at (-5.5,\yshift+1.5) {\(\chi\)};
    \fi}
\end{tikzpicture}
    \caption{Topology of the replica theory if we trace out the left twist $T_{L}$. $s$ increases from the top to the bottom.}
\end{figure}
The analysis is  identical to that for $T_L$, except now the twist boundary conditions \eqref{twist bc, TR} become, 
\begin{equation}\label{twist bc, TL}
    g_{\chi,T_{R}}^{(s)}\left(\frac{3\beta}{2},\tau_{2}\right)=g_{\chi, T_{R}}^{(s+1)}\left(\frac{\beta}{2},\tau_{2}\right), \quad\quad\quad g_{T_{R}}^{(s)}\left(\tau_{1},\frac{\beta}{2}\right)=g_{T_{R}}^{(s+1)}\left(\tau_{1},\frac{3\beta}{2}\right)\,,
\end{equation}
corresponding to a $\beta/2$ shift in both arguments in \eqref{twist bc, TR}.
Therefore the solution for the fermion correlator in the right side of the thermofield double is determined by the function,
\begin{equation}\label{replica_correlator_TR}
g_{\chi, T_{R}}^{(s)}\left(\tau_{1},\tau_{2}\right)=g_{\chi, T_{L}}^{(s)}\left(\tau_{1}+\tfrac{\beta}{2},\tau_{2}+\tfrac{\beta}{2}\right)\,.
\end{equation}
\subsection{R\'enyi entropies in the replica non-diagional saddle}
Whilst our main interest in the calculation of modular flowed fermion correlators for which we have assembled the main ingredients, let us pause to review the contribution of the replica non-diagonal saddle to the R\'enyi entropies,
\begin{equation}
    \frac{(n-1)S^{(n)}_{\chi}}{N}=\frac{I^{(n)}+n\log{Z_1}}{N}\,,
\end{equation}
where $Z_1$ is the partition function of a copy of the entire system and $I^{(n)}$ is the $n$-replica action \eqref{In-2}, which we rewrite below for convenience,
\bea\label{reminder-action}
 \frac{I^{(n)}}{N} = &&\sum_{a=\chi,\psi}\left[-\log\text{Pf}\left(\partial_\tau \delta^{\alpha\alpha'}-\Sigma_a^{\alpha\alpha'}\right) + \right.\\\nonumber
  +&&\left.  \frac{1}{2}\,\int_{C_{*}}\mathrm{d}\tau\mathrm{d}\tau'\sum_{\alpha,\alpha'}\left(\Sigma_a^{\alpha\alpha'}(\tau,\tau')G_a^{\alpha\alpha'}(\tau,\tau') - \frac{\mathcal{J}^2}{2q^{2}}\,\left[2G_{a}^{\alpha\alpha'}(\tau,\tau')\right]^q\right)\right]\\\nonumber
   -&& \frac{\mathcal{V}^2}{2q^{2}}\int_{C_{*}}\mathrm{d}\tau\mathrm{d}\tau' \sum_{\alpha\alpha'\beta\beta'}\left[2G_{\chi}^{\alpha\alpha'}(\tau,\tau')\right]^{q/2} \ g^{\alpha\beta}(\tau)g^{\alpha'\beta'}(\tau')  \left[2G_{\psi}^{\beta\beta'}(\tau,\tau')\right]^{q/2}\,.
\eea
Focussing attention on the final line of the expression, and noting that the twist boundary conditions are encoded in $ g^{\alpha\beta}(\tau)$ \eqref{twists} and that the $\psi$-correlation functions are replica diagonal, $G_{\chi}^{\beta,\beta'}(\tau,\tau')\propto\;\delta^{\beta\beta'}\text{sign}(\tau-\tau')$, 
the order ${\mathcal V}^2$ term simplifies,
\begin{align}\label{V-terms1}
   \frac{I^{(n)}}{N}=\ldots&-n\frac{\mathcal{V}^{2}}{2q^{2}}\left[\int_{C_{1}}d\tau\int_{C_{1}}d\tau'+\int_{C_{2}}d\tau\int_{C_{2}}d\tau'\right]\left(2G^{(s=0)}_{\chi}(\tau,\tau')\right)^{q/2}\left(2G_{\psi}(\tau,\tau')\right)^{q/2}\,\nonumber\\\\\nonumber
    &-n\frac{\mathcal{V}^{2}}{q^{2}}\int_{C_{2}}d\tau\int_{C_{1}}d\tau'\left(2G^{(s=1)}_{\chi}(\tau,\tau')\right)^{q/2}\left(2G_{\psi}(\tau,\tau')\right)^{q/2}\,.
\end{align}
Here,
\begin{equation}\label{Gs1}
    G^{(s=1)}_{\chi}(\tau,\tau')=G_{\chi,T_{L}}^{(s=1)}(\tau,\tau')+G_{\chi,T_{R}}^{(s=1)}(\tau,\tau')\,,
\end{equation}
and $G_{\chi}^{(s=0)}(\tau,\tau'),\,G_{\psi}(\tau,\tau')$ are the usual untwisted correlation functions \eqref{RP-sol}. Importantly, we have 
implicitly assumed the solutions \eqref{replica_correlator_v-gamma} and \eqref{replica_correlator_TR} for integer $s$ to obtain the replica non-diagonal two point function corresponding to having the two $\chi$ fermions separated by one replica unit.
Secondly, the late-time saddle is almost time-translation invariant, with the latter being broken only near the twist insertions, implying that the forward and backward real time contour contributions cancelmutually. At leading order in IR limit $\beta\mathcal{J}\gg1$ the fermion Green's function  with $s=1$ is
\begin{equation}
    G_{\chi}^{(s=1)}(\tau,\tau')\simeq \frac{1}{4}\text{sgn}(\tau-\tau')\left[\frac{\beta\mathcal{J}}{\pi}\sin{\left(\frac{\pi}{\beta}\Big|\tau-\tau'\Big|\right)}\right]^{-2/q}\,,
\end{equation}
which leadst to a simplification of the order ${\mathcal V}^2$ term \eqref{V-terms1},
\begin{equation}
\frac{I^{(n)}}{N}=\ldots    -n\frac{\mathcal{V}^{2}}{2q^{2}}\int_{0}^{\beta} d\tau\int_{0}^{\beta}d\tau'\left(2G^{(s=0)}_{\chi}(\tau,\tau')\right)^{2/q}\left(2G_{\psi}(\tau,\tau')\right)^{2/q}\,.
\end{equation}
This ${\cal O}({\mathcal V}^2)$ contribution is identical to the corresponding one  in the single replica action \eqref{unreplicated action}, and therefore the late-time R\'enyi entropy $S^{(n)}_{\chi}\propto\;( I^{(n)}+n\log{Z_1})$ is independent of $\mathcal{V}$! As a consequence, we can effectively set  $\mathcal{V}=0$ in our replica non-diagonal correlators \eqref{replica_correlator_v-gamma} so that,
\bea
\gamma_{s}=\frac{2s-1}{2}\omega\beta\,,\qquad v_{s}=1\,.
\eea
Therefore in the IR limit we can set,
\begin{equation}
    G^{\alpha\alpha'}_{\chi,T_{L}}(\tau, \tau')\simeq G^{\alpha\alpha'}_{\chi,T_{R}}(\tau, \tau')\simeq \frac{1}{2}G_{\chi}^{(s=1)}(\tau, \tau')\,.
\end{equation}
$I^{(n)}$  reduces to the action of the SYK$_q$ system, and the Rényi entropy simplifies,
\begin{align}
    \frac{(1-n)S^{(n)}_{\chi}}{N}=& (1-n)\left\{-\frac{1}{2}\text{log\,det}(\partial_{\tau}-\Sigma_{\chi})+ \right.\notag\\
    &\left.+\frac{1}{2}\int_{0}^{\beta} d\tau\int_{0}^{\beta}d\tau' \left[\Sigma_{\chi}(\tau,\tau')G_{\chi}(\tau,\tau') - \frac{\mathcal{J}^{2}}{2q^{2}}(2G_{\chi}(\tau,\tau'))^{q}\right]\right\}\,.
\end{align}
The term in curly brackets can be recognized as the high temperature thermal entropy of the SYK$_{q}$ model per degree of freedom and we obtain,
\be
    S^{(n)}_{\chi}= N\log2\,.
\ee
Together with \eqref{linear} and \eqref{linearEE} this implies a Page time for the R\'enyi and entanglement entropies. In the large $q$ regime, with weak coupling to the bath, this crossover time  between the replica-diagonal and the replica non-diagonal saddle is naturally parametrically large. In what follows below we will assume that we are in the replica non-diagonal saddle so a late time limit will be implicit; for the sake of clarity of the exposition we will however often simply keep the twist field insertions  at $t_{\rm twist}=0$. In section \ref{sec10} we explain how the results for  arbitrary $t_{\rm twist}$  can be obtained straightforwardly.

\section{Modular flowed fermion correlators}
\label{sec6}
\subsection{Single-sided correlator}
We now want to compute real time correlators, both single-sided and two-sided,  of Majorana fermions in the SYK$_\chi$ system. These can be obtained  readily from the Euclidean correlation functions, but we do need to recall the dictionary between the single Euclidean fields $\left\{\chi^j(\tau)\right\}$ we have been working with thus far, and fields of the TFD  copies,\footnote{The dictionary can be quickly checked by rewriting  the two-sided real time correlator in the ${\cal V}=0$ theory, as an ordered thermal correlator of a single field, namely $\langle {\rm TFD}|\chi_L(t_L)\chi_R(t_R)|{\rm TFD}\rangle={\rm Tr}\left(\chi(-t_L+i\beta/4) e^{-\beta H/2}  \chi(t_R+i\beta/4) e^{-\beta H/2}\right)$, where we have trivially evolved both arguments by an equal amount  $i\beta/4$ along the Euclidean thermal circle. Now writing $t_{L,R}= i\tau_{L,R}$ and viewing  the correlator as a Euclidean two-point function ${\rm Tr}\left(\chi(\tau_1) e^{-\beta H/2}  \chi(\tau_2) e^{-\beta H/2}\right)$ with $\tau_{1,2}\in[0,\beta/2]$, we obtain the dictionary \eqref{fermion_depurification}.}  
\begin{align}\label{fermion_depurification}
    \chi^{j}(\tau)=\begin{cases}
        \chi^{j}_{L}(\beta/4-\tau) \quad\quad\quad\quad &0\leq\tau\leq \beta/4\\
        \\
        i\chi^{j}_{R}(\tau-\beta/4) \quad &\beta/4\leq\tau\leq \beta/2\,.
    \end{cases}
\end{align}
\begin{figure}[ht]
\begin{center}
\begin{tikzpicture}[scale=0.8]
\draw[->,>=stealth,line width=0.45mm] (-3.5,0) arc (180:360:3.5);
\draw[<-, >=stealth, blue, line width=1mm] (3.0,0) arc (360:270:3.0);
\draw[->, >=stealth,, red, line width=1mm] (0,-3.0) arc (270:180:3.0);
\node at (-4.5, -0.3) {$\tau=0$ };
\node at (4.5, -0.3) {$\tau=\beta/2$ };
\node at (-2.5, +0.3) {\textcolor{red}{$\tau_L=\frac{\beta}{4}$} };
\node at (2.5, +0.3) {\textcolor{blue}{$\tau_R=\frac{\beta}{4}$} };
\filldraw[fill=black](0,-3) circle (0.2);
\node at (0, -2.55) {$\tau_{L,R}=0$};
\end{tikzpicture}
\caption{\footnotesize Pictorial representation of the dictionary between the single (Euclidean) field $\chi$ and the Majoranas $\chi_{L,R}$ of the thermofield double. The Euclidean times $\tau_{L,R}$ are  determined as per eq. \eqref{fermion_depurification} and analytically continued to real time by $\tau_{L,R} \to - i t_{L,R}$. }
\label{TFDdictionary} 
\end{center}
\end{figure}
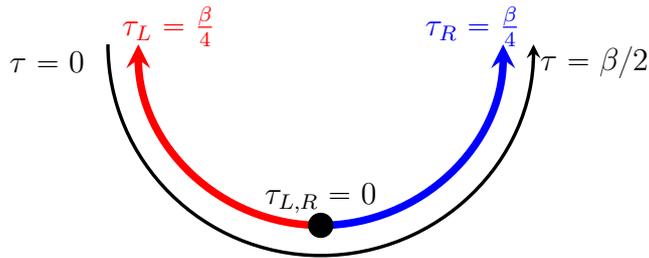
We first consider the single-sided real time correlator, 
\begin{equation}
 W_{LL}^{(s)}(0, t) \,=\, \frac{1}{N}\sum_{j=1}^{N}\text{Tr}[\rho^{1-is}\chi_{L}^{j}(0)\rho^{is}\chi_{L}^{j}(t)]\,,
\end{equation}
obtained by analytically continuing the modular flowed Euclidean two-point function ($\tau_{1}=0$ and $\tau_{2}=-it$) where  the operator insertion at $t=0$ is subjected to modular flow whilst the second operator is evolved forward to a later real time time $t$.

In the late time limit when the replica partition function is dominated by the factorised saddle point, the correlation function is given by the sum of the two independent twist field contributions discussed above,
\be
W_{LL}^{(s)}(0, t) = G_{\chi, T_L}^{(s)}(0, -it) + G_{\chi, T_R}^{(s)}(0, -it) \,,
\label{factorisedLLcorr}
\ee
where $G_{\chi, T_{L,R}}^{(s)}$ are defined via eqs. \eqref{Gchis}, \eqref{Liouville,sol2}  and their $L\to R$ counterparts. They are then readily evaluated  using \eqref{replica_correlator_v-gamma}, \eqref{gamma(s)}, \eqref{v(s)} and \eqref{replica_correlator_TR}, employing the small $\beta{\cal V}$ expansion in the IR limit ($\beta {\cal J}\to \infty$).\footnote{We will often refer to the object $\exp\left(g^{(s)}_{\chi,T_{L,R}}(\tau_{1},\tau_{2})/q\right)$ 
as the modular flowed correlator, keeping in mind that the correlator is actually $G^{(s)}_\chi(\tau_{1},\tau_{2})=G_{0,\chi}^{(s)}(\tau_{1},\tau_{2})\exp\left(g^{(s)}_{\chi, T_L}(\tau_{1},\tau_{2})/q\right)+L\to R$.
} The time dependence takes the general form,
\bea\label{one_sided_corr_microscopic}
    &&\exp\left(\tfrac1q \,g_{\chi, T_{L}}^{(s)}(0,-it)\right)\simeq \frac{\left({\pi}/{\beta\mathcal{J}}\right)^{2/q}}{\left[C^{LL}_{T_{L},1}(s)\,\cosh{\omega t} + C^{LL}_{T_{L},2}(s)\,\sinh{\omega t}\right]^{2/q}}\,,\\\nonumber\\\label{llright}
    &&\exp\left(\tfrac1q g_{\chi, T_{R}}^{(s)}(0,-it)\right)\simeq \frac{\left({\pi}/{\beta\mathcal{J}}\right)^{2/q}}{\left[C^{LL}_{T_{R},1}(s)\,\cosh{\omega t} + C^{LL}_{T_{R},2}(s)\,\sinh{\omega t}\right]^{2/q}}\,,
\eea
with $\omega = \pi/\beta +{\cal O}(1/{\cal J}\beta)$ as in \eqref{omega_IR}, and 
we have introduced  coefficients $\{C^{LL}_{T_{L(R),i}}\}$ which are functions of the modular parameter $s$. Since we solve the recurrence relations \eqref{first_rec_rel_TR} and \eqref{second_rec_rel_TR} for the SL$(2, \mathbb R)$ parameters to order $\mathcal{O}(\mathcal{V}^{2}/\mathcal{J}^{2})$, and then take the IR limit \eqref{omega_IR}, this determines  the 
$\{C^{LL}_{T_{L(R),i}}\}$ to quadratic order in the interaction strength between bath and system,\footnote{We expect that when the recursive relations are solved at higher orders in ${\cal V}^2/{\cal J}^2$, terms of $\mathcal{O}(\mathcal{V}^{2n}/\mathcal{J}^{2n})$ in \eqref{gammas}, \eqref{vs}, with $n>1$, can give rise to contributions of $\mathcal{O}(\beta^{2n}\mathcal{V}^{2n})$ when taking the IR limit. In our approach we are  taking $\beta {\cal V} \ll1$ and $\beta{\cal J}\gg 1$ with the IR  limit $\beta {\cal J}\to \infty$ to be taken first ({\em after} continuing the replica results to non-integer $s$).}
\bea
    &&C^{LL}_{T_{L},1}(s)= \left[i\sinh{\pi s}\left(1-\frac{\beta^{2}\mathcal{V}^{2}}{2\pi^{2}}\right)-\tfrac{2\pi}{\beta\mathcal{J}} s 
    \cosh{\pi s}\right]+\mathcal{O}\left(\beta^4\mathcal{V}^{4}, \tfrac{\beta{\cal V}^2}{\cal J}\right)\,, \nonumber\\\nonumber
    &&C^{LL}_{T_{L},2}(s)=\left[i\cosh{\pi s}-\tfrac{2\pi}{\beta\mathcal{J}}\,s
    %
    \sinh{\pi s}
    \right]+\mathcal{O}\left(\beta^4\mathcal{V}^{4}, \tfrac{\beta{\cal V}^2}{\cal J}\right)\,,\\
    \\\nonumber
    &&C^{LL}_{T_{R},1}(s)= i\sinh{\pi s}\left(1+\frac{\beta^{2}\mathcal{V}^{2}}{2\pi^{2}}\right)-\tfrac{2\pi}{\beta\mathcal{J}}\,s
    \cosh{\pi s}
    +\mathcal{O}\left(\beta^4\mathcal{V}^{4}, \tfrac{\beta{\cal V}^2}{\cal J}\right)\,,\\\nonumber
    &&C^{LL}_{T_{R},2}(s)=i\cosh{\pi s}-\tfrac{2\pi}{\beta\mathcal{J}}\,s
    \sinh\pi s+\mathcal{O}\left(\beta^4\mathcal{V}^{4}, \tfrac{\beta{\cal V}^2}{\cal J}\right)\,.
\eea
When the system-bath interaction is turned off ($\beta{\cal V}=0$), the coefficients are precisely such that modular flow amounts to an ordinary time translation $t\to t+\beta s$ and therefore the correlator exhibits the expected singularity when $\chi_L(0)$ is evolved by modular flow by an amount $s=-t/\beta$ and the two insertion points come together. This is what we expect, since when $\beta {\cal V}=0$, the $\chi$-system is in a pure TFD state and the single-sided correlator is a thermal correlator, whilst modular flow coincides with unitary time evolution.

When a small interaction is turned on between bath and system, two singularities in the correlator now appear when either one of the two denominators above vanishes. 
This can happen at two different real times $t=t_\pm$ along the modular flow, for sufficiently small $s$:
\be
t_\pm(s)=-s\beta \pm \frac{\beta}{\pi} \frac{\beta^2{\cal V}^2}{4\pi^2} \sinh 2\pi s\,,\qquad s\ll \frac{1}{2\pi}\log\left(\frac{8\pi^2}{\beta^2{\cal V}^2}\right)\,.
\ee
In the limit of large $s$,  the singularity at $t^{*}_-(s)$ originating from the $T_R$  sector, runs off to infinity, while the one at $t^*_+(s)$, arising from the denominator of \eqref{llright}, associated to the $T_{L}$ twist sector, approaches a finite limiting value,
\be
t_+(|s|\to\infty)=-{\rm sgn}(s)\frac{\beta}{2\pi} \log\left(\frac{4\pi^2}{\beta^2{\cal V}^2}\right)\,.
\ee
The significance of the singularities and their evolution under modular flow will become clearer when we discuss their interpretation in the context of the gravitational dual bulk description. 
\subsection{Two-sided modular flowed correlator}\label{2sidedsyk}
The two-sided version of the fermion correlator is obtained from the modular flowed Euclidean correlation function by setting $\tau_{1}=\beta/2$ and $\tau_{2}=-it$,
\bea
\label{factorisedRLcorr}
W_{RL}^{(s)}(0, t) \,=&& \frac{1}{N}\sum_{j=1}^{N}\text{Tr}[\rho^{1-is}\chi_{R}^{j}(0)\rho^{is}\chi_{L}^{j}(t)]\,,\\\nonumber =&& G_{\chi, T_L}^{(s)}(\beta/2, -it) + G_{\chi, T_R}^{(s)}(\beta/2, -it) \,,
\eea
where as previously we assume that the late time saddle dominates.
As in the single-sided situation, we use eqs. \eqref{replica_correlator_v-gamma}, \eqref{gammas}, \eqref{vs} and \eqref{replica_correlator_TR}, to obtain the two separate twist field comntributions,
\bea\label{two_sided_corr_microscopic}
&&\exp\left(\tfrac1q \,g_{\chi, T_{L}}^{(s)}(\beta/2,-it)\right)\simeq \frac{\left({\pi}/{\beta\mathcal{J}}\right)^{2/q}}{\left[C^{RL}_{T_{L},1}(s)\,\cosh{\omega t} + C^{RL}_{T_{L},2}(s)\,\sinh{\omega t}\right]^{2/q}}\,,\\\nonumber\\
    &&\exp\left(\tfrac1q g_{\chi, T_{R}}^{(s)}(\beta/2,-it)\right)\simeq \frac{\left({\pi}/{\beta\mathcal{J}}\right)^{2/q}}{\left[C^{RL}_{T_{R},1}(s)\,\cosh{\omega t} + C^{RL}_{T_{R},2}(s)\,\sinh{\omega t}\right]^{2/q}}\,.
\eea
The coefficients $\{C^{RL}_{T_{L(R)},i}\}$ closely mirror the ones found for the single-sided case,
\begin{align}
\label{c2sided}
    &C^{RL}_{T_{L},1}(s)= \cosh{\pi s}-\tfrac{2\pi}{\beta\mathcal{J}}\,
    s\sinh{\pi s}
    \,, \\
    &C^{RL}_{T_{L},2}(s)=\sinh{\pi s}\left(1+\frac{\beta^{2}\mathcal{V}^{2}}{2\pi^{2}}\right)-\tfrac{2\pi}{\beta\mathcal{J}}
    s\cosh{\pi s} 
    ,\notag\\
    &\notag\\
    &C^{RL}_{T_{R},1}(s)= \cosh{\pi s}-\tfrac{2\pi}{\beta\mathcal{J}}
    s\sinh{\pi s}
    \,, \notag\\
    &C^{RL}_{T_{R},2}(s)=\sinh{\pi s}\left(1-\frac{\beta^{2}\mathcal{V}^{2}}{2\pi^{2}}\right)-\tfrac{2\pi}{\beta\mathcal{J}}\,s
    %
    \cosh{\pi s}\,.\notag
\end{align}
Here we have also displayed terms of order $1/\beta{\cal J}$, which vanish in the strict IR limit. In fact we will also be able to arrive at these terms from the bulk dual gravity perspective. 

Modular flow has a very interesting physical consequence on the two-sided correlator. For simplicity, let us take  the IR limit $\beta {\cal J}\to \infty$. When $s=0$, $W_{RL}^{(s)}(0,t)$ has no singularities for  real time $t$. In the decoupling limit for the bath, $\beta {\cal V} =0$, there are no real time singularities for any nonzero $s$. However, for a small non-zero coupling between the system and bath, modular flow leads to  real time singularities governed by solutions to the equation,
\be
\tanh\tfrac\pi\beta t=  \coth\left(\pi s\right)\left(-1\pm\frac{\beta^2{\cal V}^2}{2\pi^2}\right)\,.\label{poleLR}
\ee
It is clear why this has no solutions for finite, real $t$ when $\beta {\cal V}=0$; the magnitudes of the left and right hand sides are less than and greater than unity respectively. When the system-bath interaction is turned on however, a singularity appears from the denominator of \eqref{two_sided_corr_microscopic} corresponding to the `+' sign choice in \eqref{poleLR} when 
\be
|s| > \frac{1}{2\pi} \log \left(\frac{4\pi^2}{\beta^2 {\cal V}^2}\right)\,.\label{scramblings}
\ee
This singularity in the correlator comes in from infinity, and in the limit of large $|s|$ approaches the constant value,
\be
t_+(|s|\to \infty)= -{\rm sgn}(s)\frac{\beta}{2\pi} \log\left(\frac{4\pi^2}{\beta^2{\cal V}^2}\right)\,.
\ee
We will explore the consequence of this striking result for the bulk gravity dual below.

\section{SL$(2,\mathbb{R})$ charge of twist operators}
\label{sec7}
In the factorised limit, the left twist field contribution to the Euclidean two-point function, labeled by the replica number difference $s$  between the two insertion points, is given in 
\eqref{Liouville_sol}. The contribution from the   twist field $T_R$ is obtained from \eqref{Liouville_sol} by shifting both time arguments by $\beta/2$.  Let us recall the result of the $T_{L}$ twist contribution for general $s\neq 0$,
\begin{align}
    \exp\left({g^{(s)}_{\chi, T_{L}}(\tau_{1},\tau_{2})}\right)=\frac{f_{s}'(\tau_{1})h'(\tau_{2})}{\mathcal{J}^{2}(1-f_{s}(\tau_{1})h(\tau_{2}))^{2}}\,,
\end{align}
where we fixed $h(\tau_{2})=\tan\omega\tau_2$, choosing to keep the fermion  insertion $\chi(\tau_{2})$ fixed in the first replica while  the second fermion $\chi(\tau_{1})$ is moved across to a different replica labelled by $s$.  As argued in section \ref{recursive} the solutions $\{f_s\}$ are related by M\"obius or SL$(2, {\mathbb R})$ transformations. In fact, these solutions to the Liouville equation are in one-to-one correspondence with a one-parameter family of points  on the group manifold of SL$(2,\mathbb{R})$.

We will now follow the method of \cite{Gao:2021tzr} to associate an
SL$(2,\mathbb{R})$ charge to each solution. This will characterize the effect of the twist operators and their SL$(2,\mathbb{R})$  charges by providing a measure of the distance between the points in the SL$(2,\mathbb{R})$ group manifold corresponding to two solutions, $f_s$ and $f_{s+1}$,  associated to replicas $s$ and $(s+1)$ respectively.  Crucially, the SL$(2,{\mathbb R})$ charges will allow us to reconstruct the dual bulk gravitational interpretation of the modular flow \cite{Maldacena:2017axo, Stanford:2020wkf}.

In order to assign a charge to each solution $f_s$ we begin with the $s=0$ solution $f_0(\tau)$, deduce its SL$(2,{\mathbb R})$ charge as a $2\times 2 $ matrix ${\bf Q}_0$ valued in the algebra $\mathfrak {sl}(2,{\mathbb R})$, then act on it with the SL$(2,{\mathbb R})$ transformation that relates $f_0$ to the solution $f_s$ associated to the $s$-th replica. The charge for the solution $f_0$ is determined by considering  SL$(2, {\mathbb R})$ transformations which leave the solution invariant. Specifically, time translation, $\tau\to \tau +\tau_0$, does not change solution and is a M\"obius transform of $f_0(\tau)$,
\be
f_0(\tau)= \tan\omega(\tau-\tfrac\beta 2) \quad\rightarrow\quad f_0(\tau+\tau_0) = \frac{\cos\omega\tau_0 \,f_0(\tau) + \sin\omega\tau_0}{-\sin\omega\tau_0\, f_0(\tau) + \cos\omega\tau_0}\,.
\ee
This is just a $U(1)$ rotation in SL$(2, {\mathbb R})$. The charge ${\bf Q}_0$ is proportional to the generator of this $U(1)$,
\be
{\bf Q}_{0}=\frac{\pi}{\sqrt{2}\beta}\begin{pmatrix}0 & 1\\
-1 & 0\end{pmatrix}\,,
\ee
where the normalisation factor is fixed by  making contact with the dual bulk picture below.
The charge ${\bf Q}_s$ of any other $s\neq 0$ solution can now be found once the SL$(2,{\mathbb R})$ transform between $f_0$ and $f_s$ is known,
\be
f_{0}\to f_{s}=\frac{a_{s} f_{0}+b_{s}}{c_{s}f_{0}+d_{s}}, \quad\quad\quad a_{s}d_{s}-b_{s}c_{s}=1,
\ee
since the charge ${\bf Q}_0$ must transform by conjugation under the action of the group,
\be
{\bf Q}_s= {\cal S}_s {\bf Q}_0 \,{\cal S}_s^{-1}\,,\qquad {\cal S}_s= \begin{pmatrix} a_s& b_s\\
c_s & d_s
\end{pmatrix}\,.
\ee
The ansatz \eqref{anstaz,fs} for $f_s$ determines the entries of this SL$(2, {\mathbb R})$ matrix,
\bea
&&a_{s}=\frac{1}{\sqrt{v_{s}}} (v_{s}\,\cos(\gamma_{s}-\gamma_{0})-u_{s}\,\sin(\gamma_{s}-\gamma_{0}))\,,\qquad c_{s}=-\frac{1}{\sqrt{v_{s}}}\sin(\gamma_{s}-\gamma_{0})\,,\nonumber\\\\\nonumber
    &&b_{s}=\frac{1}{\sqrt{v_{s}}}(u_{s}\cos{(\gamma_{s}-\gamma_{0})}+v_{s}\sin(\gamma_{s}-\gamma_{0}))\,,\qquad d_{s}=\frac{1}{{\sqrt{v_{s}}}}{\cos{(\gamma_{s}-\gamma_{0})}}\,.
\eea
This yields the SL$(2,{\mathbb R})$ charge of the solution $f_s$ corresponding to replica $s$:
\begin{equation}\label{microscopicQs}
{\bf Q}_{s}=\frac{\pi}{\sqrt{2}\beta}\begin{pmatrix}-u_{s}/v_{s} \quad &\quad  u^{2}_{s}/v_{s}+{v}_{s}\\
-1/v_{s} \quad&\quad u_{s}/v_{s}
\end{pmatrix}\,.
\end{equation}
Given that the charges are valued in $\mathfrak{sl}(2,{\mathbb R})$, they inherit the natural inner product from the algebra,
\be
({\bf Q}_1, {\bf Q}_2) = {\rm Tr} \left[{\bf Q}_1 {\bf Q}_2\right]. 
\ee
Therefore, we can define a natural distance between the charges of two consecutive replica solutions and identify this with the norm of the charge of the twist fields:
\be\label{TL-charge}
    M_{s}^{2}=\text{Tr}[({\bQ}_{s+1}-\bQ_{s})(\bQ_{s+1}-\bQ_{s})]
 = \frac{\pi^2}{\beta^2 v_s v_{s+1}} \left((u_{s+1}-u_s)^2 + (v_{s+1}-v_s)^2\right)   \,.
 \ee
We then use our perturbative solutions, \eqref{gammas} and \eqref{vs}  for $v_s$ and $\gamma_s$, and together with \eqref{self_consistency_us} arrive at a nontrivial result to leading order in ${\cal V}^2/{\cal J}^2$,
 \be\label{twistmass}
    M_s^2 \simeq \frac{4\pi^{2}}{\beta^{2}}\left(\frac{\mathcal{V}}{\mathcal{J}}\right)^{4}\tan^{2}{\frac{\omega\beta}{2}}\simeq \frac{4\mathcal{V}^{4}}{\mathcal{J}^{2}}\,.
\end{equation} 
The first point to note is that the norm of the charge of the twist operator does not depend on $s$, at least to the order we have worked in. The SL$(2,\mathbb{R})$ charge of an operator insertion on the AdS$_2$ boundary is proportional to the  mass of the corresponding bulk particle state \cite{Maldacena:2017axo} which provides a kick to the boundary particle trajectory. We expect this mass to be independent of $s$ and indeed, a non-trivial cancellation of the $s$-dependence yields the result \eqref{twistmass}. A second observation is that the IR limit needs to be taken carefully, by using the expansion $\omega\beta\simeq\pi (1- 2/\beta {\cal J})$, to obtain a finite result at this order.

Above, we focussed attention on $T_L$, the left twist contribution. The charge for the right twist contributions can be obtained by a clever trick. Recalling \eqref{replica_correlator_TR} we act on the  SL$(2,\mathbb{R})$  charge for the  left twist sector \eqref{microscopicQs} by the operation that translates the argument of the solution $f_s(\tau)$  by $\beta/2$: 
\begin{equation}
    {\cal T}\left(\tfrac{\beta}{2}\right)= \begin{pmatrix}\cos{\frac{\omega\beta}{2}} & \quad\sin{\frac{\omega\beta}{2}}\\\\
-\sin{\frac{\omega\beta}{2}} & \quad\cos{\frac{\omega\beta}{2}}\end{pmatrix}\,.
\end{equation}
The transformation acts by conjugation, hence the norms of the charges of $T_L$ and $T_R$ are the same, given by \eqref{TL-charge}. We can summarize our results in a compact form
\begin{empheq}[box=\fbox]{align}\label{microscopic-charges}
    &{\bf Q}_{s,L}={\bf Q}_{s}=\frac{\pi}{\sqrt{2}\beta}\begin{pmatrix}-u_{s}/v_{s} \quad &\quad  u^{2}_{s}/v_{s}+{v}_{s}\\
-1/v_{s} \quad&\quad u_{s}/v_{s}\\
\end{pmatrix}\,,\\\nonumber\\\nonumber
    &{\bf Q}_{s,R}= {\cal T}\left(\tfrac{\beta}{2}\right){\bf Q}_{s}{\cal T}\left(-\tfrac{\beta}{2}\right)\,,\notag\\\nonumber\\\nonumber
    &M_{s,L}^{2}=M_{s,R}^{2}=M_{s}^{2}\simeq \frac{4\pi^{2}}{\beta^{2}}\left(\frac{\mathcal{V}}{\mathcal{J}}\right)^{4}\tan^{2}{\frac{\omega\beta}{2}}\simeq \frac{4\mathcal{V}^{4}}{\mathcal{J}^{2}}\,.\notag
\end{empheq}

\section{Reconstructing the AdS$_2$ modular flow}
\label{sec8}
The infrared dynamics of the SYK theory in the TFD state is captured by the Lorentzian two-sided black hole in JT gravity. In this case the dynamics is that of  a pair of boundary particles in AdS$_2$ governed by the Schwarzian action. The solution of the Euclidean Schwarzian equations of motion is well known and describes a circular boundary particle trajectory in EAdS$_2$ with proper length proportional to $\beta$. Using the embedding space coordinates of EAdS$_2$ with $(+,+,-)$ signature metric,
\begin{equation}\label{embedding_coordinates_EAdS2}
X^\mu(\theta) = \{\sinh\rho \sin \theta,\sinh\rho \cos\theta, \cosh\rho \},\qquad \mu=1,2,3\,,
\end{equation}
satisfying the hyperboloid constraint\,,
\begin{equation}
- X_1^2 - X_2^2 + X_3^2 = 1,
\end{equation}
the boundary trajectory $X^{(0)\mu}$  at radial coordinate $\rho=\rho_0$ can be expressed as an inner product constraint,
\begin{equation}\label{eq:5.5}
X^{(0)}(\theta)\cdot Q^{(0)}  = \frac{1}{2\epsilon}, \qquad \epsilon\to 0 ,
\end{equation}
where the vector $Q^{(0)\mu}$ is the conserved SL$(2,\mathbb{R})$ Noether charge associated with the isometries of EAdS$_2$. For a particle trajectory at a fixed cutoff, this charge vector is aligned with the direction normal to the boundary curve in the embedding space, ensuring that $X\cdot Q$ is a constant along the curve. In our case for a circular trajectory satisfying \eqref{eq:5.5} the  SL$(2,\mathbb{R})$ charge is
\begin{equation}\label{sl2rcharge}
Q^{(0)\mu}= \left\{ 0,\,0,\,\frac{1}{2\epsilon\cosh\rho_0}\right\},
\end{equation}
pointing along the $X_3$ direction, corresponding to the residual $U(1)\subset {\rm SL}(2,\mathbb{R})$ symmetry preserved by the circular thermal trajectory. The radius of the circle $\rho_0$ is taken to infinity whilst simultaneously  $\epsilon$ approaches $0$ so that,
\begin{equation}\label{infiniterho_new}
2\pi\epsilon \sinh \rho_0 \to \beta\,.
\end{equation}
The boundary particle trajectory is a  finite circular path in the $(X_1,X_2)$-plane, constrained to remain at a cutoff distance $\epsilon$ from the AdS$_2$ boundary,
\begin{equation}
X_1^2 + X_2^2 \;\approx\; \left(\frac{\beta}{2\pi\epsilon}\right)^2 ,
\end{equation}
while the associated SL$(2,\mathbb{R})$ charge points in the direction orthogonal to the circle.

Modular ``flow" in the Euclidean theory is a discrete  translation between replicas and acts as an SL$(2,{\mathbb R})/{\rm U}(1)$ transformation on the embedding coordinates of the field operator insertion point \cite{Gao:2021tzr, Gao:2024gky}. The circular trajectory in the second replica can be represented as,
\begin{equation}
    X^{(1)}(\theta) = B(x,\alpha)\, X^{(0)}(\theta)\,,
    \end{equation}
where $B(x,\alpha)$ is a three dimensional matrix representation of SL$(2,{\mathbb R})$ (isomorphic to $SO(1,2)$) depending on two parameters $x$ and $\alpha$, corresponding to a boost and rotation respectively in SL$(2,{\mathbb R})$. These parameters will be functions of the two independent couplings $\beta{\cal V}$ and $(\beta{\cal J})^{-1}$  of the interacting SYK setup.  
We find that matching between the bulk EAdS$_2$ and microscopic SYK description is most conveniently achieved by choosing an ansatz for $B(x, \alpha)$ through a composition of a boost, rotation and inverse boost, 
\begin{equation}
    B(x,\alpha)=M_1(-x) \,M_3(\alpha)\,M_1(x)\,. \label{replicaboost}
\end{equation}
$M_{1,3}$  are SL$(2,{\mathbb R})$ boost and rotation matrices given explicitly as, \footnote{For completeness, the third generator is $M_2(y) =\begin{pmatrix} \cosh y & 0 & \sinh y\\
 0 & 1 & 0\\
  \sinh y & 0 & \cosh y
\end{pmatrix}$.}
\begin{align}
    M_1(x)=\begin{pmatrix}1 &0 & 0\\ 0
 & \cosh x & \sinh x\\
0  & \sinh x & \cosh x
\end{pmatrix}\,,\qquad
 M_3(\alpha)&=\begin{pmatrix}
\cos\alpha & -\sin\alpha & 0 \\
\sin\alpha  & \cos\alpha &  0 \\
 0 &0 &1 
\end{pmatrix}\,. \label{sl2rembedding}
\end{align}
The ansatz for $B$ implies that translation across $s$ replicas is achieved by the SL$(2,{\mathbb R})$ transformation,
\be
B(x, \alpha)^s = B(x, s\alpha)\,,\qquad s=0,1,2,\ldots k-1\,.
\ee
Therefore the charge operator associated to the boundary particle in the the $s$-th replica is simply
\be
Q^{(s)}=B(x, s\alpha)\, Q^{(0)}\,,
\ee
while the mass squared of the twist operator that provides kicks to the boundary particle is given by 
\begin{equation}
    (B(x,\alpha)Q^{(0)}-Q^{(0)})^{2}=-M_s^{2}\,.
\end{equation}
Matching parameters $(x, \alpha)$ in the bulk with the SYK parameters can now be achieved by matching the charges $Q^{(s)}$ with their two-dimensional representations ${\bf Q}_s$ in eq.\eqref{microscopic-charges} built from the microscopic model. The detailed procedure for matching charges and parameters is explained in appendix \ref{charges}. In particular, there are two sets of parameters, one for each  contribution from the left and right twist operators, $T_L$ and $T_R$ respectively, namely $(\alpha_L, x_L)$ and $(\alpha_R, x_R)$. We quote below the formulae relating the SYK parameters $(u_s,v_s)$  with the SL$(2,{\mathbb R})$ rotation/boost parameters  $(\alpha_L, x_L)$:
\be
u_s= v_s \sin(s\alpha_L) \sinh x_L\,,\qquad v_s= \frac{e^{x_L}}{\cosh x_L +\sinh x_L \cos (s\alpha_L)}\,,
\ee
with analogous expressions for $\alpha_R$ and $x_R$.
Within the double expansion  $\beta {\cal V}\ll 1$, $\beta {\cal J}\gg 1$ (weak bath coupling and IR limit respectively) in the SYK framework, this identification yields,
\be
\alpha_{L,R} = 2\pi\left(1 + {\cal O}\left(\tfrac{1}{\beta {\cal J}}\right)\right)\,,\qquad x_{L,R} = \pm \frac{{\beta^2}{\cal V}^2}{2\pi^2}\left(1+ {\cal O}\left(\tfrac{1}{\beta {\cal J}}\right)\right)\,,\label{alphax}
\ee
where the positive sign applies to $x_L$. 
The $1/(\beta J)$ corrections above cannot all be matched within the the AdS boost/rotation ansatz, but these go beyond the IR limit, so we do not necessarily expect agreement with  AdS bulk asymptotics. However, if we require the norm squared of the charges of the twist operators in the two parametrisations  to agree, namely \eqref{twistmass} and \eqref{adstwistmass}, we find
\be
\alpha_{L,R} = 2\pi \left(1\mp\tfrac{2}{\beta{\cal J}}\right)\,.\label{alphas}
\ee
In fact, the choice of sign for the correction term, as we show below, ensures precise matching of modular flowed fermion correlators even beyond the strict IR limit.
\subsection{Modular flowed boundary correlators in AdS$_2$}
Given the precise handle we have on the central ingredients (the charges $Q^{(s)}$) that govern modular flow in the bulk, we can now turn to fermion correlation function computations  in EAdS$_2$. In AdS$_{2}$, the two-point correlation function of a boundary operator  of conformal dimension $\Delta$ is given by, up to overall normalisation,
    \begin{equation}
        \braket{O(X)O(X')} \sim (X\cdot X')^{-\Delta}=\cosh^{-\Delta}\ell\,,
    \end{equation}
where $\ell$ is the geodesic distance between the two insertion points with embedding space coordinates $X$ and $X^\prime$.\footnote{The result is not related to the geodesic approximation, but  follows from conformal invariance.}
Real time correlators will be obtained by analytically continuing the Euclidean ones, and we will primarily be interested in the  two-sided correlator,
\be\label{bulkcorr}
W_{RL}^{(s)}(0,\tau)= \frac{1}{\left(X^{(0)}(\pi-\theta)\cdot X^{(s)}(0)\right)^{\Delta_\chi}}\,,\qquad \theta = \frac{2\pi\tau}{\beta}\,,
\ee
with ${\Delta_{\chi}}=1/q$ is the dimension of a Majorana fermion. When $\tau$ is Wick rotated to real time, the two insertions at antipodal points on the Euclidean thermal circle yield a two-sided correlation function. In the late time regime we must sum over the separate contributions from the bulk geometries corresponding to the factorised twist fields on the boundary. Therefore, we write \eqref{bulkcorr}, up to a normalisation factor, as 
\be
W_{RL}^{(s)}(0,\tau)= \sum_{i=L,R}\frac{1}{\left(X^{(0)}(\pi-\theta)\cdot B^s(x_i, \alpha_i)X^{(0)}(0)\right)^{\Delta_\chi}}\,.
\ee
Here we have used the action of $B^s$ to determine the transformed  boundary circle $X^{(0)}(\theta)$ under modular flow,
\begin{equation}
    X^{(s)}= B^s(x,\alpha)\, X^{(0)} = B(x, s\alpha)\, X^{(0)} \, ,\qquad s=0,\dots k-1 \,.\label{sl2rimages}
\end{equation}
Performing the analytical continuations $s\to i s$ and $\tau\to  -i t$, and taking the boundary limit under \eqref{infiniterho_new}, the modular flowed correlation function from the bulk is,
\bea
&&W_{RL}^{(s)}(0,t)  =
    {\cal N}\sum_{i=L,R}\left(\cosh\frac{s\alpha_i}{2}\cosh \frac{\pi t}{\beta} + e^{x_i}\sinh \frac{s\alpha_i}{2} \sinh \frac{\pi t}{\beta}  \right)^{-2/q}\,,\\\nonumber
&&    {\cal N}=\left(\pi/{\beta\mathcal{J}}\right)^{2/q}\,,
\eea
where we replaced the AdS$_2$ cutoff $\epsilon$ with $\sim 1/{\cal J}$ to fix the  normalisation in terms of SYK parameters. 
Comparing with the microscopic result \eqref{two_sided_corr_microscopic} we now see that the coefficients $\{C^{RL}_{T_{L,R},i}\}$ do precisely match the AdS result, at order $1/\beta {\cal J}$ and at leading non-trivial order in $\beta {\cal V}$:
\bea
&& C^{RL}_{T_{L,R},1}(s)= \cosh{\frac {s\alpha_{L,R}}{2}} + {\cal O}\left(\tfrac {\beta{\cal V}^2}{\cal J}\right)\,,\\\nonumber
&& C^{RL}_{T_{L,R},2}(s)= e^{x_{L,R}}\sinh{\frac {s\alpha_{L,R}}{2}} + {\cal O}\left(\tfrac {\beta{\cal V}^2}{\cal J}\right)\,.
\eea
The parameters $\alpha_{L,R}$ and $x_{L,R}$ are as in \eqref{alphax} and \eqref{alphas}.
\subsection{Singularities, maximal modular chaos}
The general form of the two-sided, modular flowed correlator for operator insertions  at arbitrary times $t_L$ and $t_R$ can be inferred straightforwardly, and we quote the effect of coupling to the bath in the IR limit,
\bea\label{LR, modular flow corr_t12}
&&W_{RL}^{(s)}(t_R,t_L)=\\\nonumber
&&   {\cal N}\sum_{i=L,R}\left[\cosh{\pi \left(\frac{t_{L}+t_{R}}{\beta}+s\right)} + (e^{x_i}-1)\sinh{(\pi s)}\sinh{\pi\left(\frac{t_{L}-t_{R}}{\beta}\right)}\right]^{-2/q}\,.
\eea
\paragraph{Lightcone singularities:} Following on from section \ref{2sidedsyk}, singularities in the two-sided modular flowed correlator can  develop when either of the two conditions below is met for some $s$ at fixed $t_L, t_R$,
\begin{equation}
    \tanh{(\pi s)}=-\frac{\cosh\frac{\pi}{\beta}(t_{L}+t_{R})}{\sinh\frac{\pi}{\beta}(t_{L}+t_{R})+ (e^{x_i}-1)\,\sinh\frac{\pi}{\beta}(t_{L}-t_{R})}\,,\qquad i= L,R\,.
\end{equation}
If we set $t_{R}=0$, the condition for the singularity simplifies to  
\begin{equation}
    \tanh{\pi s}=-e^{-x_i}\coth\tfrac{\pi}{\beta}t_{L}\,,\qquad x_{L,R}=\pm\frac{\beta^2{\cal V}^2}{2\pi^2}\,,
\end{equation}
and $|x_{L,R}|\ll1$ in the microscopic theory.
The  singularity only develops
 in the case where $x_i >0$ which is in the sector with  the $T_{L}$ twist field turned on,  and then only for a specific range of real time $t_{L}$ of the left Majorana fermion insertion:
\begin{equation}
    t_{L}\in \mathcal{\cal D}\equiv \left\{t\in \mathbb{R}\big| \;\;t>\frac{\beta}{2\pi}\log{\frac{4\pi^{2}}{\beta^{2}\mathcal{V}^{2}}}\; \cup \;t<-\frac{\beta}{2\pi}\log{\frac{4\pi^{2}}{\beta^{2}\mathcal{V}^{2}}}\right\}\,.
\end{equation}
The system has $L-R$ symmetry, so if we set $t_{L}=0$ instead, then we only have a singularity arising from the right twist field $T_{R}$  contribution  for $t_{R}\in \mathcal{D}$.  In addition, there is a minimum value \eqref{scramblings} of $|s|$ below which a real time singularity cannot appear. We will identify this below with the modular scrambling time scale.

\paragraph{Maximal modular chaos:}
The form of the correlator \eqref{LR, modular flow corr_t12} closely resembles analogous results in {\cite{Chandrasekaran:2021tkb,Chandrasekaran:2022qmq}}. In \cite{Chandrasekaran:2022qmq} modular flow was generated by the reduced density matrix for a particular subset of fermions constructed from two copies of an SYK system in the TFD state, at large $\beta\mathcal{J}$ and in the large $q$ limit. 

An interesting feature of the correlator is that it displays a signature of the so-called maximum modular chaos bound of \cite{DeBoer:2019kdj}, the modular time evolution counterpart of the Maldacena–Shenker–Stanford  chaos bound \cite{Maldacena:2015waa} for  real time evolution. In our example the non-zero interaction $\beta {\cal V}$ between the system and bath provides the necessary perturbation of the modular Hamiltonian whose matrix elements must grow exponentially for large $s$ with exponent $2\pi$  for systems saturating the bound of \cite{DeBoer:2019kdj}. Note that the first term in the  denominator of \eqref{LR, modular flow corr_t12} is the thermal result for the unperturbed system in which modular flow acts as time translation. Relative to this, the correction term shows exponential growth with exponent $2\pi$ for large $s$. To see this explicitly, let us keep  $t_R^\prime = t_R +\beta s$ fixed and $t_L\gg t_R^\prime$,   for positive $s$, 
\bea
 W_{RL}^{(s)}(t_R, t_L) \approx 
{\cal N}\sum_{i=L,R}\left[\cosh{\tfrac\pi\beta \left(t_{L}+t_{R}^\prime\right)} + \frac{e^{x_i}-1}{4}\,e^{\frac{\pi}{\beta}(t_{L}-t_{R}^\prime)}(e^{2\pi s}-1)\right]^{-2/q}\,.\label{maxmodexp}
\eea
The exponential growth of the correction term with  $s$ is a hallmark of maximal modular chaos. Away from the IR limit, we expect this growth to be submaximal with $1/(\beta {\cal J})$ corrections to the exponent. This follows from the expressions \eqref{gammas},\eqref{vs} for $(v_s, \gamma_s)$ in the microscopic theory which imply a modular growth exponent of $2\omega\beta$.  

Eq. \eqref{maxmodexp} allows to identify a  modular scrambling time scale (analogous to entanglement scrambling time scale, see e.g. \cite{Perlmutter:2016pkf}),
\be
s_{\rm scr}= \frac{1}{2\pi}\log\frac{4\pi^2}{\beta^2{\cal V}^2},\label{modscr}
\ee
at which the perturbation becomes large and comparable to the leading term. Real time light cone singularities in the two-sided correlator only develop beyond this value of the modular flow parameter.

\section{Quantum extremal surfaces and bulk modular flow}
\label{sec9}
\subsection{QES locations}
A close link exists between the signature of maximal modular chaos and the appearance of a quantum extremal surface in holographic semiclassical gravity \cite{DeBoer:2019kdj}. In semiclassical gravity, the QES is the bifurcation surface of modular flow, left invariant by the action of that flow \cite{Jafferis:2015del, Faulkner:2017vdd, Faulkner:2018faa}.
In our case in the IR limit of the microscopic system, the modular flow is the  Killing flow generated by particular isometries of AdS$_2$. In either of the two twist field sectors, the relevant SL$(2, {\mathbb R})$ transformation is,
\begin{equation}
    \Lambda_j(s)= B(x_j,2\pi is)= M_{1}(-x_j)\cdot M_{3}(2\pi is)\cdot M_{1}(x_j)\,,\qquad j=L,R\,.
\end{equation}
with
\begin{equation}
    x_{L,R}=\pm\frac{{\beta^2}{\cal V}^2}{2\pi^2}\,.
\end{equation}
Prior to the analytic continuation $s\to is$, when in EAdS$_{2}$, the transformation is given by a combination of a boost in the $X_{2}-X_{3}$ plane of magnitude $x_j$, followed by a rotation of $2\pi s$ in the $X_{1}-X_{2}$ plane and then finally a boost equal and opposite to the first one. Upon analytic continuation to Lorentzian signature and $s\to is$, the rotation  in the $X_{1}-X_{2}$ plane becomes a boost (see  Appendix \ref{appendix:AdS2}). The infinitesimal generator of the flow is
\begin{equation}
    K_j=\frac{d}{ds}\Lambda_j(s)\Bigr|_{s=0}=\begin{pmatrix}
0 & 2\pi \cosh{x_j} & 2\pi \sinh{x_j} \\
2\pi \cosh{x_j}  & 0 &  0 \\
 -2\pi \sinh{x_j} &0 &0 
\end{pmatrix}\,.
\end{equation}
Fixed points of the Killing flow are solutions to the conditions,
\begin{equation}
    K^{\mu}_j(X)\big|_{\text{QES}}=0 \quad\Longleftrightarrow  \quad (K_j X)^{\mu}\big|_{\text{QES}}=0\,.
\end{equation}
Using the representation of embedding coordinates in Kruskal gauge we find (using eq.\eqref{embedding_Kruskal} )
\bea\label{Killing_flow_SYK}
    &&(K_j X)^{\mu}=\frac{2\pi}{1+UV}\times\\\nonumber
    &&\Big((U-V)\cosh{x_j} + (UV-1)\sinh{x_j}\,,\quad
    {(U+V)\cosh{x_j}},\quad
    -{(U+V)\sinh{x_j}}\Big)\,,
\eea
and this vanishes when 
\begin{equation}
    V=-U= \frac{-\cosh{x_j}\pm 1}{\sinh{x_j}}\,.
\end{equation}
Therefore, for  each twist field sector we find two solutions, one close to the horizon in the AdS$_2$ interior,
\begin{equation}
    V_{{\rm QES},j}=-U_{{\rm QES},j}\approx -\frac{x_j}{2}\,,\qquad |x_j|=\frac{\beta^2{\cal V}^2}{2\pi^2} \ll 1\,.
\end{equation}
and the other one, at large Kruskal coordinates beyond the AdS$_2$ boundary,
\begin{equation}
    V_{\infty,j}=-U_{\infty,j}\approx -\frac{2}{x_j}\,.
\end{equation}
From the sign of $x_j$ in each twist field sector, we learn that the QES is in the left Rindler wedge  when twist  $T_L$ is turned on, and in the right Rindler wedge when $T_R$ is on. For a complete interpretation of the lightcone singularities and quantum extremal surfaces constructed above, we now examine the effect of modular flow on the right Majorana fermion in detail.

\subsection{Modular flow of fermion insertion}
In the two-sided correlation function $W_{RL}^{(s)}(t_L, t_R)$, the effect of modular flow can also be seen as keeping one fermion insertion fixed, and the other insertion point (for the right fermion, say) evolving as a function of $s$. The Lorentzian embedding space coordinates of the right fermion in each twist field sector evolve as,
 \be
    X(x_j,s)= B(x_j,2\pi is) \,X( 0) \,,\qquad j=L,R\,.\label{image1}
   \ee
At the beginning of the flow ($s=0$), the Majorana fermion is localized on the physical boundary of AdS$_2$ at a cutoff radial coordinate $\rho_0$ given by 
\be
e^{\rho_0}\approx \frac{\beta}{\pi\epsilon} \gg 1\,.\label{rho0}
\ee

\subsubsection{Flow in Rindler coordinates}
In the  Rindler space representation (appendix \ref{appendix:AdS2}) for the embedding coordinates $X^{\mu}$,
\begin{equation}
    X_{1}=\sinh{\rho}\sinh{\frac{2\pi}{\beta}t}, \quad\quad  X_{2}=\pm\sinh{\rho}\cosh{\frac{2\pi}{\beta}t}, \quad\quad X_{3}=\cosh{\rho}\,,
\end{equation}
where plus (minus) sign is for right (left) Rindler wedge, the image point \eqref{image1} of the boundary fermion insertion at modular parameter $s$ (and real time $t_R=0$) is,
\bea\label{eqXs}
    &&X_1(x_j, s)=\sinh{(2\pi s)}\left(\sinh{\rho_{0}} +x_j\cosh{\rho_{0}}\right)\,,\\
    &&X_{2}(x_j,s)=\cosh{(2\pi s)}\sinh{\rho_{0}} + 2x_j\sinh^{2}{(\pi s)}\cosh{\rho_{0}}\,,\notag\\
    &&X_{3}(x_j,s)=\cosh{\rho_{0}} -2 x_j\sinh^{2}{(\pi s)}\sinh{\rho_{0}}\,,\notag
\eea
where we have taken the small $|x_i|$ limit and linearised.
Taking  the limit of large $\rho_0$, we can deduce the radial movement  of the right fermion away from the boundary, using 
\begin{equation}
    \rho(s)=\cosh^{-1}{X_{3}(x_j,s)}\,.
\end{equation}
An interesting feature which is immediate here is that when $x_j$ is positive ($T_L$ sector) the right Majorana fermion, under the modular flow, reaches the right Rindler horizon at finite modular time,
\begin{equation}
    |s|=s_{\rm scr}\simeq\frac{1}{2\pi}\log{\left(\frac{4\pi^{2}}{\beta^{2}\mathcal{V}^{2}}\right)}\,.
\end{equation}
This is the modular timescale at which $X_3(s)$ approaches 1  and coincides with the modular scrambling time \eqref{modscr}.

\subsubsection{Flow in Kruskal-Szekeres coordinates}
For a complete global picture of the modular flow of the right Majorana fermion we turn to Kruskal-Szekeres coordinates $(U,V)$ for AdS$_2$ (see Appendix \ref{appendix:AdS2}). The embedding coordinates are
\be
    X^{\mu}=\left(\frac{V+U}{1+UV}, \,\frac{V-U}{1+UV}, \,\frac{1-UV}{1+UV}\right)\,,\label{KS}
\ee
where the horizon is at $UV=0$ and the AdS$_2$ conformal boundary at $UV=-1$. The dependence of the KS coordinates on $s$ can be straightforwardly obtained  from the embedding space coordinates \eqref{eqXs} and \eqref{KS}.

It will be important for the picture below that the physical boundary is at a fixed cutoff, away from $UV=-1$, so that at the beginning of the flow at $s=0$,\footnote{For convenience we have defined the dimensionless cutoff $\hat \epsilon$ related to $\epsilon$ in \eqref{rho0} via $\hat \epsilon= 4\pi\epsilon/\beta$.}
\be
        UV(s=0)=\frac{1-\cosh\rho_{0}}{1+\cosh\rho_{0}}=-1+\hat{\epsilon}, \quad\quad\quad \hat{\epsilon}\equiv\frac{4\pi}{\beta \mathcal{J}}\ll1\,.
\ee
At $s=0$, the right fermion is inserted at the physical boundary with embedding coordinates,
\begin{equation}
    X^{\mu}(s=0)=\left\{\sinh\rho_0\sinh{\tfrac{2\pi}{\beta}t_{R}},\, \sinh\rho_0\cosh{\tfrac{2\pi}{\beta}t_{R}},\, \cosh\rho_0\right\}\,.
\end{equation}
Here we have inserted the right fermion at some non-zero boundary time $t_R$. Physically, we want to be in the late time regime when the factorised saddle point for left and right twist fields is dominant.
We then follow the trajectory by solving for $U(s)$ and $V(s)$. 
\subsubsection{Flow in the $T_L$ twist sector}
\label{TLflownew}
In the sector with the left twist field turned on, the boost parameter $x_L= \frac{\beta^2{\cal V}^2}{2\pi ^2}$. We summarize the main features of the flow below, captured in figures \ref{figLflow} and \ref{fig8}.
\begin{itemize}
    \item The right fermion insertion starts off at the right boundary  where $UV=-1+\hat\epsilon$, and moves inward with increasing positive  $s$ with $V>0$ and $U<0$  until the first horizon crossing at $s_1^+$, eventually moving off to infinity inside the horizon at $s=s_3^+$ where $UV$ diverges, and re-emerging {\em outside} the left horizon ($U>0$, $V<0$) immediately after. Analogous behaviour for negative $s$ occurs at the respective modular times $s_1^-$ and $s_3^-$. Therefore, $UV$ is always negative except for $s\in[s^{-}_{3}, s^{-}_{1}]\cup [s^{+}_{1},s^{+}_{3}]$, where
    \begin{align}
        &s^{\pm}_{1}\simeq -\frac{t_{R}}{\beta}\pm\frac{1}{2\pi}\cosh^{-1}{\left[\cosh{\tfrac{2\pi}{\beta}t_{R}}+\left(1-\frac{\hat{\epsilon}}{2}\right)\frac{2\pi^{2}}{\beta^{2}\mathcal{V}^{2}}\right]}\nonumber\\\label{uv}\\
        &s^{\pm}_{3}\simeq -\frac{t_{R}}{\beta}\pm\frac{1}{2\pi}\cosh^{-1}{\left[\cosh{\tfrac{2\pi}{\beta}t_{R}}+\left(1+\frac{\hat{\epsilon}}{2}\right)\frac{2\pi^{2}}{\beta^{2}\mathcal{V}^{2}}\right]}\notag
    \end{align}
    \begin{figure}[ht]
    \centering
    \includegraphics[width=0.49\linewidth]{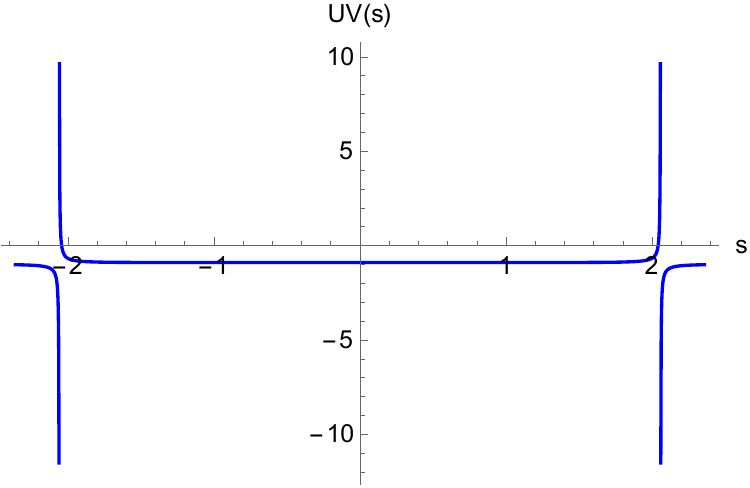}
    \includegraphics[width=0.49\linewidth]{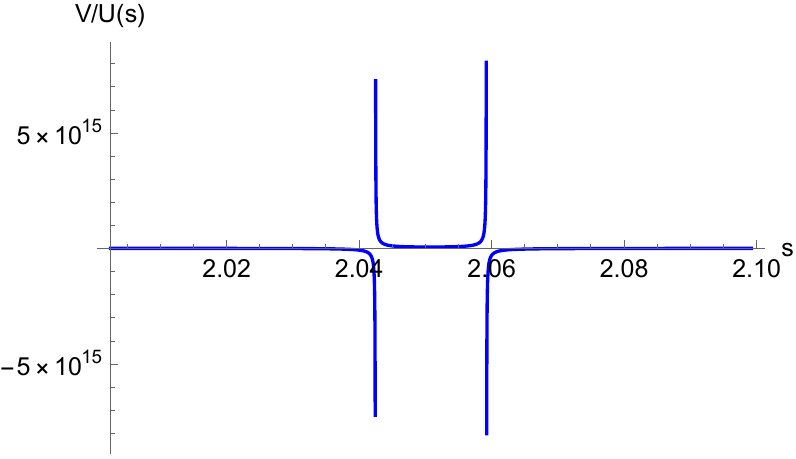}
    \caption{Plots of combinations of KS coordinates $UV$ and $V/U$ as functions of $s$, for  $\beta=1$, $\mathcal{V}=10^{-2}$, $\hat{\epsilon}=0.1$ and $t_{R}=0$.  Points where $UV$ vanishes correspond to horizon crossings. We reversed the direction of the flow via $s\to -s$ for convenience.}
    \label{fig8}
\end{figure}
Some intuition for these modular time scales can be gained if we consider small enough insertion times $t_R$ such that $t_R\ll \frac{\beta}{2\pi}\ln|4\pi^2/{\beta^2{\cal V}^2}|$. In this regime,
\bea
&&s_1^\pm\approx-\frac{t_R}{\beta}\pm \frac{1}{2\pi}\log\left[(2-\hat \epsilon)\frac{2\pi^2}{\beta^2{\cal V}^2}\right]\,,\\\nonumber
&&s_3^\pm\approx-\frac{t_R}{\beta}\pm \frac{1}{2\pi}\log\left[(2+\hat \epsilon)\frac{2\pi^2}{\beta^2{\cal V}^2}\right]\,,
\eea
so flowing forward, we encounter the past horizon a little before $(s+t_R/\beta)$ approaches the modular scrambling time $s_{\rm scr}$, the trajectory then moving off to infinity in the interior a little after, and  emerging outside the horizon on the left. 
 Figure \ref{fig8} shows the evolution of the KS coordinates $U(s)$ and $V(s)$, confirming the picture above. 
 At              $s=s^{+}_{1}$ i.e. horizon crossing, $V$ vanishes and subsequently at $s=s^{+}_{3}$, $U$ diverges whilst $V$ remains finite, reappearing outside the future horizon on the left. Reversing the direction of modular flow we enounter at $s=s^{-}_{1}$ the future horizon on the right where $U$ vanishes and at $s=s^{-}_{3}$ the coordinate $V$ diverges.
    \item Inside the horizon, there is a special point in the evolution with $UV=1$ where,
    \begin{equation}
        s= s^{\pm}_{2}= -\frac{t_{R}}{\beta}\pm\frac{1}{2\pi}\cosh^{-1}{\left[\cosh{\tfrac{2\pi}{\beta}t_{R}} \,+\,\frac{2\pi^{2}}{\beta^{2}\mathcal{V}^{2}}\right]}\,.
    \end{equation}
    In addition $s=s^{-}_{2}$ also represents a stationary point of $V(s)/U(s)$ in the interval $[s^{-}_{3},s^{-}_{1}]$. Interestingly, this half-way point in the evolution is independent of the cutoff  parameter $\hat \epsilon$.
    
    \item The modular time spent by the right Majorana fermion in its flow inside the interior can be evaluated as the difference $|s^{\pm}_{3}-s^{\pm}_{1}|$, which vanishes with $\hat{\epsilon}$, as the physical boundary recedes towards the conformal boundary of AdS$_2$.
    
    \item The endpoints of the flow, in Kruskal coordinates, are finite and located 
    \bea
\left(U(s),\,V(s) \right) \simeq \begin{cases}
    \left(\dfrac{4\pi^{2}}{\beta^{2}\mathcal{V}^{2}},\,-\dfrac{\beta^{2}\mathcal{V}^{2}}{4\pi^{2}} \right)\quad\quad s\to +\infty\\
    \notag\\
    \left(\dfrac{\beta^{2}\mathcal{V}^{2}}{4\pi^{2}} ,\,-\dfrac{4\pi^{2}}{\beta^{2}\mathcal{V}^{2}} \right)\quad\quad s\to -\infty
\end{cases} 
\eea
\end{itemize}

\subsection{Flow in $T_{R}$ twist sector}
\label{TRflownew}
The two factorised sectors yield distinct contributions with different singularity structures, following from the sign of $x_i$ the boost parameter associated to the modular flow generators. 

Modular flow  generated in the $T_R$ twist sector is nontrivial and rich, probing different portions of the bulk geometry including behind the right horizon, depending on the insertion time $t_{R}$ for the  right Majorana fermion.  The KS coordinates of the fermion insertion point as a function of $s$ are obtained by applying the boost operator $B(x_R, 2\pi i s)$ (with $x_R = -{{\beta^2}\cal V}^2/2\pi^2$) on the initial position $X(0)$ and using \eqref{KS}. The key events along the flow trajectory are horizon crossings $UV=0$, and modular times at which the flow goes off to infinity $|UV|\to\infty$. These special values of $s$ can be determined straightforwardly by applying the formulae \eqref{uv} with the replacement ${\beta^2}{\cal V}^2/2\pi^2 \to - {\beta^2}{\cal V}^2/2\pi^2$.

\begin{figure}[ht]
\centering
\includegraphics[width=1.8in]{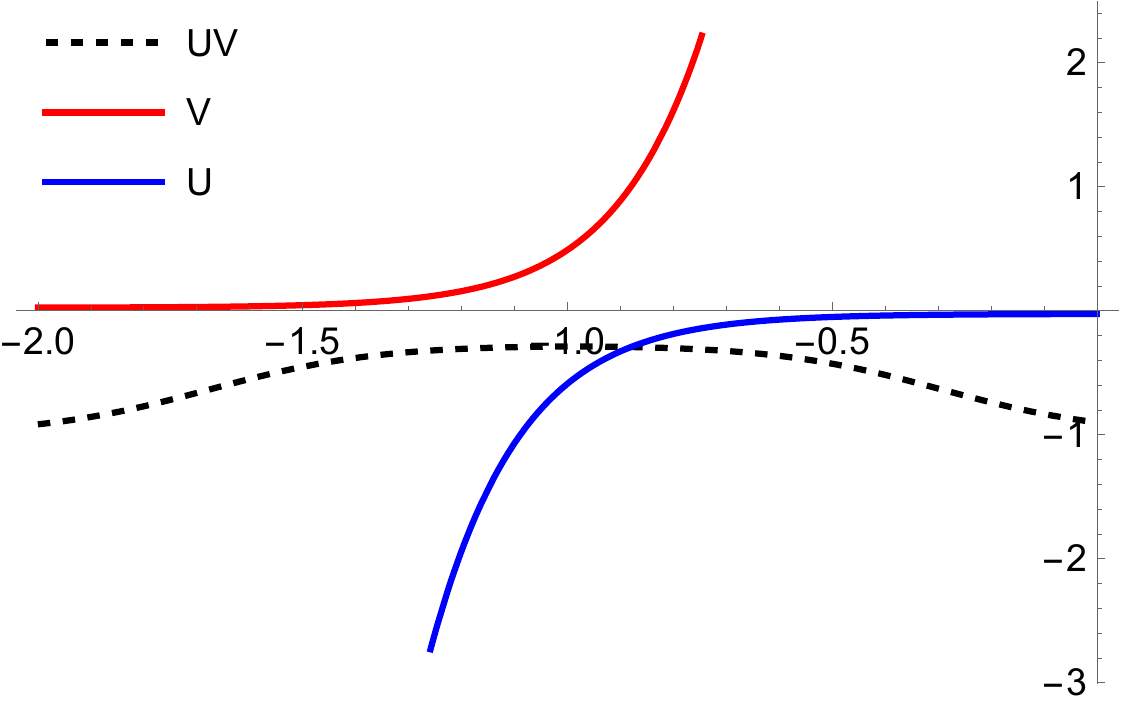}\hspace{0.1in}
\includegraphics[width=1.8in]{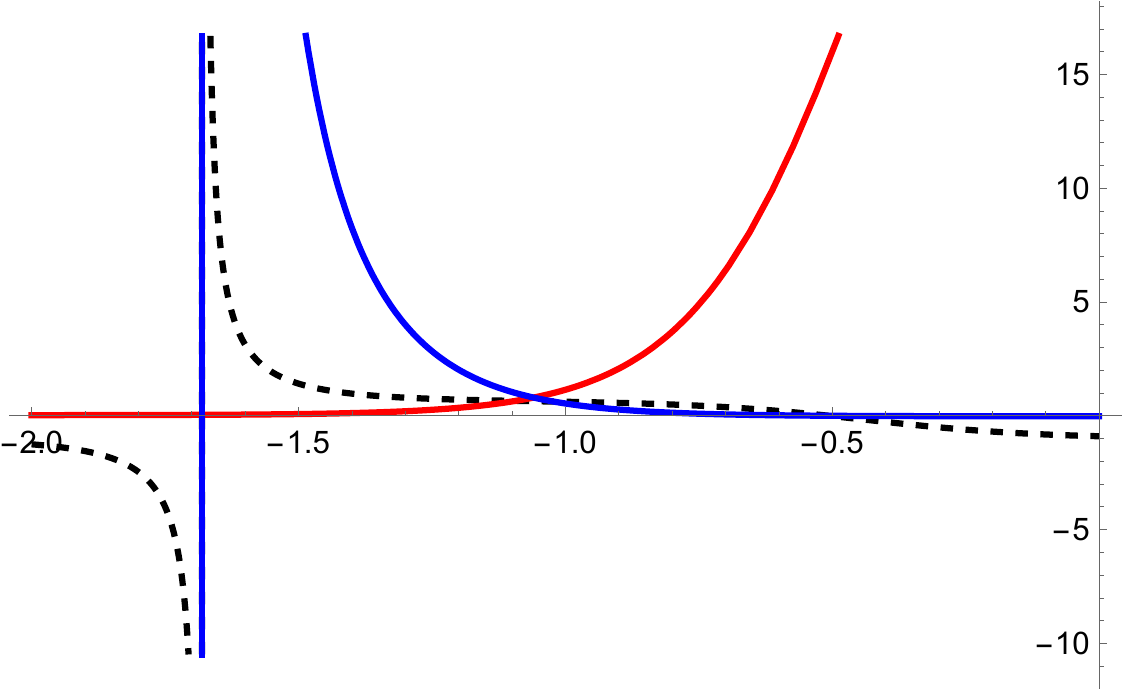}\hspace{0.1in}
\includegraphics[width=1.8in]{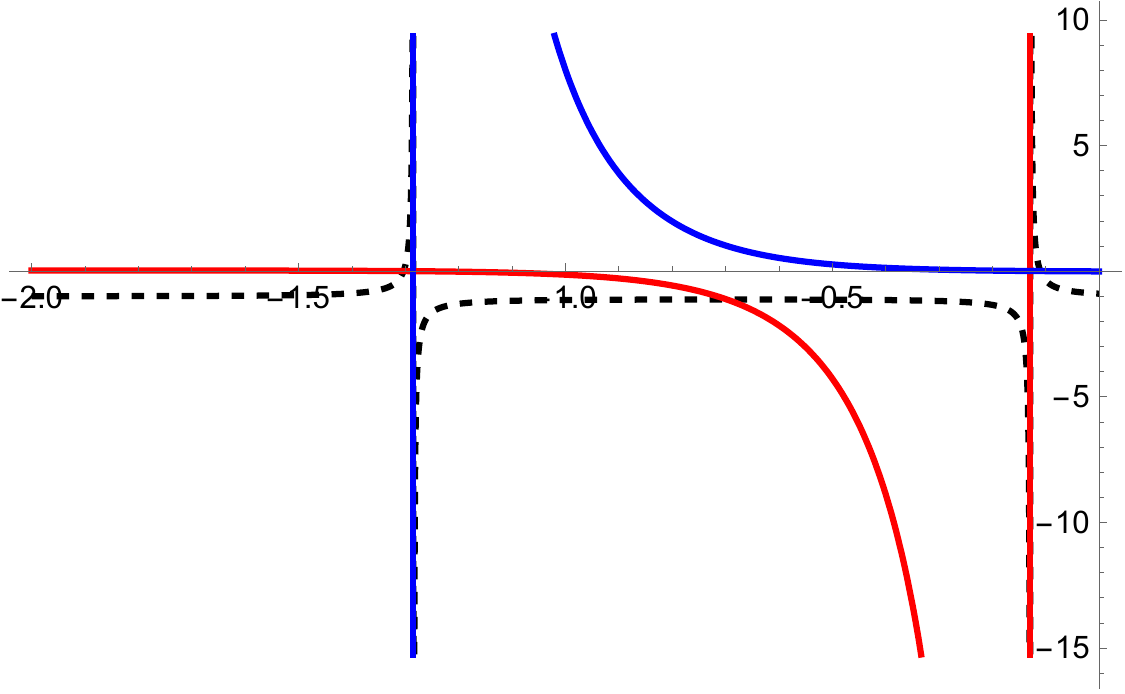}
\caption{\footnotesize{{\it Left:} Plot of KS coordinates of fermion insertion point as function of (negative) $s$ in early time regime I, starting and ending at the physical boundary on the right (green curve in figure \ref{figRflow}). {\it Centre:} Intermediate time regime II. Trajectory plunges into  horizon at $U=0$, approaching  left horizon from inside and diverging at finite $s$, and re-emerging in left exterior (blue curve of figure \ref{figRflow}).  {\it Right:} Late time, regime III -- after crossing into the future horizon $U=0$, the coordinate diverges and emerges outside the past left horizon approaching the left boundary asymptotically.  All plots produced with $\beta=2\pi$, ${\beta^2}{\cal V}^2/2\pi^2 = 0.05$, $\hat \epsilon =0.1$.}}
\label{Rflow}
\end{figure}

We find distinct behaviours across three temporal regimes for the insertion time $t_R$, with a fourth type separating the first two regimes. The details summarised below, are depicted in figures \ref{figRflow} and \ref{Rflow}.

\paragraph{Regime I (early time):} When the insertion time is small i.e., 
\begin{equation}
    t_{R}<\frac{\beta}{2\pi}\log{\left[\frac{4\pi^{2}}{\beta^{2}\mathcal{V}^{2}}\,\left(1-\frac{\hat{\epsilon}}{2}\right)\right]}\,,
\end{equation} 
$UV$ stays negative, without poles and zeros, and asymptotes to $-1$, returning to the right boundary as $|s|$ becomes large, in both flow directions.  It is important that $\hat \epsilon$, parametrising the location of the boundary is non-vanishing, so the time scale above is slightly smaller than the scale set by the modular scrambling time $s_{\rm scr}$.

The flow is constrained to lie in the entanglement wedge bounded by the QES located at $V_{{\rm QES},R}=-U_{{\rm QES},R}=\frac{\beta^{2}\mathcal{V}^{2}}{4\pi^{2}}$, which in turn is contained within the right Rindler wedge. The endpoint of the fermion modular flow into the boundary is at the tip of the entanglement wedge:
 \bea
   \label{end,flow, TR}
\left(U(s),\,V(s) \right) \simeq \begin{cases}
    \left(-\dfrac{\beta^{2}\mathcal{V}^{2}}{4\pi^{2}}\,,\dfrac{4\pi^{2}}{\beta^{2}\mathcal{V}^{2}} \right)\quad\quad s\to +\infty\\
    \\
    \left(-\dfrac{4\pi^{2}}{\beta^{2}\mathcal{V}^{2}}\,,\dfrac{\beta^{2}\mathcal{V}^{2}}{4\pi^{2}} \right)\quad\quad s\to -\infty
\end{cases} 
\eea
The flow trajectory in this  regime is depicted in green in figure \ref{figRflow}.
\paragraph{Edge of the wedge:}
If the right Majorana fermion is prepared exactly at the edge of the wedge, i.e. at 
\begin{equation}
    t_{R}=\frac{\beta}{2\pi}\log{\left[\frac{4\pi^{2}}{\beta^{2}\mathcal{V}^{2}}\left(1-\frac{\hat{\epsilon}}{2}\right)\right]} \,,
\end{equation}
the flow follows the straight line $U=-\frac{\beta^{2}\mathcal{V}^{2}}{4\pi^{2}}$, and depending on the sign of  $s$, it either moves towards the right boundary, or approaches the QES, which is a fixed point of the flow. This is depicted in red in figure \ref{figRflow}.

Given the QES coordinates, it follows that for any excitation inserted on the right boundary ($\hat\epsilon\to0$), the real time required to leave the entanglement wedge of the right boundary i.e. the SYK$_{\chi}$ system is 
\begin{equation}
     t_{\rm scr} = \frac{\beta}{2\pi}\log{\left(\frac{4\pi^{2}}{\beta^{2}\mathcal{V}^{2}}\right)}\,,\label{tscr}
\end{equation}
which is  then interpreted as the scrambling time. This is the  time after which a message falling into the black hole starts to be reconstructable from the complementary system i.e. the bath. This is a manifestation of the Hayden-Preskill protocol \cite{Hayden:2007cs}.

Let us provide an argument to explain why \ref{tscr} is the scrambling time, by recalling the general definition \cite{Leichenauer:2014nxa}\,,
\begin{equation}
    t_{\text{scr}}=\frac{\beta}{2\pi}\log{\left(\frac{E}{\delta E}\right)}\,,
\end{equation}
where $E$ is the energy of the unperturbed system, in our case the SYK$_{\chi}$ model, and $\delta E$ is the energy of the perturbation, which can be thought of as induced by the coupling between the two SYK systems. At low energies, the  energy $E$ of an SYK with $q$-fermon random couplings is, 
\begin{equation}
    E=\frac{N}{q^{2}}\frac{1}{\mathcal{J}}\frac{2\pi^{2}}{\beta^{2}}\,,
\end{equation}
while the energy of the perturbation can be estimated by the expectation value of  from the on-shell interaction Hamiltonian \eqref{DeltaI},
\begin{equation}
    \delta E\sim\frac{N}{q^{2}}\frac{\mathcal{V}^{2}}{\mathcal{J}}\,.
\end{equation}
This explains the pramatric dependence of the scrabling time on $\beta {\cal V}$.

\paragraph{Regime II (just outside the wedge):}
When the insertion time $t_{R}$ sits inside the narrow interval
\begin{equation}
    \frac{\beta}{2\pi}\log{\left[\frac{4\pi^{2}}{\beta^{2}\mathcal{V}^{2}}\left(1-\frac{\hat{\epsilon}}{2}\right)\right]}< t_{R} \leq \frac{\beta}{2\pi}\log{\left[\frac{4\pi^{2}}{\beta^{2}\mathcal{V}^{2}}\left(1+\frac{\hat{\epsilon}}{2}\right)\right]}\,,
\end{equation}
adjacent to the edge of the wedge, the KS coordinate $V(s)$ does not display zeroes or poles and stays between the boundary and the $V= -\frac{\beta^{2}\mathcal{V}^{2}}{4\pi^{2}}$, the $V$-coordinate of the QES. On the other hand  $U(s)$  and $UV(s)$ reveal the flow crossing the  horizon and entering the future interior,  going deep into it, eventually   re-emerging on the right boundary. The endpoints of the flow are the same as in the early time regime \eqref{end,flow, TR}. The modular flow in this regime is  depicted in blue in figure \ref{figRflow}.

\paragraph{Regime III (late times):}
At late time, specifically,
\begin{equation}
    t_{R} > \frac{\beta}{2\pi}\log{\left[\frac{4\pi^{2}}{\beta^{2}\mathcal{V}^{2}}\left(1+\frac{\hat{\epsilon}}{2}\right)\right]} \,,
\end{equation}
we find two horizon crossings and two divergences under modular flow, shown in orange in \ref{figRflow}. Each of $U$ and $V$ diverges once and crosses the horizon once (figure \ref{Rflow}).
The fermion insertion, initially located outside the entanglement wedge bounded by the QES on the right, crosses the horizon at large positive Kruskal coordinate $V$, following which  $V$ diverges and the fermion is carried on the left side of the Penrose diagram as the modular parameter is dialed to large (negative in our convention) values. At this point the fermion seems to leave the AdS$_2$ (although the trajectory can still be formally plotted outside the boundary), reappearing behind the past horizon, crossing it  and terminating the flow on the right boundary. Reversing the sign of $s$ takes the fermion into the right conformal boundary at the edge of entanglement wedge. The flow endpoints are the same as in the early time regime \eqref{end,flow, TR}.

The comprehensive picture of modular flowed trajectories automatically implies the  appearance of new light-cone singularities in the boundary-to-bulk AdS$_2$ propagator,
\be
\langle O(X_{\rm bdry}) O(X(s))\rangle \sim \left(X_{\rm bdry}\cdot X(s)\right)^{-\Delta}\,,\label{bulktob}
\ee
where $X_{\rm bdry}$ is a point on either boundary of AdS$_2$ and $X(s)$ is the modular flowed coordinate \eqref{eqXs} of a bulk operator insertion at the cutoff  boundary with radial coordinate $\sinh\rho_0 = \beta/(2\pi \epsilon)$.  It is easy to check that the propagator above has bulk cone singularities due especially to modular flowed trajectories that cross the horizon (figure \ref{figRflow}).

Within the SYK$_\chi$ system, the correlation function \eqref{bulktob} can be obtained by applying modular flow to a correlator where one fermion is smeared with an HKLL propagator to obtain a local bulk insertion displaced slightly from the boundary \cite{Hamilton:2006az, Lensky:2020ubw}
\be
\langle O(X_{\rm bdry}) O(X(s))\rangle=\int_{D} dt^\prime K_\Delta(\rho_0; t_1,  t^\prime )\,{\rm Tr}\left[\rho_r  \rho_r^{-is} \chi(t^\prime)\rho_r^{is} \, \chi(t_2)\right]\,,
\ee
where $D$ is the boundary time-strip which is space-like separated from $(\rho_{0},t_{1})$ . (Here $\rho_r$ is the reduced density matrix, not to be confused with the radial position of the insertion $\rho_0$.)

\section{Summary and future questions}
\label{sec10}
In this work we studied the emergence of a bulk gravity description from a microscopic setup involving two SYK systems interacting with each other via all-to-all $q$-random interactions prepared in a TFD state, and the reduced state of one SYK subsystem obtained by tracing out the other, viewed as a bath.

This model was originally proposed in \cite{Penington:2019kki} in order to mimic a setup in which two black holes in thermal equilibrium  weakly interact with each other, exchanging Hawking radiation. The bulk problem, involving  two black holes in JT gravity interacting via bulk CFT matter, 
was analysed analytically to yield $n$-replica saddle points  in the $n\to 1$ limit to extract the locations of the QES. The focus on the microscopic interacting SYK system there was for the second R\'enyi entropy which was attacked numerically.

\begin{figure}[ht]
\begin{center}
\begin{tikzpicture}[scale=1.2, >=stealth]

  \draw[->, very thick] (-6,-0.5)--(0.5,5.5) node[anchor=west] {$V$};
  \draw[->, very thick] (0.5,-0.5)--(-6,5.5) node[anchor=east] {$U$};

  \fill[yellow, opacity=0.5] (-2,2.5)--(0,0.5)--(-0.4,1.4)--(-0.8, 2.5)--(-0.4,3.6)--(0,4.5)--cycle;
  \fill[yellow, opacity=0.5] (-3.5,2.5)--(-5.5,0.5)--(-5,1.5)--(-4.7,2.5)--(-5,3.5)--(-5.5,4.5)--cycle;
  \fill[red, opacity=0.5] (-3.5,2.5)--(-2.75,3.25)--(-2,2.5)--(-2.75,1.75)--cycle;

  \draw (-2,2.5) node[circle,fill,inner sep=1.8pt]{};
  \node at (-1.5,2.5) {$a_{R}$};

  \draw (-3.5,2.5) node[circle,fill,inner sep=1.8pt]{};
  \node at (-4,2.5) {$a_{L}$};

  \draw [gray, line width=0.7mm, dashed] plot[smooth]coordinates {(0.3, 0) (0,0.5) (-0.8, 2.5) (0, 4.5) (0.3, 5)};
  \draw [gray, line width=0.7mm,dashed] plot[smooth]coordinates {(-5.8,0) (-5.5,0.5) (-4.7,2.5) (-5.5, 4.5) (-5.8,5)};
\end{tikzpicture}
\end{center}
\caption{\small Entanglement wedge structure in the post Page time phase, obtained by patching the reconstructed bulk geometries in the two twist sectors.}
\label{island}
\end{figure}
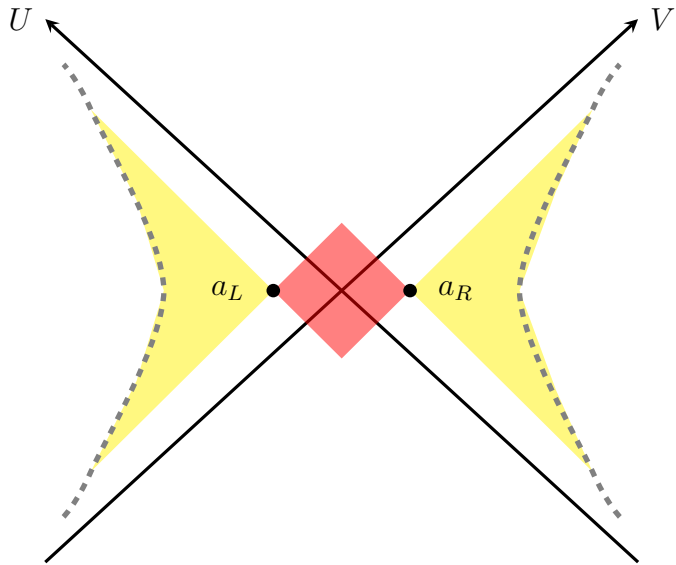

In this paper we have employed the large-$q$ limit to make the microscopic model analytically tractable, and  explored the late time ``island" saddle to extract modular flowed correlation functions in the limit $n=1$. The  SL$(2,\mathbb{R})$ transformations that ratchet through the replicas allowed us to  reconstruct bulk modular flows through an appropriate composition of  AdS$_2$ boosts. The fixed point of these flows gave us the QES locations demarcating the island (figure \ref{island}). 
The bulk modular flow trajectories we found, for operator insertions close to the boundary, are rich and probe the interior of the horizon of the  black hole dual to the SYK$_\chi$. 

However, as expected, bulk modular flow trajectories cannot explore the island. In fact, moving back to the gravity picture of the coupled black holes, before the bath system was integrated out, it is an interesting to question to understand if, and how, modular flowed operators in the second black hole (the bath) appear in the island region behind the horizon. In this context, a very instructive toy model for a black hole coupled to a bath is provided by BCFT which can be viewed as an IR proxy for SYK coupled to a CFT bath \cite{Hollowood:2021wkw}. In appendix \ref{BCFT} we show how an operator inserted in the radiation bath generates a second image that appears and stays within the island upon implementing modular flow. This is due to the bi-locality of the modular Hamiltonian in this case. The BCFT modular flow, since it can be carried out exactly, is also useful to understand how the late  time effective factorised picture can be patched together to yield a consistent description of the island.

The most challenging question in our framework is the experience of an infalling observer in the evaporating black hole. In order to implement this along the lines of \cite{Gao:2021tzr} we need to couple our interacting SYK system to an additional reference system that plays the role of the observer. In the non-evaporating case it has been shown that modular time coincides with proper time of the infalling observer. For the evaporating setup, modular flow must  be generated by the reduced density matrix obtained by tracing out both the bath and the observer system.  There are several fascinating questions to then pursue. What is the nature of the island if any, within such a microscopic framework? Can it be explored by the infalling reference system?  Is the experience of the infalling observer in the island region consistent with semiclassical expectations? We hope to address these questions in the near future.

Some additional questions naturally follow from the analysis presented in this work.  One concerns the perturbative solution to the replica recursion relations which we have solved to leading nontrivial order in the small ${\beta }{\cal V}$ limit. It would be interesting to understand how this can be extended to higher orders, and indeed to all orders in ${\beta }{\cal V}$. The bulk boost transformations that implement the modular flows are valid for any $\beta {\cal V}$, and provide a potential all orders answer to match with the microscopic calculation.  It would also be interesting to see explicitly how correctly smeared  SYK$_\chi$ correlators, upon modular flow, reproduce the bulk cone singularities implied by the AdS$_2$ modular flow trajectories we have found.

\acknowledgments 
The authors would like to thank Tim Hollowood, Carlos N\'u\~nez, Andy O'Bannon, Luke Piper, Giuseppe Policastro and  Neil Talwar for discussions and Andrea Legramandi for discussions and comments on early drafts of the paper. SPK would like to acknowledge support from STFC Consolidated Grant Award ST/X000648/1. NB is supported by the STFC DTP grant reference no. ST/Y509644/1. SPK would like to thank the Isaac Newton Institute for Mathematical Sciences, Cambridge, for support and hospitality during the programme ``Quantum field theory with boundaries, impurities, and defects", where work on this paper was undertaken; this work was supported by EPSRC grant EP/Z000580/1. 
\\
\\
{\small {\bf Open Access Statement} - For the purpose of open access, the authors have applied a Creative Commons Attribution (CC BY) licence to any Author Accepted Manuscript version arising. 

Data access statement: no new data were generated for this work.}

\newpage
\appendix
\section{Some discrete sums}
\label{app:sums}
We list below discrete sums that are used to obtain closed form expressions for $v_s$ and $\gamma_s$ in section \ref{sec:recursionsolve}. Taking $N$ to be odd as required we have,
\begin{align}
    &\sum_{n=0}^{N}{(-1)}^{n}\sin{na}= \text{Im}\left[\sum_{n=0}^{N}e^{i(\pi+a)n}\right]= -\frac{\sin\left(\frac{N+1}{2}a\right)\cos\left(\frac{N}{2}a\right)}{\cos\left(\frac{a}{2}\right)}\\
    &\sum_{n=0}^{N}{(-1)}^{n}\cos{na}=\text{Re}\left[\sum_{n=0}^{N}e^{i(\pi+a)n}\right]= \frac{\sin\left(\frac{N}{2}a\right)\sin\left(\frac{N+1}{2}a\right)}{\cos\left(\frac{a}{2}\right)}\\
    &\sum_{n=1}^{N}n(-1)^{n}\sin{an}= -\text{Re}\left[\frac{d}{da}\sum_{n=1}^{N}e^{i(\pi+a)n}\right] = -\frac{N\sin{[(N+1)a]} + (1+N)\sin{(Na)}} {4\cos^{2}\left(\frac{a}{2}\right)}\\
    &\sum_{n=1}^{N}n(-1)^{n}\cos{an}= \text{Im}\left[\frac{d}{da}\sum_{n=1}^{N}e^{i(\pi+a)n}\right] = -\frac{1+(1+N)\cos{(Na)}+N\cos{[(N+1)a]}} {4\cos^{2}\left(\frac{a}{2}\right)}
\end{align}

\section{AdS$_{2}$/EAdS$_{2}$: embeddings and coordinate systems}\label{appendix:AdS2}
\paragraph{Euclidean AdS:} We can describe the EAdS$_{2}$ as the hypersurface 
\begin{equation}
    -X_{3}^{2}+X_{1}^{2}+X_{2}^{2}=-\ell^{2}\,,
\end{equation}
in a three-dimensional embedding space with metric
\begin{equation}
    ds^{2}_{\rm ambient}=-dX_{3}^{2}+dX_{1}^{2}+dX_{2}^{2}\,.
\end{equation}
Parameterizing this hyperplane as
\bea
    X_{1}=\ell\sinh{\rho}\sin{\theta}\,,\quad
    X_{2}=\ell\sinh{\rho}\cos{\theta}\,,\quad
    X_{3}=\ell\cosh{\rho}\,,
\eea
yields the EAdS$_{2}$ metric
\begin{equation}
    ds_{\rm E}^{2}=\ell^{2}(d\rho^{2}+\sinh^{2}{\rho}d\theta^{2})\,.
\end{equation}
This metric describes the usual Euclidean disk, where $\rho\in[0,+\infty)$ is the radial coordinate and $\theta\sim \theta+2\pi$ is the angular one, with horizon at $\rho=0$ and conformal boundary at $\rho\to+\infty$.

\paragraph{Lorentzian AdS:} Lorentzian AdS$_{2}$ is  obtained by the  Wick rotation,
\begin{equation}
    \theta\to i\frac{2\pi}{\beta}t\,,
\end{equation}
and can be described by the hyperplane
\begin{equation}
    -X_{3}^{2}-X_{1}^{2}+X_{2}^{2}=-\ell^{2}\,,
\end{equation}
in  three-dimensional ambient space with metric
\begin{equation}
    ds^{2}_{{\rm ambient}}=-dX_{3}^{2}-dX_{1}^{2}+dX_{2}^{2}\,.
\end{equation}
Parameterizing this hypersurface as,
\bea
X_{1}=\ell\sinh{\rho}\sinh{\frac{2\pi}{\beta}t}\,,\quad X_{2}=\ell\sinh{\rho}\cosh{\frac{2\pi}{\beta}t}\,,\quad X_{3}=\ell\cosh{\rho}\,,
\eea
yields the AdS$_{2}$ metric 
\begin{equation}\label{Rindler, AdS2}
    ds^{2}=\ell^{2}(d\rho^{2}\,-\,\tfrac{4\pi^2}{\beta^2}\,\sinh^{2}{\rho}\,dt^{2})\,,
\end{equation}
which is just the usual Rindler patch of AdS$_{2}$, where now $t\in \mathbb{R}$ is non-compact. Setting $\ell=1$ for simplicity, the right and left Rindler wedges are respectively described by
\begin{equation}
    X_{2}=\pm\sinh{\rho}\cosh{\frac{2\pi}{\beta}t}\,.
\end{equation}
The two signs can each be understood as the analytic continuations via $\theta\to i\tfrac{2\pi}{\beta}t$  and $\theta \to \pi+i\tfrac{2\pi}{\beta}t$ from the Euclidean disk.

Another useful coordinate system is the AdS$_{2}$ metric written in Schwarzschild coordinates, obtained from \eqref{Rindler, AdS2} by defining the radial Schwarzschild coordinate $r=\cosh{\rho}$, 
\begin{equation}
    ds^{2}=-(r^{2}-1)dt^{2} + \frac{dr^{2}}{r^{2}-1}
\end{equation}
with horizon at $r=1$ and conformal boundary at $r\to +\infty$. Introducing the tortoise radial coordinate
\begin{equation}
    dr_{\ast}=\frac{dr}{r^{2}-1}\implies r_{\ast}(r)=\frac{1}{2}\log{\left|{\frac{r-1}{r+1}}\right|}\,,
\end{equation}
and the null coordinates
\begin{equation}
    u= t-r_{\ast}\,,\quad\quad\quad v= t+r_{\ast}\,.
\end{equation}
We can define the corresponding Kruskal-Szekeres (KS) coordinates,
\begin{equation}
    U=-e^{-u}, \quad\quad\quad V=e^{v}\,.
\end{equation}
The horizon is at $UV=0$, and the following relations hold between KS and Schwarzschild coordinates,
\bea
UV= -\frac{r-1}{r+1}\implies r=\frac{1-UV}{1+UV}\,,\qquad\quad
    \frac{V}{U}=-e^{2t}\implies t=\frac{1}{2}\log{\left(-\tfrac{V}{U}\right)}\,.
\eea
The AdS$_2$ metric in Kruskal coordinates becomes
\begin{equation}
    ds^{2}=-\frac{4dUdV}{(1+UV)^{2}}
\end{equation}
with the conformal boundary at $UV=-1$.
The embedding coordinates of AdS$_{2}$ in the KS coordinates are straightforwardly 
\be\label{embedding_Kruskal}
X_{1}=\frac{V+U}{1+UV}\,,\qquad X_{2}=\frac{V-U}{1+UV}\,,\qquad X_{3}=\frac{1-UV}{1+UV}\,.
\ee

\section{Microscopic-Bulk matching}
\label{charges}
In order to perform the matching between microscopic SYK and bulk gravity SL$(2,{\mathbb R})$ charges, we  introduce the $2\times2$ matrix built out of the embedding-space coordinates $X_\mu=(X_1,X_2,X_3)$,
\begin{equation}\label{Xmatrix}
\mathbf X \;=\frac{1}{\sqrt{2}}\; \begin{pmatrix}
X_1 & -X_2 + X_3 \\
-X_2 - X_3 & -X_1
\end{pmatrix} \,,
\end{equation}
so that ${\bf X}$ transforms by conjugation under SL$(2,{\mathbb R})$, while $X_\mu$ transforms like an SO$(2,1)$ vector.
Introducing the matrix basis,
\begin{equation}\label{basis matrices}
E_1=\frac{1}{\sqrt{2}}\begin{pmatrix}1&0\\0&-1\end{pmatrix},\qquad
E_2=\frac{1}{\sqrt{2}}\begin{pmatrix}0&-1\\-1&0\end{pmatrix},\qquad
E_3=\frac{1}{\sqrt{2}}\begin{pmatrix}0&+1\\-1&0\end{pmatrix},
\end{equation}
we can then write \eqref{Xmatrix} as
\begin{equation}
\mathbf X = X_1 E_1 + X_2 E_2 + X_3 E_3 .
\end{equation}
Noting that,
\begin{align}
\text{Tr}(\mathbf X\mathbf Y) = \big(X_1Y_1+X_2Y_2 - X_3Y_3\big) \,,
\end{align}
the scalar product in the embedding space  is realized by
\begin{equation}
X\cdot Y = -\text{Tr}(\mathbf X\mathbf Y).
\end{equation}
Therefore the inner product constraint $X\cdot Q = \frac{1}{2\epsilon}$, defining the boundary circular trajectory, becomes
\begin{equation}
-\text{Tr}(\mathbf X\mathbf Q) = \frac{1}{2\epsilon}.
\end{equation}
The SL$(2,\mathbb{R})$ charge
\begin{equation}
Q^{(0)\mu} = \{0,0,q_3\},\qquad q_3\equiv\frac{1}{2\epsilon\cosh\rho}\,,
\end{equation}
 becomes
\begin{equation}\label{Q0}
\mathbf Q_{0} = q_3 E_3 = \frac{q_3}{\sqrt{2}}\begin{pmatrix}0& 1\\-1&0\end{pmatrix}\,,
\end{equation}
in the two dimensional representation. The group elements corresponding to the generators of  SL$(2,\mathbb{R})$  in this representation are obtained by exponentiating the basis matrices \eqref{basis matrices} of the corresponding $\mathfrak{sl}(2,\mathbb{R})$ algebra\,,
\begin{align}
g_1(x) &= \exp\Big(-\frac{x}{\sqrt{2}}E_1\Big) = \begin{pmatrix}e^{-\frac{x}{2}} & 0 \\ 0 & e^{+\frac{x}{2}}\end{pmatrix},\\
g_2(y) &= \exp\Big(\frac{y}{\sqrt{2}}E_2\Big) = \begin{pmatrix}\cosh\frac y2 &\quad -\sinh\frac y2 \\-\sinh\frac y2 &\quad \cosh\frac y2\end{pmatrix},\\
\label{eq:g3}
g_3(\alpha) &= \exp\Big(\frac{\alpha}{\sqrt{2}}E_3\Big) = \begin{pmatrix}\cos\frac{\alpha}{2} &\quad +\sin\frac{\alpha}{2} \\-\sin\frac{\alpha}{2} &\quad \cos\frac{\alpha}{2}\end{pmatrix}.
\end{align}
The transformation between replicas can be then be represented as, 
\begin{equation}
B(x,\alpha) = M_1(-x)\,M_3(\alpha)\,M_1(x)\quad\Longleftrightarrow\quad g_B(x,\alpha)=g_1(-x)\,g_3(\alpha)\,g_1(x).
\end{equation}
Carrying out the matrix multiplication gives the closed form
\begin{equation}
g_B(x,\alpha)=\begin{pmatrix} \cos\frac{\alpha}{2} & e^{x}\,\sin\frac{\alpha}{2}\\ -e^{-x}\,\sin\frac{\alpha}{2} & \cos\frac{\alpha}{2} \end{pmatrix}\,.
\end{equation}
Its $s$-th power is obtained by replacing $\alpha\mapsto s\alpha$:
\begin{equation}
g_B^s(x,\alpha) = \begin{pmatrix} \cos\frac{s\alpha}{2} & e^{x}\,\sin\frac{s\alpha}{2}\\[6pt] -e^{-x}\,\sin\frac{s\alpha}{2}& \cos\frac{s\alpha}{2} \end{pmatrix}\,.
\end{equation}
The group action on embedding matrices is by conjugation:
\begin{equation}
\mathbf X \mapsto \mathbf X' = g\,\mathbf X\,g^{-1},\qquad g\in {\rm SL}(2,{\mathbb R}).
\end{equation}
Therefore the charges transform as
\begin{equation}
\mathbf Q_{s, B} = g_B^s\,\mathbf Q_{0}\,g_B^{-s}\,.
\end{equation}
yielding
\begin{equation}
\mathbf Q_{s, B} = \frac{q_3}{\sqrt{2}}\begin{pmatrix}
-\sinh x\,\sin(s\alpha) & \cos^2(\frac{s\alpha}{2}) + e^{2x}\,\sin^2\left(\frac{s\alpha}{2}\right)\\
-\cos^2(\frac{s\alpha}{2}) - e^{-2x}\,\sin^2\left(\frac{s\alpha}{2}\right) & \sinh x\,\sin(s\alpha)
\end{pmatrix} \;.
\end{equation}
This charge is expected to match with the one found in the microscopic picture \eqref{microscopic-charges}, both for the left and right twist field contributions. Explicitly, for the left twist field contributions, we can make the identification,
\be
u_s= v_s \sin (s\alpha_L) \sinh x_L\,,\qquad v_s= \frac{e^{x_L}}{\cosh x_L + \sinh x_L \cos (s\alpha_L)}\,.
\ee
The norm squared of the twist operator using the formula \eqref{TL-charge} in this parametrisation is,
\be\label{adstwistmass}
M_s^2= \frac{4\pi^2}{\beta^2}\sin^2\left(\tfrac12\alpha_{L,R}\right)\sinh^2 x_{L,R}\,.
\ee
Matching with SYK parameters yields,
\begin{equation}
\begin{cases}
    \displaystyle\alpha_{L}=\alpha_{R}\simeq 2\pi \\\\\nonumber
     \displaystyle x_{L,R}\simeq\pm\frac{\beta^{2}\mathcal{V}^{2}}{2\pi^{2}}
\end{cases}
\end{equation}

\section{Modular flow in BCFT in a TFD state}\label{BCFT}
CFT on the half line, or BCFT provides a very interesting toy model with many of the ingredients required for a quantitative and qualitative match with the IR physics of a CFT (proxy for radiation bath) coupled to the SYK model \cite{Hollowood:2021wkw, Piper:2025}, in particular in the limit of high temperature and/or large interval lengths in the radiation bath, so the  scrambling time scale is negligible in comparison. In this limit, positions of QES can be related to BCFT image points obtained via the doubling trick in BCFT.

Let us  consider a BCFT on the half-line and prepare two copies BCFT$_{L}$ and BCFT$_{R}$ in a TFD state. The respective spatial coordinates $x$ are positive, and we are interested in subregion $V$ to be $V=[a,+\infty]_{L}\cup [a,+\infty]_{R}$. In the gravity picture, the Hartle-Hawking state and its time evolution is described by a two-sided black hole. Let us now define Kruskal-Szekeres (KS) coordinates $w^{\pm}$ covering the full space, as well as the null coordinates $x^{\pm}$ which are Minkowski-like in the baths and Schwarzschild-like in the AdS$_{2}$ region
\begin{equation}\label{Kruskal coordinates}
    w^{\pm}=\pm e^{\pm\frac{2\pi}{\beta}x^{\pm}}
\end{equation}
The Schwarzschild time in the left region includes a sign change and an imaginary shift with respect to the coordinate on the right side
\begin{equation}
    L:\quad x^{\pm}=-t\pm x +\frac{i\beta}{2}, \quad\quad R:\quad x^{\pm}=t\pm x
\end{equation}
where $x\geq0$ in both left and right Minkowski baths. The imaginary shift on the left implements the necessary analytic continuation to obtain correlators in the TFD state from Euclidean thermal correlators. 

For the  problem with semi-infinite radiation bath intervals,  in the setup with  CFT propagating on the AdS$_2$ black hole coupled to a non-gravitating bath in the TFD state, the late time positions of the quantum extremal surfaces can be easily shown to be \cite{Hollowood:2021wkw},
\be
w_{Q, L}^\pm= \mp e^{\pm 2\pi/\beta(-t\mp a) + t_{\rm scr}}\,,\qquad 
w_{Q, R}^\pm= \pm e^{\pm 2\pi/\beta(t\mp a ) + t_{\rm scr}}
\ee
If we neglect $t_{\rm scr}$ in the exponent, it follows immediately that these coordinates can be interpreted as the images of the reflections of the point $x_{L,R} =a$ about the origin in BCFT$_{L,R}$. In this sense we can identify QES positions in gravity with BCFT image points associated to interval endpoints where twist fields are inserted.

\subsection{Modular flow of a single fermion}
With the above picture and correspondence in mind, let us take the simplest BCFT, namely free massless Dirac fermions on the half-line, denoting incoming (into the boundary) chirality modes as $\psi_{+}(x^{+})$ and the outgoing modes as $\psi_{-}(x^{-})$. The two-point function of spinors of generic chirality is given by\,,
\begin{equation}\label{2pf_BCFT}
    G(x^{i},y^{j})\equiv\braket{\psi_{i}(x^{i})\psi_{j}(y^{j})}=\frac{1}{2i\beta\sinh{\left[\frac{\pi}{\beta}(x^{i}-y^{j})\right]}}\,, \quad\quad\quad i,j\in\{+,-\}\,.
\end{equation}
Now consider a Cauchy slice $\Sigma$ at constant time and a subregion $V\subset \Sigma$ consisting of $N$ disjoint intervals $V=\cup_{l}(a_{l}, b_{l})$ with $l=1,\dots, N$. Given
\begin{equation}
    \rho_{V}=\text{Tr}_{V^{c}}(\rho_{\Sigma})\,,
\end{equation}
we wish to compute the modular flow of a incoming mode $\psi_{+}(x)$
\begin{equation}
    \sigma_{s}(\psi_{+}(x))\equiv e^{-isK}\psi_{+}(x)e^{isK}
\end{equation}
where $K=-\log{\rho_{V}}$ is the modular Hamiltonian associated to the subregion $V$ and $s$ is the modular time.

Using the resolvent method, as shown in \cite{Erdmenger:2020nop, Reyes:2021npy}, the modular flow of the incoming mode $\psi_{+}(x)$ can be computed as
\begin{equation}
    \sigma_{s}(\psi_{+}(x))=\int_{V}dy \Sigma_{ij}(x,y) \psi_{j}(y)\,,
\end{equation}
with $i,j\in\{+,-\}$ and 
\bea
    &&\Sigma_{ij}(x,y)=2 i \sinh{(\pi s)}\;\delta(Z(x^{i})-Z(y^{j})-s)G_{ij}(x,y)\,,\\\nonumber\\
    &&Z(x)=\frac{1}{2\pi}\log\Omega(x), \quad\quad \Omega(x)=-\prod_{l}\frac{G(x,b_{l}^{+})}{G(x,a_{l}^{+})}\frac{G(x,a_{l}^{-})}{G(x,b_{l}^{-})}\,.\label{cross_ratio_BCFT}
\eea
In our case of interest, the cross ratio $\Omega(x)$ is, 
\begin{equation}
    \Omega(x)=-\frac{G(x,a_{R}^{+})G(x,a_{L}^{-})}{G(x,a_{R}^{-})G(x,a_{L}^{+})}\,,
\end{equation}
where we have denoted the (finite) endpoints of the two semi-infinite intervals as $a_{R}$ and $a_{L}$. 

Since we are considering the union of two semi-infinite intervals in the TFD state, with only one pair of twist operator insertions (one for each BCFT), we can use the fact that correlators in the TFD state can be viewed as analytic continuations of thermal correlators in a single BCFT. Employing the explicit expression of the correlation functions \eqref{2pf_BCFT}, $\Omega(x)$  can be written explicitly as 
\begin{align}
    \Omega(x)&=-\frac{\sinh{\frac{\pi}{\beta}\left[x-(t-a)\right]}\sinh{\frac{\pi}{\beta}\left[x-(-t+\frac{i\beta}{2}+a)\right]}}{\sinh{\frac{\pi}{\beta}\left[x-(t+a)\right]}\sinh{\frac{\pi}{\beta}\left[x-(-t+\frac{i\beta}{2}-a)\right]}}\\
    &=-\frac{\sinh{\frac{\pi}{\beta}\left[x-(t-a)\right]}\cosh{\frac{\pi}{\beta}\left[x+t-a\right]}}{\sinh{\frac{\pi}{\beta}\left[x-(t+a)\right]}\cosh{\frac{\pi}{\beta}\left[x+t+a\right]}}\\
    &=-\frac{\left(X+\frac{A}{T}\right)\left(X-\frac{T}{A}\right)}{\left(X-AT\right)\left(X+\frac{1}{AT}\right)}\,,
\end{align}
where we introduced the change of coordinate 
\begin{equation}
    K=e^{\frac{2\pi}{\beta}k}, \quad\quad k=x,a,t\,.
\end{equation}
To understand the locality properties of the flow, we must examine the number and nature of solutions to
\begin{equation}
    Z(x^{i})-Z(y)-s=0\,,
\end{equation}
or, in terms of the new coordinates
\begin{equation}
    -\frac{\left(Y+\frac{A}{T}\right)\left(Y-\frac{T}{A}\right)}{\left(Y-AT\right)\left(Y+\frac{1}{AT}\right)}=\Omega(x^{i})e^{-2\pi s}\,.
\end{equation}
\subsection{Images for the two intervals case}
First, take  the case $i=+$, for which we have
\begin{equation}\label{positive_chirality_eq}
    -\frac{\left(Y+\frac{A}{T}\right)\left(Y-\frac{T}{A}\right)}{\left(Y-AT\right)\left(Y+\frac{1}{AT}\right)}=-e^{-2\pi s}\frac{\left(XT+\frac{A}{T}\right)\left(XT-\frac{T}{A}\right)}{\left(XT-AT\right)\left(XT+\frac{1}{AT}\right)}\,.
\end{equation}
This is a quadratic equation for $Y$ and it is immediate that the two roots which we label $Y_1^+$ and $Y_1^-$, must satisfy,
\be
Y_1^+(s) Y_2^-(s)=-1\,.
\ee
The labelling of the roots indicates that one solution is the modular flow of the position of the chirality $+$ fermion that was inserted, and the other is a negative chirality image as a result of the refecting boundary in the BCFT.
In particular, taking $s\to0$, we have
\begin{equation}
    Y^{+}_{1}(s=0)=XT, \quad\quad\quad Y^{-}_{2}(s=0)=-\frac{1}{XT}\,.
\end{equation}
Since the positive chiral component of $Y_{1}$ approaches $X^{+}=XT$ as $s\to0$, we will refer to it as the “equal chirality" image, whereas $Y_{2}$ will be denoted as the “opposite chirality" image. Thus, solving the equation with respect to $x^{+}$, we will find $Y^{+}_{1}$ and $Y^{-}_{2}$, while solving with respect to $x^{-}$, we will find $Y^{-}_{1}$ and $Y^{+}_{2}$.\\
Solving equation \eqref{positive_chirality_eq}, we get the solutions $Y^{+}_{1}$ and $Y^{-}_{2}$ with the following patterns, depending on whether $X$ is larger or smaller than $A$,
\begin{figure}[H]
            \centering
            \includegraphics[width=0.4\linewidth]{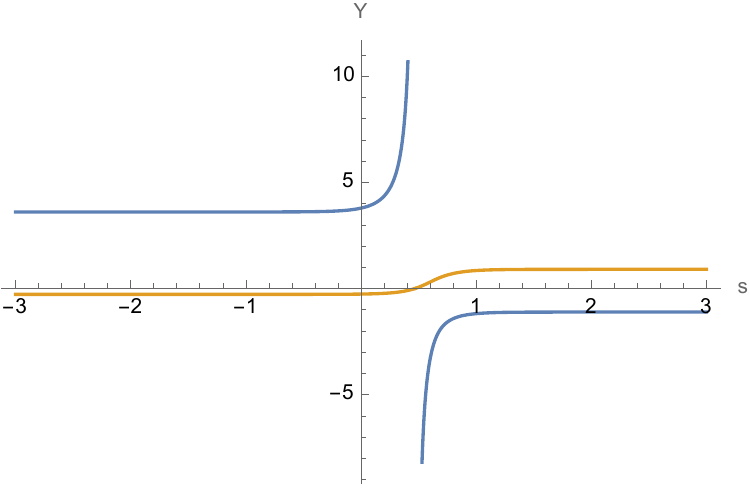}
            \includegraphics[width=0.4\linewidth]{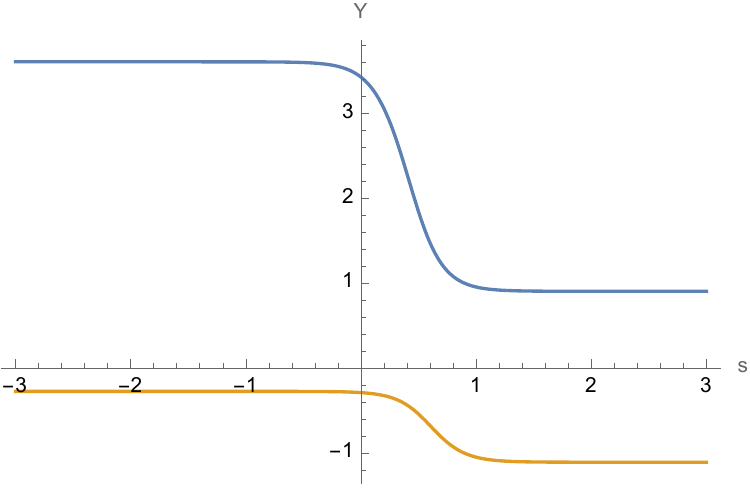}\\
            \includegraphics[width=0.4\linewidth]{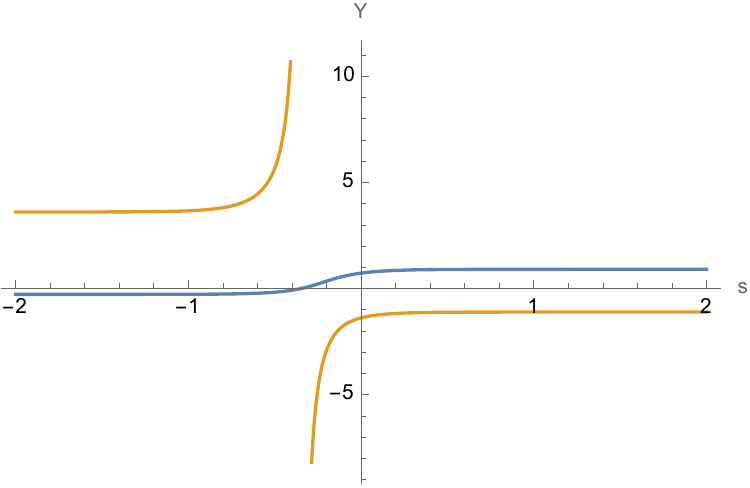}
            \includegraphics[width=0.4\linewidth]{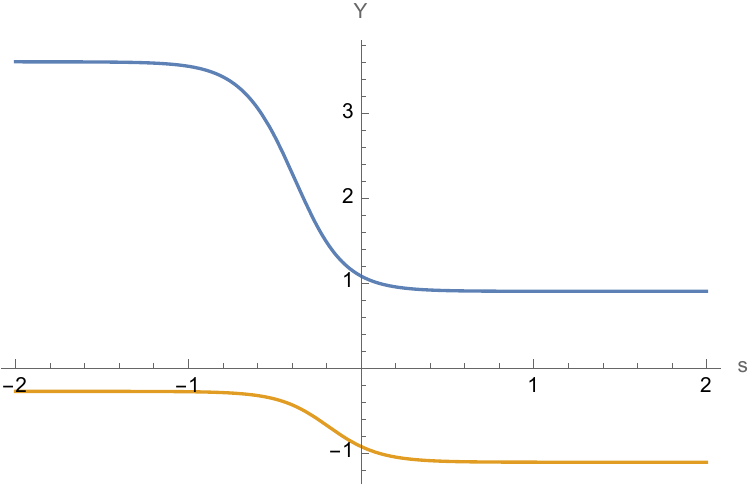}
            \caption{$Y^{+}_{1}$ denoted in blue and $Y^{-}_{2}$ in orange. $A=2$, $T=1.8$, $X=2.1$ (top left),  $X=1.9$ (top right), $X=0.4$ (bottom left), $X=0.6$ (bottom right).}
\end{figure}
\
\\
\
and with the following asymptotic behaviour for the two images
\begin{align}
    &Y_{1}^{+}(s\to+\infty)=\begin{cases}
        -\frac{A}{T} \quad\quad\quad X>A\\
        \frac{T}{A}  \quad\quad\quad\;\; X<A
    \end{cases} \quad &&Y_{2}^{-}(s\to+\infty)=\begin{cases}
        \frac{T}{A} \quad\quad\quad\;\;\; X>A\\
        -\frac{A}{T}  \quad\quad\quad X<A
    \end{cases}\\
    &Y_{1}^{+}(s\to-\infty)= \begin{cases}
        AT \quad\quad\quad X>\frac{1}{A}\\
        -\frac{1}{AT}  \quad\quad\, X<\frac{1}{A}
    \end{cases} \quad\quad\quad &&Y_{2}^{-}(s\to-\infty)= \begin{cases}
        -\frac{1}{AT} \quad\quad\, X>\frac{1}{A}\\
        AT\quad\quad\quad X<\frac{1}{A}
    \end{cases}
\end{align}
The singularities/zeros of the solutions, occurring in the cases $X>A$ and $X<1/A$, are located at
\begin{equation}
    1+e^{-2\pi s}\Omega(x^{+})=0 \implies s_{1}=\frac{1}{2\pi}\log{[-\Omega(x^{+})]}\,.
\end{equation}
The physical interpretation of this singularity and the change of sign of the images will be explained in a moment. \\
Solving the equation for the case $i=-$, we get solutions with the following patterns, again depending on whether $X$ is larger or smaller than $A$
\begin{figure}[H]
            \centering
            \includegraphics[width=0.4\linewidth]{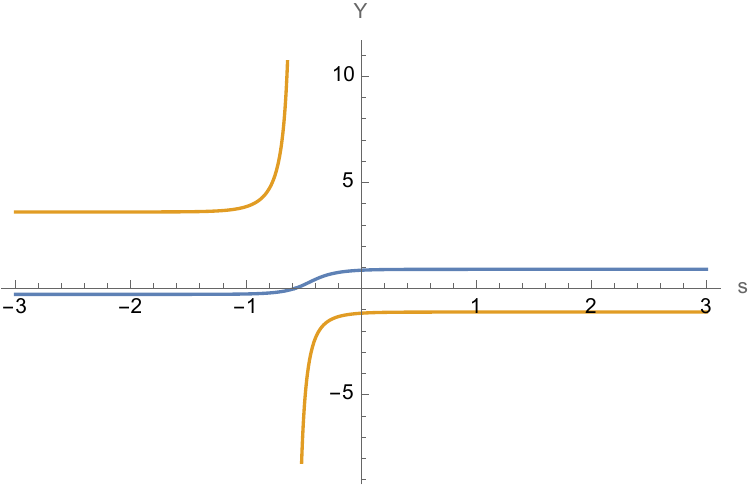}
            \includegraphics[width=0.4\linewidth]{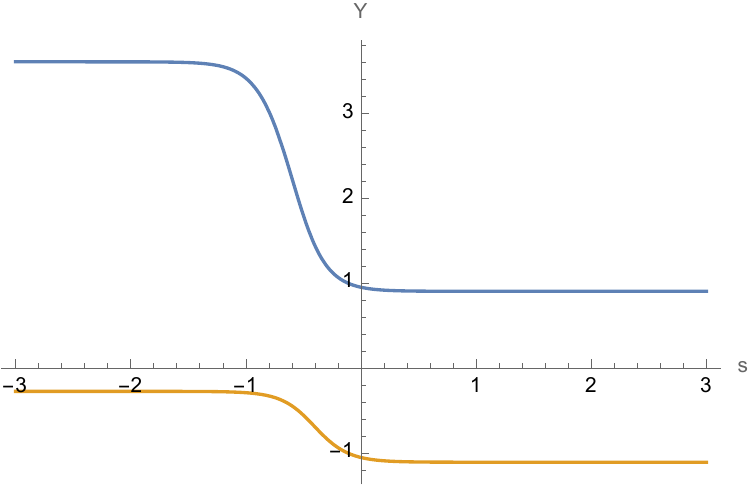}\\
            \includegraphics[width=0.4\linewidth]{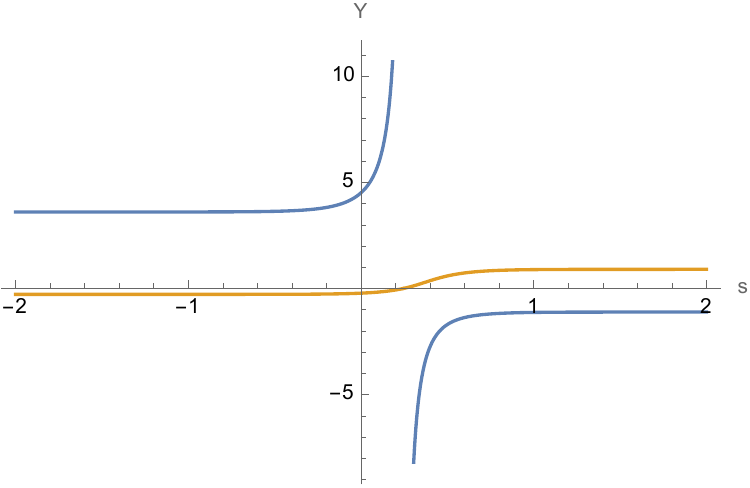}
            \includegraphics[width=0.4\linewidth]{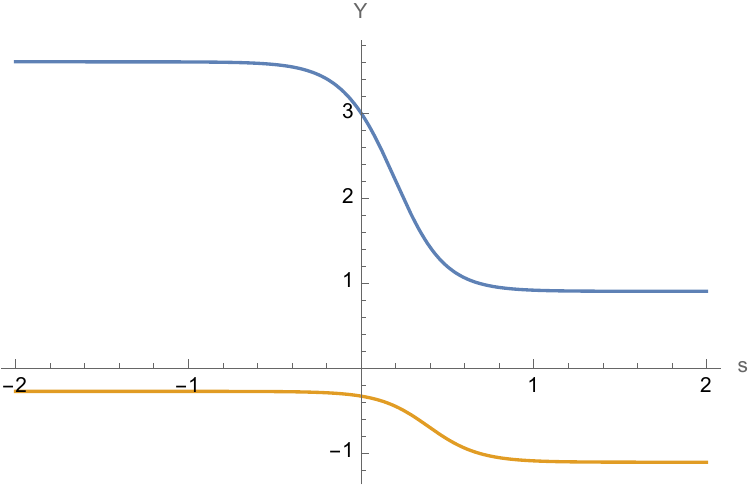}
            \caption{$Y^{-}_{1}$ denoted in blue and $Y^{+}_{2}$ in orange. $A=2$, $T=1.8$, $X=2.1$ (top left),  $X=1.9$ (top right), $X=0.4$ (bottom left), $X=0.6$ (bottom right).}
\end{figure}
In this case, we get the negative chiral component of the 'equal chirality' solution and the positive chiral component of the 'opposite chirality' solution, i.e. $Y^{-}_{1}$ and $Y^{+}_{2}$ as explained above. They satisfy
\begin{align}
    &Y_{1}^{-}(s=0)=\frac{T}{X} \quad\quad\quad &&Y_{2}^{+}(s=0)=-\frac{X}{T}\\
    &Y_{1}^{-}(s\to+\infty)=\begin{cases}
        \frac{T}{A} \quad\quad\quad\quad\;\;\; X>\frac{1}{A}\\
        -\frac{A}{T}  \quad\quad\quad\quad X<\frac{1}{A}
    \end{cases} \quad\quad\quad &&Y_{2}^{+}(s\to+\infty)=\begin{cases}
        -\frac{A}{T} \quad\quad\quad\quad X>\frac{1}{A}\\
        \frac{T}{A}  \quad\quad\quad\quad\;\;\, X<\frac{1}{A}
    \end{cases}\\
    &Y_{1}^{-}(s\to-\infty)=\begin{cases}
        -\frac{1}{AT} \quad\quad\quad X>A\\
        AT  \quad\quad\quad\;\; X<A
    \end{cases} \quad\quad\quad &&Y_{2}^{+}(s\to-\infty)=\begin{cases}
        AT \quad\quad\quad\;\;\; X>A\\
        -\frac{1}{AT}  \quad\quad\quad X<A
    \end{cases}
\end{align}
and, for the cases $X>A$ and $X<1/A$, there is still a pole/zero for solutions,
\begin{equation}
    1+e^{-2\pi s}\Omega(x^{-})=0 \implies s_{2}=\frac{1}{2\pi}\log{[-\Omega(x^{-})]}\,.
\end{equation}
For large $X$, both $s_{1}$ and $s_{2}$ becomes small and close to $s=0$.\\
At this point, the modular flow of the two images can be fully understood using the \textit{doubling trick}, which enables to view the BCFT in the half-line as a chiral CFT in the complex plane, and the Kruskal-Szekeres coordinates \eqref{Kruskal coordinates}. It turns out
\begin{align}
    \text{image 1}:\quad\left(w^{+}_{1},w^{-}_{1}\right)=\left(Y_{1}^{+},-\frac{1}{Y_{1}^{-}}\right)\,,\\
    \text{image 2}:\quad\left(w^{+}_{2},w^{-}_{2}\right)=\left(Y_{2}^{-},-\frac{1}{Y_{2}^{+}}\right)\,.
\end{align}
In the bulk geometry, the flows turn out to be of the following form:
\begin{figure}[H]
            \centering
            \includegraphics[width=0.4\linewidth]{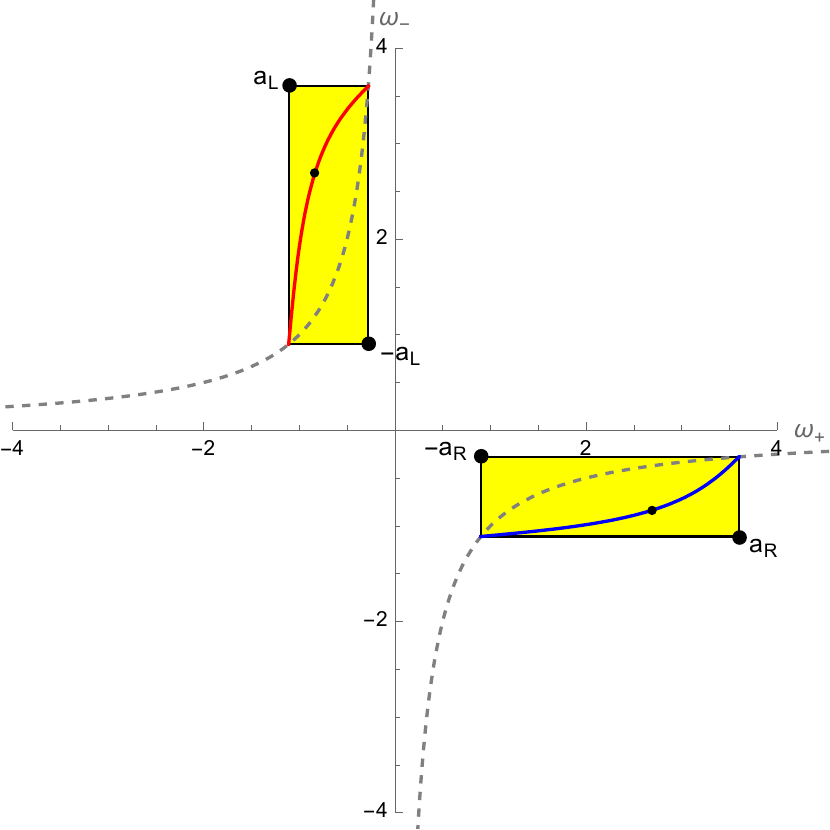}
            \includegraphics[width=0.4\linewidth]{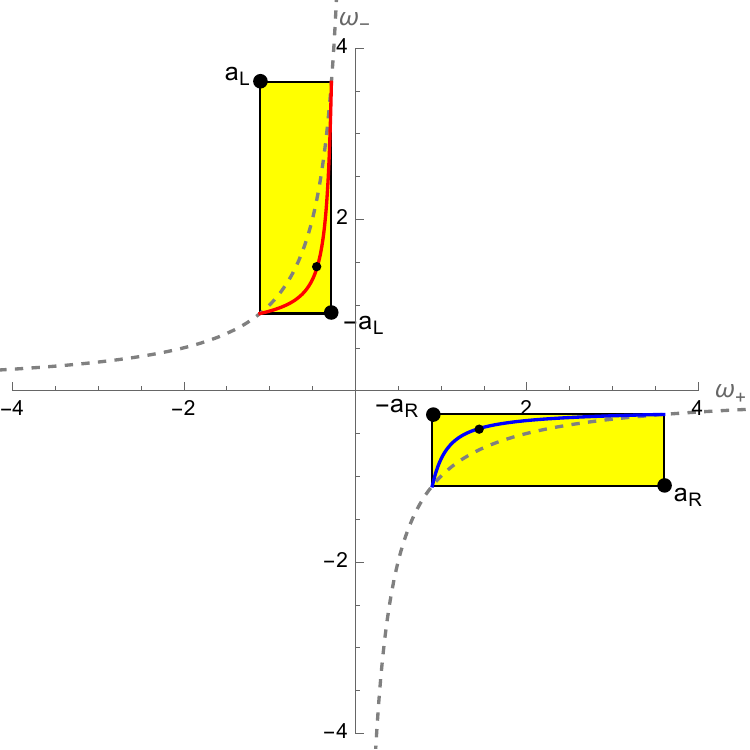}\\
            \
            \\
            \includegraphics[width=0.4\linewidth]{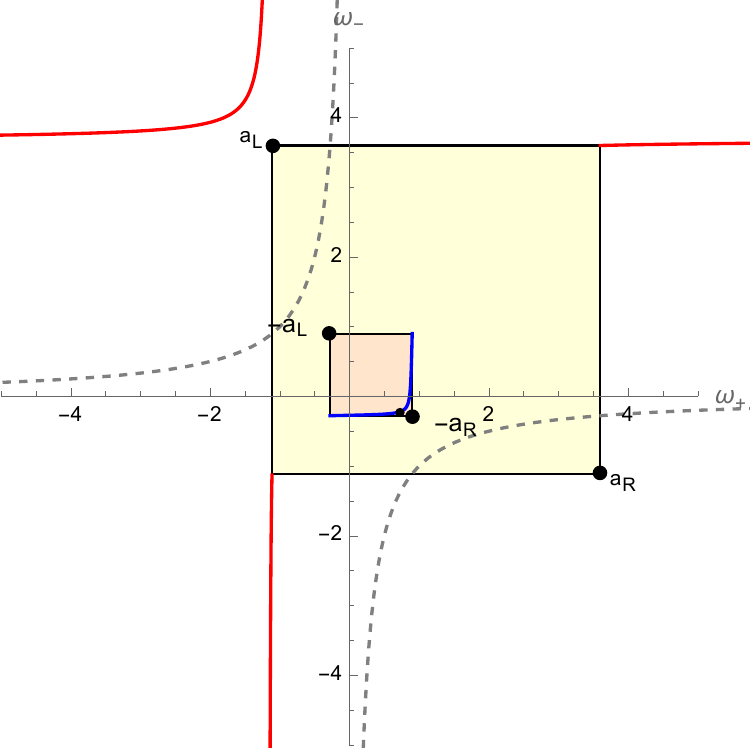}
            \includegraphics[width=0.4\linewidth]{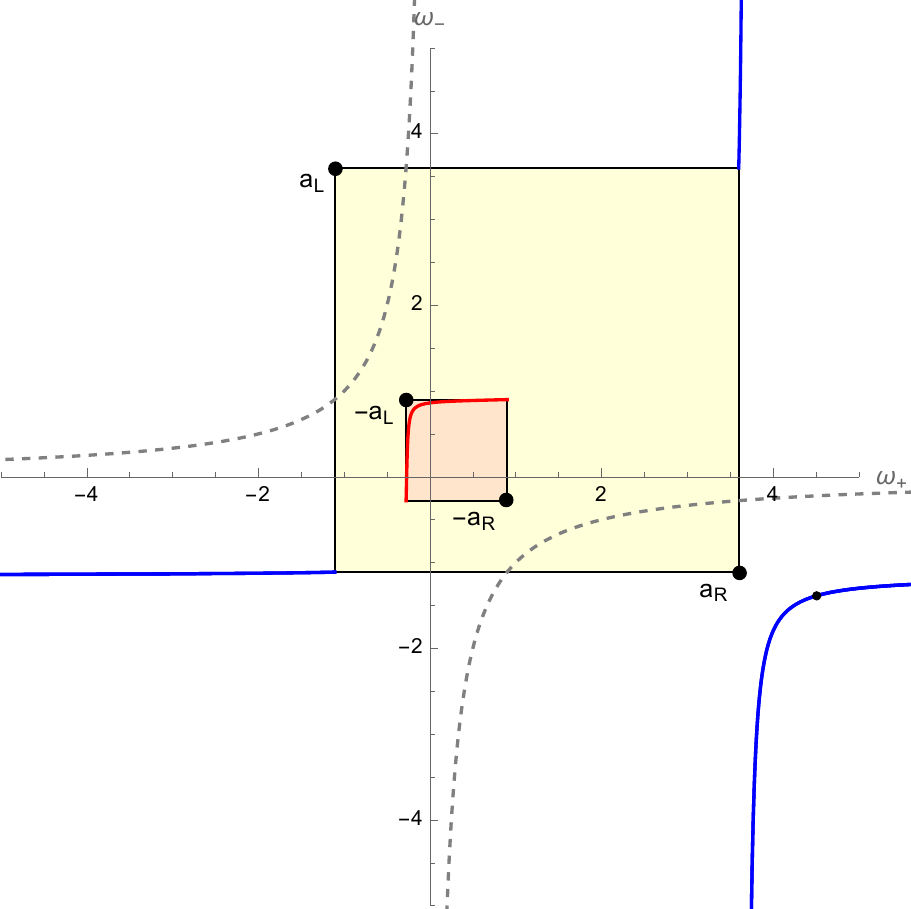}
            \caption{The modular flow of the two images in Kruskal coordinates, where the image $Y_{1}$ is depicted in blue and $Y_{2}$ in red. $A=2$, $T=1.8$, $X=1.5$ (top left),  $X=0.8$ (top right), $X=0.4$ (bottom left) and $X=2.5$ (bottom right). AdS$_2$ boundaries are denoted by the gray dashed lines and the entanglement wedges of the relevant intervals are depicted in yellow.}
            \label{flow_BCFT}
\end{figure}
The first row of Figure \ref{flow_BCFT} refers to the case in which $1/A<X<A$: the equal-chirality image and opposite-chirality one result constrained within the entanglement wedge of the two intervals $[-a_{R},a_{R}]$ and $[-a_{L},a_{L}]$ respectively (shaded in yellow) and both of them end on the boundary at $s\to \pm \infty$; in addition, if $0<X<A$ each image lies entirely in its respective bath region, if $1/A<X<1$ they lie entirely inside the AdS$_2$ black hole and if $X=1$ each image will flows along its respective boundary.\par
The second row of of Figure \ref{flow_BCFT} reports the flow of the images in the cases $X<1/A$ and $X>A$ respectively. \\
In the case we modular flow a fermion in the island (i.e. $X<1/A$), we get the equal-chirality image flowing in the entanglement wedge of the island\footnote{Analogous results have been found in \cite{Chen:2019iro}. The difference is that, while in \cite{Chen:2019iro} the island is included as part of the entanglement wedge of Hawking radiation from the outset, here the island and the fermion's image flowing inside it emerge in the BCFT picture by employing the doubling trick.}  (shaded in orange), and the opposite-chirality image flowing in the entanglement wedge of the two semi-infinite intervals $[a_{L},+\infty]\cup [a_{R},+\infty]$, complement to the entanglement wedge of $[a_{L},a_{R}]$ (shaded in light yellow). The opposite-chirality image sits at $s=0$ in the left semi-infinite interval and flow into the black hole's future/past interior, without crossing the horizon, rather by going to infinity at modular times $s_{1,2}$, which are the times at which the island image crosses the horizons. Interestingly, when the island image spans the black hole interior, the semi-infinite interval image lies in the interior as well.\\ 
In the case we modular flow a fermion in the right semi-infinite interval (i.e. $X>A$), we get the analogous scenario emerging for $X<1/A$, with the equal-chirality image flowing in the entanglement wedge of $[a_{L},+\infty]\cup [a_{R},+\infty]$ and the opposite-chirality one flowing in the island's entanglement wedge.

\paragraph{Special case: $\boldsymbol{T \gg A}$}\
\\
\
If we consider $T \gg A>1$, the two solutions to \eqref{positive_chirality_eq} will simplify as
\begin{align}\label{solutions_largeT_BCFT}
    &Y_{a}(s)\simeq \frac{T}{A}\frac{(1-A^{2}C)-\sqrt{(1-A^{2}C)^{2}}}{2(1-C)}\\
    &Y_{b}(s)\simeq \frac{T}{A}\frac{(1-A^{2}C)+\sqrt{(1-A^{2}C)^{2}}}{2(1-C)}
\end{align}
where we defined 
\begin{equation}
    C = -e^{-2\pi s}\Omega(x^{+})\simeq e^{-2\pi s}\frac{X-\frac{1}{A}}{X-A}
\end{equation}
Notice that we have denotes the two solutions with a general notation $Y_{a,b}(s)$, since according to whether $X>A$, $1/A<X<A$ or $X<1/A$, they will describe the equal or opposite chirality image.\\
At this point, defining 
\begin{equation}
    s_{\star}=\frac{1}{2\pi}\log{\left[-A^{2}\,\Omega(x^{+})\right]}\simeq \frac{1}{2\pi}\log{\left(A\frac{AX-1}{X-A}\right)}
\end{equation}
for the cases $X<1/A$ and $X>A$, for which $s_{\star}$ is real, we get, neglecting $\mathcal{O}(1/T)$ terms,
    \begin{align}
    &\boxed{Y_{a}(s)\simeq\begin{cases}
        0\quad\quad\quad\quad\quad\;\;\; s\geq s_{\star}\\
        \frac{T}{A}\frac{1-A^{2}C}{1-C}\quad\quad\;\;\, s< s_{\star}
    \end{cases}
    }\label{image1_late_time_outew}
\\\notag
    \
    \\
    \
    &\boxed{Y_{b}(s)\simeq\begin{cases}
        \frac{T}{A}\frac{1-A^{2}C}{1-C}\quad\quad\;\;\; s\geq s_{\star}\\
        0\quad\quad\quad\quad\quad\;\;\;\, s< s_{\star}
        \end{cases}
        }\label{image2_late_time_outew}
    \end{align}
with $Y_{a}(s)=Y_{1}^{+}(s)$ for $X>A$ and $Y_{a}(s)=Y_{2}^{-}(s)$ for $X<1/A$, and consequently vice versa for $Y_{b}(s)$.\\
For the remaining case $1/A<X<A$, there is no sign ambiguity in simplifying the square root in \eqref{solutions_largeT_BCFT}, thus we get, neglecting $\mathcal{O}(1/T)$ terms,
   \begin{align}
    &\boxed{Y_{2}^{-}(s)=Y_{a}(s)\simeq0
    }\label{image1_late_time_inew}
\\\notag
    \
    \\
    \
    &\boxed{Y_{1}^{+}(s)=Y_{b}(s)\simeq
        \frac{T}{A}\frac{1-A^{2}C}{1-C}
        } \label{image2_late_time_inew}
    \end{align}

\subsection{Modular flowed fields}
The modular flow of $\psi_{+}(x)$ will take the form
\begin{equation}\label{modular_flow_BCFT}
    \sigma_{s}(\psi_{+}(x))=2\sinh{(\pi s)}\left[G_{++}(x,y_{1}^{+})\psi_{+}(y_{1}^{+})\left|\frac{2\pi\Omega(y)}{\Omega'(y)}\right|_{y=y_{1}} + G_{+-}(x,y_{2}^{-})\psi_{-}(y_{2}^{-})\left|\frac{2\pi\Omega(y)}{\Omega'(y)}\right|_{y=y_{2}}\right]
\end{equation}
where we use, since we are dealing with a static mirror,
\begin{equation}
    G_{ij}(x,y)=G(x^{i},y^{j}), \quad\quad i,j\in\{+,-\}
\end{equation}
Let's prove that the modular flow is unitary, by showing that the norm of $\psi_{+}(x)$ is preserved. First of all, the spinors on RHS, i.e. $\psi_{+}(y_{1}^{+})$ and $\psi_{-}(y_{2}^{-})$, must be evaluated in $x^{+}$ and $x^{-}$ respectively to correctly check the preservation of the norm, thus let's perform the transformation back with the proper Jacobian
\begin{align}
    &\psi_{+}(y_{1}^{+}(x))=\left(\frac{dx^{+}}{dy_{1}^{+}}\right)^{1/2}\psi_{+}(x^+(s)) = \left(\left.\frac{\partial\Omega(X)}{\partial X}\right|_{X=XT}\left.\frac{\partial Y}{\partial\Omega(Y)}\right|_{Y=Y_{1}^{+}}\frac{XT}{Y_{1}^{+}}e^{-2\pi s}\right)^{-1/2}\psi_{+}(x^+(s))\\
    &\psi_{-}(y_{2}^{-}(x))=\left(\frac{dx^{-}}{dy_{2}^{-}}\right)^{1/2}\psi_{-}(x^{-}(s))=\left(\left.\frac{\partial\Omega(X)}{\partial X}\right|_{X=XT}\left.\frac{\partial Y}{\partial\Omega(Y)}\right|_{Y=Y_{2}^{-}}\frac{XT}{Y_{2}^{-}}e^{-2\pi s}\right)^{-1/2}\psi_{-}(x^{-}(s))
\end{align}
thus putting all together, the two coefficients of the two spinors on the right hand side of \eqref{modular_flow_BCFT}, written in “capital" variables, turn out to be
{
\footnotesize
\begin{align}
    & c_{++}(s)=2\sinh{(\pi s)}\frac{1}{i\beta}\frac{\sqrt{XTY_{1}^{+}}}{XT-Y_{1}^{+}}\left(\left.\frac{\partial\Omega(X)}{\partial X}\right|_{X=XT}\left.\frac{\partial Y}{\partial\Omega(Y)}\right|_{Y=Y_{1}^{+}}\frac{XT}{Y_{1}^{+}}e^{-2\pi s}\right)^{-1/2}\left(\frac{\beta}{Y_{1}^{+}}\Omega(Y_{1}^{+})\left.\frac{\partial Y}{\partial\Omega(Y)}\right|_{Y=Y_{1}^{+}}\right)\\
    & c_{+-}(s)=2\sinh{(\pi s)}\frac{1}{i\beta}\frac{\sqrt{XTY_{2}^{-}}}{XT-Y_{2}^{-}}\left(\left.\frac{\partial\Omega(X)}{\partial X}\right|_{X=XT}\left.\frac{\partial Y}{\partial\Omega(Y)}\right|_{Y=Y_{2}^{-}}\frac{XT}{Y_{2}^{-}}e^{-2\pi s}\right)^{-1/2}\left(\frac{\beta}{Y_{2}^{-}}\Omega(Y_{2}^{-})\left.\frac{\partial Y}{\partial\Omega(Y)}\right|_{Y=Y_{2}^{-}}\right)
\end{align}
}
where the first factor represents $2\sinh{(\pi s)}G_{+\pm}(x,y)$ and the last one account for $\left|\frac{2\pi\Omega(y)}{\Omega'(y)}\right|$ which is invariant under $Y\to -1/Y$; thus \eqref{modular_flow_BCFT} can be written compactly
\begin{equation}
    \sigma_{s}(\psi_{+}(x))= c_{++}(s)\psi_{+}(x^+(s))+c_{+-}(s)\psi_{-}(x^-(s))
\end{equation}
where the coefficients turn out to be, using $Y_{1}^{+}=-1/Y_{2}^{-}$ and \eqref{positive_chirality_eq},
\begin{align}
    &c_{++}(s)=i\frac{1+TX\,Y_{1}^{+}(s)}{\sqrt{(1+T^{2}X^{2})(1+(Y_{1}^{+}(s))^{2})}}\\
    &c_{+-}(s)=-i\frac{TX-Y_{1}^{+}(s)}{\sqrt{(1+T^{2}X^{2})(1+(Y_{1}^{+}(s))^{2})}}
\end{align}
yielding 
\begin{equation}
    |c_{++}(s)|^{2}+|c_{+-}(s)|^{2}=1
\end{equation}
Notice that $c_{+-}(s)$ vanishes at $s=0$ as expected.
\begin{figure}[H]
            \centering
            \includegraphics[width=0.45\linewidth]{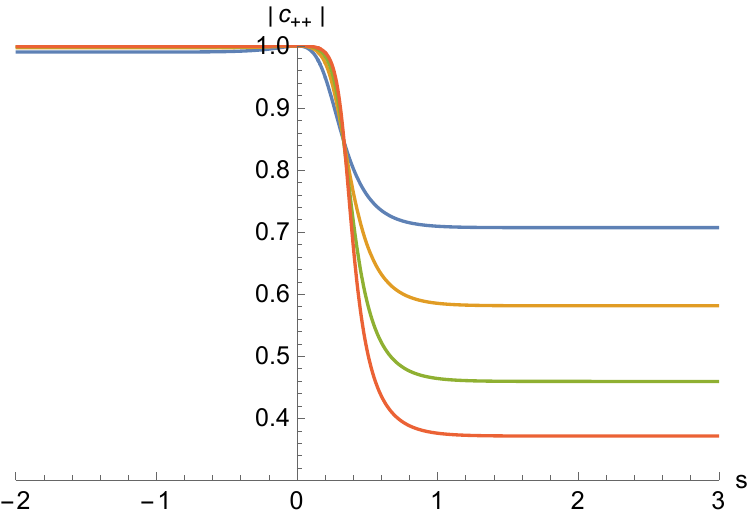}
            \includegraphics[width=0.45\linewidth]{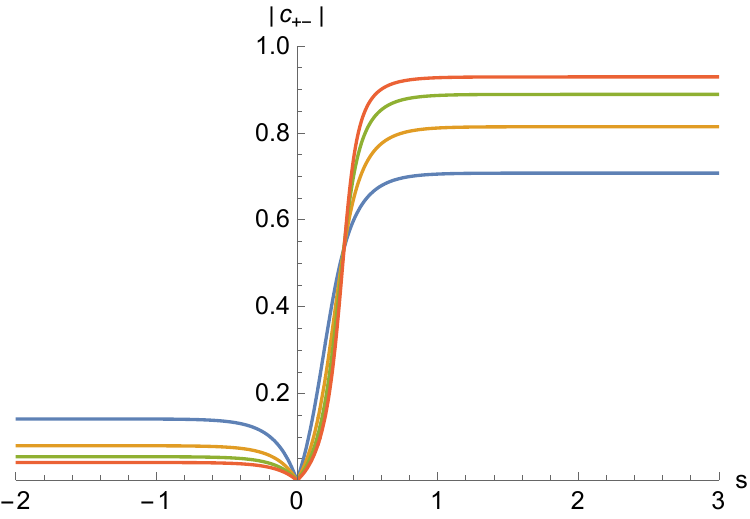}
            \caption{Plot of $|c_{++}(s)|$ and $|c_{+-}(s)|$ for $A=2$, $X=3$ and $T=1$ (blue), $T=2$ (yellow), $T=3$ (green), and $T=4$ (red).}
\end{figure}
We can also analyze the behavior of the two coefficients in the large $T$ limit, using the simplified expression for the images found previously.\\
For the case $1/A<X<A$, using \eqref{image1_late_time_inew}, \eqref{image2_late_time_inew}, the coefficients become
\begin{equation}
    |c_{++}(s)|\simeq
        1 \quad\quad\quad\quad\quad\quad |c_{+-}(s)|\simeq\mathcal{O}\left(\frac{1}{T}\right)
\end{equation}
thus, at late time, the modular flow of an incoming spinor will be fully described by an incoming spinor, and the same reasoning will hold if we consider the modular flow of an outgoing spinor $\psi_{-}(x^{-})$.\\  
For the case $X>A$, using \eqref{image1_late_time_outew}, \eqref{image2_late_time_outew}, we get
\begin{align}
    |c_{++}(s)|\simeq\begin{cases}
        \mathcal{O}\left(\frac{1}{T}\right)\quad\quad\quad &s\geq s_{\star}\\
        1 \quad\quad\;\;\, &s< s_{\star}
    \end{cases}\quad\quad\quad\quad |c_{+-}(s)|\simeq\begin{cases}
        1\quad\quad\quad &s\geq s_{\star}\\
        \mathcal{O}\left(\frac{1}{T}\right) \quad\quad\;\;\, &s< s_{\star}
    \end{cases}
\end{align}
while for the case $X<1/A$ the two branch swap consistently with the observation done around \eqref{image1_late_time_outew}, \eqref{image2_late_time_outew}. \\
It seems that the only way that a modular evolved left-moving spinor can be fully described by a left-moving one (excluding at $s=0$) or a right-moving one along its modular flow is considering a late-time regime. An analog reasoning holds if we consider the case of a modular evolved right-moving fermion $\psi_{-}(x^{-})$.\\
In order to show how the strength of the island image changes across the Page time, below we plot $|c_{+-}|$ for the case in which we modular flow a fermion in the semi-infinite bath and in the high-temperature limit $\beta\ll1$. 
\begin{figure}[H]
            \centering
            \includegraphics[width=0.6\linewidth]{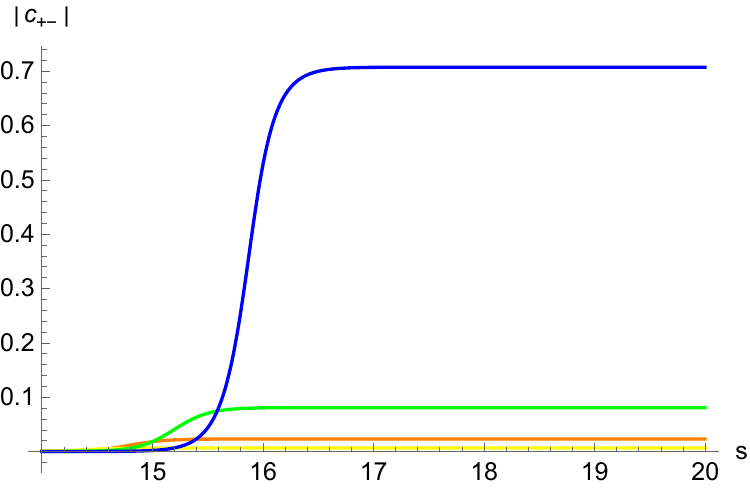}
            \caption{Plot of $|c_{+-}(s)|$ for $a=4$, $x=5$, $\beta=0.5$ and $t=3.6$ (yellow), $t=3.7$ (orange), $t=3.8$ (green), and $t=4$ (blue).}
\end{figure}


\end{document}